\documentclass[12pt,twoside]{amsart}

\usepackage{amsmath}
\usepackage{amssymb}
\usepackage{latexsym}
\usepackage{epsfig}
\usepackage{graphicx}
\usepackage{psfrag}

\newcommand{\proofbox}{\hfill{$\Box$}}

\newcommand{\mmltext}{} 
\newcommand{\mllltext}{}%
\newcommand{\mltext}{} 
 
\newcommand{\nnntext}{}

\newcommand{\weakto}{\rightharpoonup}

\newcommand{\HOX}[1]{}

\newtheorem{definition}{Definition}[section]
\newtheorem{theorem}[definition]{Theorem}
\newtheorem{lemma}[definition]{Lemma}

\newtheorem{proposition}[definition]{Proposition}
\newtheorem{corollary}[definition]{Corollary}

\newcommand{\glim}[1]{\underset{ #1}{\hbox{\rm  $\Gamma-$lim}\, }}
\newcommand{\spec}{\hbox{spec}\,}
\newcommand{\R}{{\mathbb R}}
\newcommand{\D}{{\mathbb D}}
\newcommand{\C}{{\mathbb C}}
\renewcommand{\S}{{\mathbb S}}
\newcommand{\N}{{\mathbb N}}
\renewcommand{\proofbox}{{$\Box$}}

\def\ttilde{}
\def\A{ {\mathcal A} }
\def\B{ {\mathcal B} }

\def\D{{\mathcal D}}
\def\hat{\widehat}
\def\tilde{\widetilde}
\def \bfo {\begin {eqnarray*} }
\def \efo {\end {eqnarray*} }
\def \ba {\begin {eqnarray*} }
\def \ea {\end {eqnarray*} }
\def \beq {\begin {eqnarray}}
\def \eeq {\end {eqnarray}}

\def \dist {\hbox{dist}}

\def \det {\hbox{det}}

\def \e {\varepsilon}
\def \p {\partial}
\def \a {\alpha}
\def\M{{\mathcal M}}

\def\Z{{\Bbb Z}}
\def\p{\partial}

\def\R{\mathbb R}
\def\la{{\lambda}}
\def\N{\mathbb N}
\def\ie{\emph{i.e.}}
\def\eg{\emph{e.g.}}
\def\cf{\emph{cf. }}
\def\etal{\emph{et al.}}

\title{Approximate quantum and acoustic cloaking}

\author[Greenleaf, Kurylev,  Lassas, Uhlmann]
{Allan Greenleaf,\, Yaroslav Kurylev, \\ Matti Lassas and
Gunther Uhlmann}

\date{}

\begin{document}
\maketitle

\begin{abstract}

For any $E\ge 0$,
we construct a sequence of bounded potentials
$V^E_{n},\, n\in\N$, supported in
{an annular region $B_{out}\setminus B_{inn}\subset\R^3$,}
which act as  approximate cloaks for solutions of Schr\"odinger's equation
at
energy $E$:
For any  potential $V_0\in L^\infty(B_{inn})$ {such that
$E$ is not a Neumann eigenvalue of $-\Delta+V_0$ in $B_{inn}$},
the scattering amplitudes $a_{V_0+V_n^E}(E,\theta,\omega)\to 0$ as
$n\to\infty$.
The $V^E_{ n}$ thus not only
form a family of approximately transparent potentials, but also function as
approximate invisibility cloaks
  in quantum mechanics.
{On the other hand, for $E$ close to interior eigenvalues,
resonances develop and there exist {\it almost trapped states} concentrated
in $B_{inn}$.} 
We derive the $V_n^E$  from singular, anisotropic transformation
optics-based cloaks  by a de-anisotropization procedure, which we call
\emph{isotropic
transformation optics}. This technique uses truncation, inverse
homogenization and
spectral theory to produce
nonsingular, isotropic approximate cloaks.
As an intermediate step, we also obtain approximate cloaking for a general
class
of equations including the acoustic equation.

\end{abstract}

\section{Introduction}\label{sec-intro}

A fundamental problem  is to describe the scattering
of waves by
a potential, as governed by the
time-independent Schr\"odinger equation at energy $E\ge 0$,
\beq\label{scattering}
& &(-\Delta +V(x) )\psi(x) =E\psi(x),\quad x\in \R^d,\\
& &\psi(x)=\exp(iE^{1/2}x\cdotp \theta)+\psi_{sc}(x),\nonumber
\eeq
where $\theta\in \mathbb S^{d-1}$ and $\psi_{sc}(x)$ satisfies
the Sommerfeld radiation condition.
  We restrict ourselves in this paper  to  compactly
supported  potentials
$V$,
so
\ba
\psi_{sc}(x)=  C_d E^{\frac{d-3}4} \frac{a_V(E,
\frac {x}{|x|},\theta)}{|x|^{\frac{d-1}2}}
\exp(iE^{1/2} |x|)+{o}\Big({|x|^{-\frac{d-1}{2}}}\Big),\quad
\hbox{as }|x|\to \infty.
\ea
The function $a_{V}(E,\omega,\theta)$ is the \emph{scattering amplitude}
at energy $E$
of the potential $V$. The associated
inverse scattering problem consists of trying to determine  $V$ from the
scattering amplitude, {or measurements of waves at the boundary of some
 region $\Omega$ containing the support of $V$.}

Recently, Zhang, \etal, \cite{Zhang}  described quantum
mechanical cloaking
at any fixed energy $E$. Their construction starts with a homogeneous,
isotropic mass
tensor and a potential $V\equiv 0$, and  subjects this pair to
a singular change of variables. This ``blowing up a point'' transformation
{had been} used in \cite{GLU2,GLU3} to produce conductivities that hide
objects
from detection by electrostatic measurements\footnote{The
2D version
 has the same electrostatic cloaking  property \cite{KSVW}.}, and was
subsequently used to describe the same phenomenon for electromagnetic waves
\cite{PSS};
one
now refers to  a specification of material parameters having this effect as
a \emph{cloak}.  The
cloaking Schr\"odinger equation in \cite{Zhang}, which has an anisotropic,
singular
mass tensor,  is  equivalent with the Helmholtz equation (at frequency
$\omega=\sqrt{E})$ for an associated singular Riemannian metric, and thus
covered
by cloaking for the Helmholtz equation in 3D, as we analyzed in
\cite[Sec. 3]{GKLU1}.
Similarly,  cloaks for acoustics in 3D have been described in
\cite{Ch3,Cu}\footnote{See \cite{CuSc} for the 2D case, and
\cite{MBW,Norris} for more regarding elastic and acoustic cloaking.}; again,
these
are in fact  direct consequences of cloaking for the Helmholtz equation in
3D,
\cf
\cite{GKLU5}.

Thus, for ideal 3D cloaking in each of scalar optics, quantum mechanics and
acoustics, one knows from \cite{GKLU1} that any  finite energy
distributional solution
decouples into
a sum of a wave on the exterior of the cloak,  unaffected (in terms
of
scattering or boundary measurements)  by the  cloak,  and a wave  within
the
cloaked region  satisfying the Neumann boundary condition at the cloaking
surface. Hence, if $E$ is
not a
Neumann eigenvalue, then the wave must vanish within the cloaked region and
cloaking works as advertised. On the other hand, if E is an eigenvalue, the
cloaked
region supports interior resonances, or {\it trapped states}. 
{This is an unphysical situation, since the Dirichlet problem on $\Omega$ 
no longer has unique solutions,
and this can be considered as a failure of cloaking, {\it per se}. 
However, what emerges from this failure  in the setting of approximate
cloaking 
described below is, we believe, quite useful.}

In this paper, we construct,  for each energy $E$, a
family
$\{V^E_{ n}\}_{n=1}^\infty$ of potentials, supported in
{an annulus $B_{out}\setminus B_{inn}\subset\R^3$,} which are
not only almost transparent, in that the scattering amplitudes
$a_{V^E_n}(E,\cdot,\cdot)\to 0$ as
$n \to\infty$, but also   act as  \emph{approximate
cloaks} for potentials supported in the inner ball $B_{inn}$:
for any potential $V_0\in L^\infty\left(B_{inn}\right)$
{for which $E$ is not an eigenvalue of $-\Delta+V_0$ in
$B_{inn}$}, the scattering
amplitudes
$a_{V_0+V^E_n}(E,\cdot,\cdot)\to 0$ as $ n \to\infty$, as well.
There are also approximate versions of the interior resonances supported by
the
ideal cloak: there exist energies close to the Neumann eigenvalues of the
cloaked
region $B_{inn}$,  near which there are waves largely concentrated in
$B_{inn}$, which
we call
{\it almost trapped states}.  Furthermore, the quality of the approximate
cloaking
degrades as we move towards these energies, with
{waves being able to penetrate $B_{inn}$, and } the presence of this
region
and $V_0$
detectable by scattering or boundary measurements. In addition, we are
able
to include a magnetic potential  in the Schr\"odinger equation; this allows
one to
switch between the approximate cloak and almost trapped state alternatives
by
application of a suitable homogeneous magnetic field. We have given physical
applications of this elsewhere \cite{GKLU6,GKLU7}.

To put these results in a mathematical context, recall that in  several
dimensions
numerous uniqueness results are available for the inverse scattering and
boundary
value problems, subject to some regularity assumptions,  in terms of the
scattering
data at a fixed energy
$E$ and the closely
related Dirichlet-to-Neumann (DN) operator.
{Physically, these correspond to far- and near-field measurements, resp.} In
fact,
for compactly supported potentials the
scattering data is equivalent to the DN operator measured on the boundary of
a
domain containing
the support of the potential \cite{Be}. In dimensions $d\ge 3$, Sylvester
and Uhlmann
\cite{SyU} showed that a smooth  potential  is determined by the
associated DN map, and this was extended to $L^\infty$ in \cite{NSU}.
Lavine and Nachman
\cite{LaNa} and Chanillo \cite{Chanillo} generalized this further, showing
that, if a potential $V\in L^{\frac{d}2}(\R^d)$ or  has small norm in
the
Fefferman-Phong space $F_p,\, p>\frac{d-1}2$, resp.,
then the DN map determines
$V$. For $d=2$, Nachman \cite{Nach2} proved  that uniqueness holds
if $V$ has
the special form $V=-\Delta({\sqrt{\gamma}})/{\sqrt{\gamma}}$ arising
from the
gauge transformation of the conductivity equation to the Schr\"odinger
equation, with
$\gamma\in
W^{2,p},\, p>1$, while  Sun and Uhlmann \cite{SuU}
showed  that uniqueness holds for a generic
class of potentials.
Very recently, Bukhgeim\cite{Buk} has shown that
uniqueness holds for all potentials in $L^p_{comp}(\R^2),\, p>2$.

On the other hand, for each positive energy $E$, Regge \cite{Re}
 constructed
(noncompactly supported) potentials
which are transparent, \ie, for which the scattering amplitude
$a_V(E,\theta,\omega)\equiv
a_0(E,\theta,\omega)=0$. See also \cite{New, Sab,GrinNov}, where the last
gives a
construction of transparent potentials in the Schwartz class
$\mathcal S(\R^2)$. In the physics literature, Hendi, Henn and
Leonhardt \cite{HHLe}
have
described central, \ie, radial,  potentials which are transparent on the
level of
the ray geometry.

We construct the families $\{V_{n}^E\}$ of approximately
cloaking potentials
by means of a result  of independent interest.
The use of changes of variables to produce novel optical effects on waves
or to facilitate computations has been considered in the physics literature,
(see\footnote{We thank A.
Kildishev for this reference.}, \eg, Dolin \cite{Dolin} or more recently
Ward and Pendry \cite{ward}), and is now generally
referred to as {\it transformation optics} (TO).
However, to produce
cloaking and other extreme effects, nonsingular changes of variables 
are insufficient.
The recently proposed plans for cloaking  are based on singular
transformations and consist of medium parameters which are
  both {\sl anisotropic} and {\sl singular}\footnote{By {\it singular},
 we mean that at least one of the
eigenvalues  goes to zero or infinity at some points.},
 whether for the conductivity (electrostatics)
\cite{GLU2,GLU3}; index of refraction
(Helmholtz) \cite{Le,GKLU1,KOVW};
permittivity and permeability (Maxwell) \cite{PSS,GKLU1}; mass density
(acoustic)
\cite{CuSc,Ch3,Cu,GKLU5}; or effective mass (Schr\"odinger)\cite{Zhang}.
Physical
realization
of such designs is now potentially feasible due to the rapidly developing
area of \emph{metamaterials},
but the singularity and extreme anisotropy make   characterizing and
fabricating the materials to implement such designs an enormous obstacle to
manufacturing invisibility devices, These same remarks are valid for other
TO designs, such as
\cite{LePhil,ChenChanRot,Luo,GKLU2,GKLU4,Rahm}, some of which are singular.

We propose here a  general
method, which we refer to by the oxymoronic {\sl isotropic transformation
optics},
for dealing with both the anisotropy and singularity of TO material
parameters.
We describe this in  detail in the context of cloaking, but it should be
applicable to a wider range of TO designs.
We in fact derive the quantum mechanical approximate cloaks
from approximate cloaks for a general class of equations that includes the
acoustic
equation.
Using ideas from Nguetseng \cite{Ngu}, Allaire \cite{Allaire}, Cherkaev
\cite{Cherka} and elsewhere,  we show how  to find
cloaking material parameters that are at once both  isotropic and
nonsingular, at the
price of replacing perfect (ideal) cloaking with  {\it approximate} cloaking
(of
arbitrary accuracy).
This  method, starting with  transformation optics-based
designs and constructing  approximations to them, first by {\it
nonsingular}, but still anisotropic,
material parameters, and
then by nonsingular {\it  isotropic}  parameters,  appears to be a very
flexible tool for
creating physically realistic theoretical designs, easier to  implement than
the ideal ones due
to the relatively tame nature of the materials needed,
yet (up to an arbitrarily small error) essentially capturing the desired
effects on wave
propagation.

We review the ideal electrostatic cloak of \cite{GLU2}, and  extend this in
Sec.
\ref{sec
ideal gen}  to a  class of equations which will allow us  to deal
with
both general acoustic  and magnetic Schr\"odinger equations. These equations
have
coefficients which, borrowing the terminology from acoustics, we refer to as
mass density and bulk modulus   and which are singular at
the cloaking surface 
$\Sigma$, \ie, the interface between the cloaked and uncloaked regions.
The desingularization process begins in Sec. \ref{sec trunc},
  by truncating the mass density  away from  $\Sigma$, on the outer side of
which the ideal
cloaking parameters are singular.
(Similar truncations have been considered
before in the
context of cylindrical or 2D cloaking, \cf \cite{RYNQ,GKLU3,Ch4,KSVW}.) We
show
in Sec. \ref{sec Gamma}
that the Dirichlet forms for the ideal cloaks can be well-approximated (in
the sense of
$\Gamma$-convergence) by these truncations.
We then
desingularize the bulk modulus in Sec. \ref{density}.

So far, the approximately cloaking mass densities are still
anisotropic. As is  well known
in effective medium theory, homogenization of isotropic
material parameters  may lead to anisotropic ones
\cite{Milt}; in Secs. \ref{sec-approx second} and
\ref{isotropic-to-singular}, we use this phenomenon in
reverse, showing
that the
Dirichlet forms obtained in Sec. \ref{sec-first approx}  can be
well-approximated by
those for certain nonsingular {\it isotropic} conductivities,  which thus
provide
approximate cloaks for the general class of acoustic-like equations.
These then allow us
 to obtain   in Thm. \ref{quantum-cloaking} families of approximate quantum
cloaks.
 In Sec.  \ref{sec exceptional} 
we study failure of cloaking near exceptional energies,
{mirroring the failure of the { existence of unique solutions
for the} 
 ideal cloak at Neumann eigenvalues of
the
cloaked region; see the Remark at the end of the section.
Further physical applications, including a new
type of ion trap,   can be found in  \cite{GKLU6,GKLU7}.
Finally, numerical simulations are presented in Sec. \ref{sec numerics}. }

AG and GU were supported by the US NSF, YK by EPSRC, ML by Academy of
Finland (CoE project 213476), and GU by a
Walker Family Professorship. The authors thank A. Cherkaev and
V. Smyshlyaev for useful discussions on homogenization.

\section{Approximating  cloaking material parameters by
nonsingular anisotropic
parameters}\label{sec-first approx}

{We will denote by  $B(a,R)$   the ball of radius $R$ centered at
$a$ in $ \R^3$, sometimes denoted simply $B(R)$  when centered at the
origin, ${\it O}$.}
{Let $M_1={\overline B}({\it O_1},3)\subset \R^3$ and $M_2={\overline
B}({\it
O_2},1)$, with $ {\it O_1}, \it {O_2}$ being  two copies of ${\it
O}$,    considered as disjoint compact manifolds with boundary; set
$M=M_1\cup M_2$.
Also, let
$\Omega={\overline B}(3)\subset
\R^3$,
and
$F^1:M_1\setminus
\{{\it O_1}\}
\to \Omega\setminus  \overline B(1)\subset\R^3$ be the map}
\beq \label{1a}
& &F^1(x)=x,\quad\hbox{for }2<|x|\le 3,\\ \nonumber
& &F^1(x)=\left(\frac {|x|}2+1\right)\frac x{|x|},\quad\hbox{for }
0<|x|\le 2.
\eeq
Define also $F^2: M_2 \to \overline B(1)$ as the identity map,
\beq \label{1d}
F^2(x)=x.
\eeq
Together, these form a surjective map $F=(F^1, F^2)$ from 
{the {\it cloaking manifold} (or virtual space) $M\setminus\{{\it
O_1}\}$ to the {\it cloaking device} (or physical space) 
$\Omega$.}
By a conductivity  we mean a measurable map with
values in the symmetric non-negative $\R^{3\times 3}$ matrices.
  Let $\gamma_0=1$ be the constant isotropic conductivity
on $\R^3$ and define the conductivity $\sigma_1$ on $\Omega$
as
\begin{equation} \label{ideal1}
\sigma_1=F^1_*\gamma_0, \quad x \in \Omega\setminus \overline B(1),
\qquad
\sigma_1=2 \,F^2_* \gamma_0, \quad x \in B(1),
\end{equation}
{which has a singularity on the {\it cloaking surface}  $\Sigma :=\p B(1)$,
both in that one of the eigenvalues (corresponding to the radial direction)
tends to
$ 0$ as
$r\searrow 1$ and that there is a jump discontinuity across the sphere
$\Sigma$.
This conductivity
$\sigma_1$ is, up to the radius of $\Omega$ and the factor
$2$ in the second of formulae  (\ref{ideal1}), used here for
technical reasons, the one introduced in
\cite{GLU2, GLU3} and shown
to be indistinguishable from $\gamma_0$, {\it vis-a-vis} electrostatic
boundary
measurements at $\p \Omega$. In fact, $\sigma_1|_{B(1)}$ can be
replaced by
\emph{any}
smooth, non-degenerate anisotropic conductivity tensor and its values will
be
undetectable at $\p \Omega$. The same construction of
$\sigma_1|_{\Omega\setminus B(1)}$, applied instead to the electric
permittivity $\epsilon$ and magnetic
permeability $\mu$ in
Maxwell's equations, was proposed in \cite{PSS} (see also \cite{Le}) to
{\it cloak} the region $B(1)$
from
observation by electromagnetic waves at positive frequency; we thus refer to
$\sigma_1$
as a \emph{cloaking} conductivity,}  and following the physics
literature, we will refer to
$(\Omega, \sigma_1)$ as the {\it ideal} cloak.

This gives rise to the Dirichlet problem for the singular conductivity
equation,
\beq \label{a1}
\nabla \cdot \sigma_1 \nabla u =0, \quad \hbox{in}\,\, \Omega,
\quad u|_{\p \Omega}=h,
\eeq
and to the corresponding singular conductivity operator
${\mathcal A}$,
\beq \label{a2}
{\mathcal A}u:= -g^{-1/2} \nabla \cdot \sigma_1 \nabla u,
\eeq
that we consider with Dirichlet boundary condition $u|_{\p \Omega}=0.$
{Here, we use the  singular Riemannian metric
$(g_{jk})_{j,k=1}^3$ 
associated to the conductivity $\sigma_1$, namely,
\beq \label{g-sigma}
g^{1/2} g^{jk}= \sigma_1^{jk}, \quad g =|\hbox{det}[g^{ij}]|^{-1}=
|\hbox{det}[\sigma_1^{ij}]|^2,
\eeq
\cf \cite{GLU2, GLU3}. We denote by $g$
both the metric and the corresponding scalar function, the  meaning being
clear from the
context.

A rigorous definition of the meaning of (\ref{a1})
and the operator ${\mathcal A}$ is given in the following sections.
In particular, the operator ${\mathcal A}$ is self-adjoint on
$L^2_g(\Omega)$, the weighted $L^2$-space
defined using the weight $g^{1/2}$.}
For a general weight $w(x) \geq 0$, we
denote by $L^2(\Omega, w dx)=L^2(w dx)$ the weighted
space,
\ba
L^2(\Omega, w dx)=\{u:\Omega\to \C \,
\hbox{ measurable}, \|u\|^2_{L^2(w dx)}:=\int_{\Omega}
 |u|^2\,w\, dx<\infty\}.
\ea
For simplicity, we denote $L^2_g=L^2_g(\Omega):=L^2(\Omega,g^\frac12 dx)$,
the natural
$L^2$ space
for the metric $g$,  and  the norm in
this space  by 
$\|\cdot\|_g$.
Note that
\beq \label{inclusion}
L^2(\Omega) \subset L^2_g(\Omega), \quad \|u\|_g \leq \sqrt 8\, \|u\|.
\eeq
We also use the Sobolev spaces
\ba
& &\hspace{-5mm}H^1_g(\Omega)=\big\{u\in L^2_g(\Omega) 
:\ u|_{\Omega\setminus \Sigma}\in H^1_{loc}(\Omega\setminus \Sigma),\
\int_{\Omega\setminus\Sigma}
\sigma_1^{jk}\p_ju\,\overline {\p_k u}\,dx<\infty\big\},\\
& &\hspace{-5mm}H^1_{0,g}(\Omega) =\{v \in H^1_{g}(\Omega):\, v|_{\p \Omega}=0\}.
\ea
{Here and below, we  use Einstein summation convention, 
summing over indices $j$ and $k$ appearing both as sub- and super-indices.}
Observe that $H^1(\Omega) \subset H^1_g(\Omega)$ and 
\beq \label{inclusionH}
\|u\|_{H^1_g}
=(\|u\|_{L^2_g}^2+\int_{\Omega}\sigma_1^{jk}\p_ju\,\overline {\p_k u}\,dx)^{\frac 12}
 \leq \sqrt 8\, \|u\|_{H^1}.
\eeq
{Throughout, we also use the following standard  terminology: 
when considering convergence of sequences $\{x_n\}_{n=1}^\infty$ in a
Hilbert space
$H$,
we say that $x_n$ converges strongly to $x$ in $H$, 
{if $\|x_n-x\|_H \to 0$, as $n\to \infty$,  while $x_n$ converges weekly to
$x$, 
if $\left(x_n-x, y\right)_H \to 0$, as $n\to \infty$, for any $y \in H$. }

\subsection{Ideal cloaking for more general equations}\label{sec ideal gen}

In this paper we treat  equations  more general than (\ref{a1}), which are
important for physical applications,
cf. \cite{GKLU6,GKLU7}. To this end,  consider a family of equations that
simultaneously
includes both the magnetic Schr\"odinger equation  and the acoustic
equation; for simplicity, we will refer to equations of this type as {\it
acoustic}. Let $q$ and
$b=(b_1, b_2, b_3)$ be 
{scalar- and vector-valued functions, which eventually will
represent the 
electric and magnetic potentials;}
we will assume that
\bfo
b \in L^\infty(\Omega; \R^3), \quad q \in L^\infty(\Omega; \R), \quad q \geq
0.
\efo
We note that the condition $q \geq 0$ is  merely a convenience, since for
general  $ q \in L^\infty(\Omega; \R)$ we can always add  a constant to
achieve positivity,
and of course this just shifts the spectrum.

To deal rigorously with the elliptic boundary value problem
\beq
\label{a4}
- g^{-1/2} (\nabla+ib)\cdot \sigma_1 (\nabla+ib)u + q u -\la u =f,
\quad  u|_{\p \Omega}=h,
\eeq
which has singular coefficients, we consider the corresponding quadratic
form.
In the following, we use the notation
\ba
\nabla^b=\nabla+ib(x).
\ea

\begin{lemma}\label{lem:closed form}
The quadratic form
\beq \label{a5}
a_1[u, u]=  \int_{\Omega \setminus \Sigma} \sigma_1 \nabla^b u \cdot
{\overline{\nabla^b u}} dx+ \int_{\Omega \setminus \Sigma} q g^{1/2}|u|^2\,
dx,
\eeq
defined in the domain ${\mathcal D}(a_1)=H^1_{0, g}(\Omega)$ is closed.
Moreover,
the embedding  ${\mathcal D}(a_1)\hookrightarrow L^2_g(\Omega)$ is compact,
and  there is  $C_b>1$ so  that
\beq \label{a17}
C_b^{-1}\|u\|^2_{H^1_{0,g}(\Omega)} \leq \|u\|_g^2+a_1[u, u]
\leq
C_b\|u\|^2_{H^1_{0,g}(\Omega)}.
\eeq
\end{lemma}

{\bf Proof.}
To prove the assertion, we need to show two facts. First, we need to prove
that
 $a_1[u, u] < \infty$ for $u\in H^1_{0, g}(\Omega)$.
Second, we have to prove the closedness of the form $a_1$
on $H^1_{0. g}(\Omega)$.

We start with the fact \cite{GKLU1} that
the map $F_*$ is unitary from  $L^2\left(M_1\right)
\oplus
L^2(M_2)$ to $L^2_g(\Omega)$ and from
$H^1_0(M_1)  \oplus
H^1(M_2)$ to $H^1_{0, g}(\Omega)$, respectively.
For $v=(v_1,\,v_2)$ and $u=F_*v$,
we have
\beq \label{a6}
& &(v_1,\,v_2) \in L^2\left(M_1\right) \oplus L^2(M_2) \quad \hbox{iff}\,\,
u \in L^2_g(\Omega);
\\ \nonumber
& &
(v_1,\,v_2) \in H^1_0(M_1)  \oplus
H^1(M_2) \quad \hbox{iff}\,\, u \in H^1_{0, g}(\Omega).
\eeq
Hence, as
$H^1(M_j)$, $j=1,2$ is compactly embedded into $L^2(M_j)$,
we see that the space
$H^1_g(\Omega)$ is compactly embedded into $L^2_g(\Omega)$.

Using  definition (\ref{ideal1}) and the transformation rule for $1-$forms,
we see that
\beq \label{a7}
& &a_1[u, u]= \a[v, v]= \a^1[v_1, v_1]+\a^2[v_2, v_2];
\\ \nonumber
& &
\a^1[v_1, v_1]= \int_{M_1}|\nabla v_1 +i \beta_1\, v_1|^2dx+
\int_{M_1}\kappa_1\, |v_1|^2 dx,
\\ \nonumber
&&
\a^2[v_2, v_2]= 2\int_{M_2}|\nabla v_2 +i \beta_2\, v_2|^2dx+
8\int_{M_2}\kappa_2\,|v_2|^2 dx.
\eeq
Here the $1-$forms $\beta_1=\beta|_{M_1}$, $\beta_2=\beta|_
{M_2}$, and functions $\kappa_1=\kappa|_{M_1}$, $\kappa_2=\kappa|_{M_2}$,
are given by {$b= F_*\beta$, $q=F_* \kappa$, that is,}
\ba
b_j(x)= \frac {\p F^k}{\p y^j}(y)\beta_k(y), \quad q(x)=\kappa(y),
\quad x=F(y)\in \Omega\setminus \Sigma.
\ea
It  follows from (\ref{1a}), (\ref{1d}) that $\beta_2,\kappa_1$ and
$\kappa_2$ are bounded, but
$\beta_1$ has a singularity at $x=0$, of the order $1/|x|$, and
\beq \label{a9}
|\beta_1(x)| \leq \frac{C \|b\|_{L^\infty}}{|x|}.
\eeq

Let us consider $\a^2[v_2, v_2]$
as an unbounded non-negative quadratic form in $L^2(M_2)$,
with domain $\D(\a^2)=H^1(M_2)$. Then $\a^2$ is closed.
The quadratic form $\a^1[v_1, v_1]$ requires further
analysis. We consider  $\a^1[v_1, v_1]$
as an unbounded non-negative quadratic form in $L^2(M_1)$
having the domain
\beq \label{a10}
\quad \quad {\mathcal D}(\a^1) =\{v_1 \in L^2(M_1):\, \nabla v_1 +i \beta_1 v_1 \in
L^2(M_1),\,
v_1|_{\p M_1}=0\}.
\eeq
Note that the condition $\nabla v_1 +i \beta_1 v_1 \in L^2(M_1)$ implies
that the trace $v_1|_{\p M_1}$ is well-defined.
By \cite{LeSi,simon}, $C^\infty_0(M_1)$ is dense on
${\mathcal D}(\a^1)$ and $\a^1$ on ${\mathcal D}(\a^1)$
is a closed, non-negative quadratic form.

By Hardy's inequality \cite{Kufner}, it follows from (\ref{a9}) that
\beq
\label{a11}
\|\beta_1 v_1\|_{L^2(M_1)} \leq C \|b\|_{L^\infty} \, \|v_1\|_{H^1(M_1)},
\eeq
so that 
\beq
\label{14.09.1 A}
H^1_0(M_1)  \subset {\mathcal D}(\a^1)\quad\hbox{is dense}.
\eeq

Let $v_1 \in {\mathcal D}(\a^1)$. Then,
by \cite{LeSi}
$|v_1| \in H^1_0(M_1)$.
Using  (\ref{a11}) again, we obtain
\bfo
\|\beta_1 v_1\|_{L^2(M_1)}= \|\, \beta_1 |v_1|\, \|_{L^2(M_1)} < \infty.
\efo

Thus, by definition (\ref{a10}), this yields that  $\nabla v_1 \in L^2
(M_1)$,
and hence $v_1 \in H^1_0(M_1)$.
{This shows that
$\D(\a^1) \subset H^1_0(M_1)$
{which, together with (\ref{14.09.1 A}), implies that
$\D(\a^1)= H^1_0(M_1)$}. Thus,  the domain of the closed
form $\alpha$ is $\D(\a)= H^1_0(M_1)\oplus H^1(M_2)$. 
Using the transformation rule (\ref{a7}) for $u=F_*v$, and  (\ref{a6}), 
we conclude that
$a_1[u, u] < \infty$ for all $u\in H^1_{0, g}(\Omega)$ and that the
quadratic form
$a_1$ with domain ${\mathcal D}(a_1)= H^1_{0, g}(\Omega)$ is closed.} 
The inequality (\ref{a17}) follows
from this by the open mapping theorem.
\hfill \hfill\proofbox

By the theory of quadratic forms \cite{Kato},
the closed quadratic form $a_1[u, u]$ defines
an associated  non-negative self-adjoint operator
$\A_1$ in $L^2_g(\Omega)$, 
having domain
\beq
\label{a13}
{\mathcal D}(\A_1):= \{u\in H^1_{0, g}(\Omega)&:&\, \hbox{there is $f\in
L^2_g(\Omega)$ such that}\\
& & \nonumber
a_1[u, v]=(f, v)_g
\,\, \hbox{for}\,\, v \in H^1_{0, g}(\Omega)\},
\eeq
that is, for $u\in {\mathcal D}(A_1)$ we have, for all $v\in H^1_{0,g}
(\Omega)$,
\beq \label{a14}
 \int_{\Omega \setminus \Sigma} \sigma_1 \nabla^b u \cdot
{\overline{\nabla^b v}} dx+
\int_{\Omega \setminus \Sigma} q g^{1/2}u \overline{v}\, dx
=  \int_{\Omega \setminus \Sigma} g^{1/2} f \overline{v}\, dx
\eeq with
$\A_1 u =f\in L^2_g(\Omega)$.
{By Lemma \ref{lem:closed form}, the spectrum of $\A_1$
consists of discrete eigenvalues with finite dimensional eigenspaces.}

Observe that, formally integrating by parts in (\ref{a14}),
we come to  (\ref{a4}) with $h=0$.
To  better understand the nature of the operator ${\mathcal A_1}$, we give
also  an alternative definition in the case when $b \in C(\Omega; \R^3)$.
In this case, we use the fact that the map
$D_\sigma^j:\phi\mapsto \sigma_1^{jk} (\p_k \phi +i b_k \phi)$, defined
initially for
$\phi\in C^\infty_0(\Omega)$,
has a bounded extension
\ba
D^j_{\sigma}:H^1_{0,g}(\Omega)\to \M(\Omega;\R^3),
\ea
where $ \M(\Omega;\R)$ denotes the space of   Borel
measures on $\Omega$, \cf \cite[Lem.\ 3.2]{GKLU1}.
For $b \in C(\Omega; \R^3)$, an 
equivalent definition of the operator
${\mathcal A}_1$ is then
\beq
\label{Laplace-type-1}
& &{\ttilde {\mathcal A}}_1(u):= -{g}^{-1/2} (\p_j+ib_j) D_\sigma^j u+qu,
\\ \nonumber
& &
{\mathcal D}({\ttilde {\mathcal A}}_1)=\{u \in H^1_{0, g}(\Omega): \, \,
(\p_j+ib_j) D_\sigma^j u\in L^2(\Omega, g^{-1/2} dx)
 \}.
\eeq

\subsection{Approximate cloaking by truncation}\label{sec trunc}

For any $1<R<2$, consider the nonsingular truncations of (\ref{a4}),
\beq\label{eq: conductivity eq in B(0,R)}
-  \nabla^b\cdotp\,\sigma_R \nabla^b u +q g^{1/2}u -
\la g^{1/2}u&=&0 \quad \hbox{in }\Omega,
\\ u|_{\p \Omega}=h, \nonumber
\eeq
where  $\sigma_R$ are measurable anisotropic conductivities in $\Omega$
satisfying
\beq\label{eq: sigma_R conditions}
& &\sigma_R|_{\Omega\setminus B(0,5/2)}=\gamma_0,\quad
\sigma_R|_{B(0,1)}=2\gamma_0,\\
\nonumber
& &\lim_{R \searrow 1}\sigma_R(x)=\sigma(x),
\quad c_1 (R-1)\gamma_0\leq \sigma_R(x)\leq c_2 \gamma_0, 
\quad\hbox{for }x\in \Omega,
\\
\nonumber
& &
\sigma_{R_2}(x)\leq \sigma_{R_1}(x),\quad \hbox{for }R_2\geq R_1,
\eeq
{for some $c_1 \leq 1/2, \, c_2 \geq 2$.}
For instance, we can choose
\beq \label{R-ideal}
\sigma_R(x)=
\begin{cases}F_*\gamma_0, \quad x \in \Omega \setminus \overline B(R),
\\
2 \gamma_0, \quad x \in B(R).
\end{cases}
\eeq
We note  that, by smoothing the conductivities (\ref{R-ideal}),
it is possible to construct conductivities $\sigma_R\in C^2(\overline
\Omega;\R^2)$ , {which we use in Sec. \ref{schrodinger},}
satisfying (\ref{eq: sigma_R conditions}).

{We denote the solution of (\ref{eq: conductivity eq in B(0,R)})
by $u=u^h_R$.}
Note that, for $b=0,\,q=0$,  equation (\ref{eq: conductivity eq in B(0,R)})
is 
an {\it acoustic}
equation
with  {\it mass density} $\sigma_R$ and {\it bulk modulus}
  $g^{1/2}$  with $k=\sqrt{\la}$;
in the quantum mechanical setting, $\sigma_R$
corresponds to the
effective mass
and $(q-\la) g^{1/2}$ to the potential.
By abuse of notation, even for $b\ne 0$ and $q\not =0$, we will
refer to
$\sigma_R$
as the mass density.

 Observe that, for each $R>1$, the mass density $\sigma_R$ is nonsingular,
\ie,
is bounded from above and below with, however, the lower bound going to $0$
as
$R \searrow 1$. Moreover, for any
$x\in \R^3$, the symmetric
matrix valued function $R\mapsto \sigma_R(x)$ is increasing
\footnote{Note that, due to the behavior of the eigenvalues of $\sigma_1$,
this simple but important monotonicity property
fails in
the 2D case, which we are not treating.} as a function of
$R$, and
therefore decreases as $R \searrow 1$. Nonsingular regularizations or
truncations of singular
ideal cloaks have  previously been considered in \cite{RYNQ,GKLU3,KSVW}.

{To motivate the treatment here,  consider for $R>1$} the
Dirichlet-to-Neumann (DN) map
  $\Lambda_R^\la:H^{1/2}(\p \Omega)\to H^{-1/2}(\p \Omega)$ 
that maps
\beq\label{Dirichlet-to-Neumann map}
\Lambda_R^\la:u|_{\p \Omega}\mapsto (\p_\nu u +i\nu \cdot b u)|_{\p \Omega},
\eeq
where $u$ solves (\ref{eq: conductivity eq in B(0,R)}).
The DN  map corresponds to the Dirichlet-to-Neumann
quadratic form, which by abuse of notation is denoted also by
$\Lambda_R^\la$,
\beq
\label{DN_map}
\Lambda_R^\la[h]=\int_{\p \Omega}(\Lambda_R^\la h)(x)
\overline{h(x)}\,dS(x),
\eeq
where we denote $h=u|_{\p \Omega}\in H^{1/2}(\p \Omega)$;
for $\la\leq 0$, 
the Dirichlet-to-Neumann form}
{may be also represented as}
\beq\label{D-to-N}
\Lambda_{R}^\la[h]=\inf \left(a_R[u,u]-\la\|u\|^2_g \right),
\eeq
where infimum is taken over all $u\in H^1(\Omega)$ with $u|_{\p \Omega}=h$.
{However, to treat general $\la$, and  the general class of equations 
(\ref{eq: conductivity eq in B(0,R)}), we will use the definition
(\ref{29.09.1}) below.}

{Returning to equation (\ref{eq: conductivity eq in B(0,R)}), note
that for $\la<0$ and $R>1$ the solution can be 
obtained 
from
the minimization problem for the quadratic functional associated to 
the sesquilinear functional
\beq\label{05.10.1}
\quad \quad a_{R}[u,v]=
\int_{\Omega}\sigma_R(x) \nabla^b u(x)\cdotp
\overline{\nabla^b v(x)}
\,dx +\int_{\Omega} q(x) g^{1/2}(x) u(x)\overline{v(x)}\, dx\hspace{-1cm}
\eeq
Moreover, we have
\ba
u^h_R=
{\hbox{argmin}}\left( \,a_R[u,u] -\la\|u\|^2_g\,
\right),
\ea
where minimization is taken over $u\in H^1(\Omega)$ such that $u|_{\p
\Omega}=h$.
Observe that the DN form (\ref{D-to-N}) is related to $a_R$ defined by
(\ref{05.10.1}).}

Thus, the solution $u^h_R$ of the acoustic equation
(\ref{eq: conductivity eq in B(0,R)}) with the
Dirichlet boundary value $h$ is the minimizer of $a_R[u,u]-\la\|u\|^2_g$.

{Now consider  the quadratic form $a_{R}[u,u]$ restricted to $u \in
H^1_0(\Omega)$.

\begin{lemma} \label{a_R_closed}
 For $R>1$, the sesquilinear form (\ref{05.10.1})
with domain ${\mathcal D}(a_R)= H^1_0(\Omega)$, is closed
 {and positive definite} on $L^2_g(\Omega)$.
\end{lemma}

{\bf Proof.} {{Clearly $a_R$ is non-negative.} The fact that $a_R$ is closed
on ${\mathcal D}_R=H^1_0(\Omega)$
follows from the same considerations as those in Lemma \ref{lem:closed
form}. 
Next we show that $\nabla^b u=0$ iff $u=0$
which, due to the compactness of the embedding $H^1_0(\Omega)
\hookrightarrow L^2_g$,
implies
the positive-definiteness of $a_R$.
Assume, on the contrary, that there is
a non-zero $\psi \in H^1_0(\Omega)$ such that
$\nabla^b \psi=0$ in $\Omega$.
Continue $b$ and  $\psi$ by $0$ to  $\R^3\setminus \Omega$.
and also extend $\sigma_R$
to  $\R^3\setminus \Omega$ as $\gamma_0=1$ in $\R^3\setminus B(3)$.}
Then 
$\psi \in H^1(\R^3)$ and $\nabla^b\psi=0$ in $\R^3$, and hence
$\nabla^b \cdot \nabla^b\psi=0$  in $\R^3\setminus \{{\it O}\}$.
Using unique continuation for elliptic equations with non-smooth first order
terms,
 see \cite{Aron,KochTat}, we deduce that $\psi=0$ in  $\R^3\setminus \{{\it
 O}\}$.
\hfill \hfill\proofbox}

{As $\sigma_R$ is bounded from below, 
Lemma \ref{a_R_closed} implies that  there is $c_R>0$ such that}
\beq \label{1.06.11}
  \int_\Omega  \sigma_R \nabla^b u \cdot 
{\overline {\nabla^b u}}\, dx \geq c_R \|u\|^2_{L^2(\Omega)},\quad
  \hbox{ $u\in H^1_0(\Omega)$.}
\eeq

Similarly to (\ref{a13}), (\ref{a14}), 
the sesquilinear form $a_R$ defines a self-adjoint operator 
${\mathcal A}_R$ in $L^2_g$, and, analogously to (\ref{Laplace-type-1}), we
have
\beq \label{AR-operator}
& &{\ttilde {\mathcal A}}_R(u):= -{g}^{-1/2} \nabla^b \cdot( \sigma_R
\nabla^b u)+qu,
\\ \nonumber
& &
{\mathcal D}({\ttilde {\mathcal A}}_R)=\{u \in H^1_{0}(\Omega): \, \,
\nabla^b \cdot( \sigma_R \nabla^b u) \in L^2(\Omega, g^{-1/2} dx)
   \}.
\eeq

Using the operator ${\ttilde {\mathcal A}}_R$, $R>1$ we see that, for 
$\la \notin \hbox{spec}({\ttilde {\mathcal A}}_R)$,
\beq \label{29.09.1}
& &
\Lambda_R^\la h=(\p_\nu+i \nu \cdot b)\left(u_0+v^\la_R\right)|_{\p \Omega};
\\ \nonumber
& &
v^\la_R:=  {\mathcal R}_R(\la)\left( \nabla^b
\cdot(\sigma_R \nabla^b u_0)
-q u_0 +\la u_0\right).
\eeq
Here $u_0\in H^1(\Omega)$ with $ \hbox{supp}(u_0) \subset \Omega \setminus
{\overline
{B({\it O}, 2)}}$ satisfies $u_0|_{\p \Omega}=h$, while ${\mathcal
R}_R(\la)$ is the
resolvent,
\beq\label{eq: resolvent}
{\mathcal R}_R(\la)=({\ttilde {\mathcal A}}_R-\la I)^{-1}:\, H^{-1}(\Omega)
\to L^2(\Omega).
\eeq
We will use the notation (\ref {eq: resolvent}) for $R=1$ as well as  $R>
1$,
{in which case ${\mathcal R}_1(\la):\, H^{-1}_g(\Omega) \to L^2_g(\Omega).$
Note that the right-hand side in (\ref{29.09.1}) should be understood in the
sense of the pairing
$H^{-1/2}(\p \Omega) \times H^{-1/2}(\p \Omega)$,
\ba
\int_{\p \Omega} ((\p_\nu+i \nu \cdot b)(u_0+v^\la_R))  {\overline \psi}\,
dS=
a_R[u_0+v^\la_R,\, \psi] - \la (u_0+v^\la_R, \psi)_{L^2_g}, 
\ea
for all $\psi \in H^1(\Omega),$
and we again consider $a_R$ on the whole of  $H^1(\Omega)$.
}

{Next,  consider the DN map in the ideal case $R=1$.
Following \cite{GKLU1}, we say that $u$ is a {\it finite energy solution} of
the boundary value problem (\ref{a4})
with $h\in H^{1/2}(\p \Omega)$ and $f\in H^{-1}_g(\Omega)$,  if
$u\in H^1_{g}(\Omega)$ and
\beq \label{form equation}\quad\quad
a_1[u-u_0, \phi]=-\int_{\Omega}(
\sigma_1 \nabla^b u_0 \cdot
{\overline{\nabla^b \phi}}+(q g^{1/2}u_0-\la
g^{1/2}u-f)\overline \phi\,
dx, \hspace{-1cm}
\eeq
for every $\phi\in  H^1_{0,g}(\Omega)$.
Here $H^{-1}_{g}(\Omega)$ is the dual space to $H^1_{0,
g}(\Omega)$.

On the other hand,
\beq \label{push forward of equations 1}
& &-(\nabla+i\beta_1)\,\cdotp (\nabla+i\beta_1)v_1
+\kappa_1v_1-\la v_1=\tilde
f_1,\quad\hbox{in }M_1,\\
\nonumber
& &\quad \quad v_1|_{\p M_1}=h,\\
  \label{push forward of equations 3}
& &-\frac14 (\nabla+i\beta_2)\,\cdotp (\nabla+i\beta_2)v_2 +\kappa_2 v_2
-\la v_2=\tilde
f_2,\quad\hbox{in }M_2,\\
\nonumber
& &\quad \quad (\p_\nu +i\nu\cdotp \beta_{2})v_2|_{\p M_2}=0,
\eeq
are satisfied in the {\it weak sense} if
\beq \label{form equation 2}
  \a[v-v_0, \psi]=-\sum_{j=1}^2  \int_{M_j}(
 c_j\nabla^\beta v_0 \cdot
{\overline{\nabla^\beta \psi}} +(\kappa_j u_0 -\la v-\tilde f_j)
\overline \psi)\, dx, \hspace{-1.5cm}
\eeq
for all $\psi\in H^1_0(M_1)\oplus  H^1(M_2)$,
{where $c_1=1,\, c_2 = \frac14$.}  Here $v_0\in H^1(M_1)\oplus
H^1(M_2)$ is supported in $M_1\setminus {\overline {B( 2)}}$
and satisfies $v_0|_{\p M_1}=h$.

\begin{lemma}\label{lem: DN equivalence}
Let  $h\in H^{1/2}(\p \Omega)$ and $f\in H^{-1}_g(\Omega)$.
A function $u\in H^1_{g}(\Omega)$
is a finite energy solution in the sense (\ref{form equation})  of the
boundary
value problem 
(\ref{a4}) if and only if
$v=(v_1,v_2)\in H^1(M_1)\oplus  H^1(M_2)$, $u=F_*v$ satisfies
  equations (\ref{push forward of equations 1}) and 
(\ref{push forward of equations 3}) in the weak sense (\ref{form equation
2})
with $f=F_*(\tilde f_1,\tilde f_2)$.

In
particular, for any $\la \in \R$
the Cauchy data, on $\p \Omega$,  of solutions  to (\ref{eq: conductivity eq
in B(0,R)})
satisfy
\beq\label{equiv. formula}
& &\{(u|_{\p \Omega}, (\p_\nu +i\nu \cdotp b)u|_{\p
\Omega}):\hspace{-.1cm}
-\nabla^b \cdotp \sigma_1 \nabla^bu+qg^{1/2}u=g^{1/2}\la u\hbox{ in
}\Omega\}\\  \nonumber
&= &
\{(v_1|_{\p \Omega}, (\p_\nu +i\nu \cdotp \beta_1)v_1|_{\p
\Omega}):\hspace{-.1cm}
-\nabla^{\beta_1}\cdotp  \nabla^{\beta_1}v_1+\kappa_1 v_1=\la v_1\hbox{ in
}\Omega\}
\\ \nonumber
& & \subset H^{1/2}(\p \Omega) \times H^{-1/2}(\p \Omega) \nonumber
\eeq
\end{lemma}

{\bf Proof.} 
By Lemma \ref{lem:closed form} and formulae (\ref{a6}) and (\ref{a7})
we see that $u\in \D(a_1)$ if and only if $u=F_*v,\, v=(v_1, v_2)\in
\D(\a)$,  and
\beq \label{a7 copy}
a_1[\tilde \psi, \tilde \phi]= \a[\psi,  \phi]\quad\hbox{for } \phi,\psi\in
\D(\a), \tilde 
\phi=F_*\phi,
\tilde \psi=F_*\psi.
\eeq
Let $v_0\in H^1(M_1)\oplus
H^1(M_2)$ be supported in $M_1\setminus {\overline {B( 2)}}$
and satisfy $v_0|_{\p M_1}=h$, and $u_0\in  H^1(\Omega)$ be such that
$u_0=F_*v_0$.
Using formula  (\ref{a7 copy}) with $\tilde \psi=u-u_0$ and $\psi=v-v_0$,
we see that
$u\in H^1_{g}(\Omega)$ is a finite energy solution if and only if
$v=(v_1,v_2)\in H^1(M_1)\oplus  H^1(M_2)$, $u=F_*v$ satisfies
(\ref{form equation 2}) 
for all $\phi\in H^1_0(M_1)\oplus  H^1(M_2)$, that is,
$v$ satisfies equations (\ref{push forward of equations 1}) and 
(\ref{push forward of equations 3}) in the weak sense.
\hfill \hfill\proofbox\smallskip

Assume next that $\la\not\in \hbox{spec}(\A_1)$. 
{Then the solution to equations (\ref{a4}), in
  the sense of definition
(\ref{form equation 2}),  may be found in terms of the resolvent
  ${\mathcal R}_1(\la)$ of ${\mathcal A}_1$, \cf (\ref{eq: resolvent}).
  Indeed,
 comparing 
(\ref{a13}), (\ref{a14}) with equation (\ref{form equation}), we see that
its solution $u$ has the form
  \beq \label{29.09.2}
& &u=u_0+{\mathcal R}_1(\la)\left(f +
  \nabla^b \cdot \sigma_1 \nabla^b u_0-qu_0+\la u_0
\right),
  \\ \nonumber
& &  u_0|_{\p \Omega}=h, \,\, \hbox{supp}(u_0) \subset \Omega \setminus
  {\overline {B( 2)}},
\eeq
  at least when $f
 \in
L^2_g(\Omega)$ and $h \in H^{3/2}(\p \Omega)$ so that  $u_0 \in
H^2(\Omega)$.
 Since ${\mathcal D}(a_1)= {\mathcal D}({\mathcal A}_1^{1/2})=H^1_{0,
  g}(\Omega)$,
we see that $H^{-1}_{ g}(\Omega)=  {\mathcal D}({\mathcal A}_1^{-1/2})$.
  Therefore, the operator ${\mathcal R}_1(\la)$ can be extended by
  continuity
 to a bounded operator
from $H^{-1}_{ g}(\Omega)$ onto $H^{1}_{0, g}(\Omega)$. This makes it
possible to generalize
  (\ref{29.09.2}) for all $u_0 \in H^1_g(\Omega)$, i.e. $h \in H^{1/2}(\p
 \Omega)$,
and $f \in H^{-1}_g(\Omega)$.}
{Observe that the right-hand side in (\ref{equiv. formula}) is  related to
the 
unbounded selfadjoint operator
 $A_{out}$ in $L^2(\Omega)$, $\,{\mathcal D}(A_{out})
\subset H^1_0(\Omega)$, 
associated with the form $\a_1$, see (\ref{a7}) where we use $\Omega=M_1$.
More precisely, this operator is the unbounded selfadjoint operator 
in $L^2(\Omega)$ given by
\beq \label{A_out}
A_{out}&=&-\nabla^{\beta_1} \cdot  \nabla^{\beta_1}+\kappa_1, \\
\D(A_{out})&= &\{v_1\in H^1(\Omega):\
\nabla^{\beta_1} \cdot \nabla^{\beta_1} v_1\in L^2(\Omega),\ 
 v_1|_{\p \Omega}=0\};\nonumber
\eeq
Moreover, when $b\in C^1(\overline\Omega;\R^3)$, the selfajdoint
operator associated to the form $\a_2$ on $B(1)=M_2$ is the operator
\beq \label{A_in}
A_{in}&=&-\frac 14\nabla^{\beta_2} \cdot  \nabla^{\beta_2}+\kappa_2,
 \\
\D(A_{in})&=&\{v_2\in H^2(B(1)):\ \p_\nu v|_{\p B(1)}=0\}.\nonumber
\eeq

When $\la \notin \hbox{spec}(A_{out})$, the set}
(\ref{equiv. formula})  coincides with the graph of the DN-map
\beq\label{Dirichlet-to-Neumann map R=1}
\Lambda_{out}^\la:v_1|_{\p \Omega}\mapsto (\p_\nu +i\nu\cdotp \beta_1
)v_1|_{\p \Omega},
\eeq
where $v_1$ solves equation (\ref{push forward of equations 1}) with $\tilde
f_1=0$.

Note that if $b=0,\, q=0$, then $\lambda=0$ is an eigenvalue of $\A_1$
with the corresponding eigenfunctions of the form (\cf \cite{GLU3,GKLU1})
\beq\label{eq: interior eigenstate}
u(x)=
\left\{\begin{array}{cl} 0,& \quad \hbox{for }x \in \Omega \setminus
\overline B(1),
\\
c_0& \quad \hbox{for } \quad x \in B(1),
\end{array}\right.\quad c_0 \neq 0.
  \eeq
However, as follows from Lemma \ref{lem: DN equivalence} even,
in this case the Cauchy data on $\p \Omega$ of solutions of the equation
(\ref{a4}) with $\la=0$}
coincide with the Cauchy data of the solutions of  $\Delta v_1=0$ on  $\p
M_1= \p \Omega$.

\subsection{$\Gamma-$convergence and spectral convergence}\label{sec Gamma}

{In this section we  establish
  $\Gamma-$convergence and spectral convergence results
for  $a_{R}$ as $R \searrow 1$}.
{{To that end, following the comment in the previous section, } observe that
since 
$\sigma_{R_1}\geq \sigma_{R_2}$ for $R_1\geq R_2$, one thus has
\beq \label{1.07.11}
\quad a_{R_1}[v,v]\geq a_{R_2}[v,v], \,\, v \in H^1(\Omega), \quad
\,\, \hbox{if}\,\,R_1\geq R_2.
\eeq
This implies}  that $R\mapsto a_{R}[v,v]$ is decreasing as $R\searrow 1$.
For $R>1$,
consider  non-linear (quadratic) functionals
$a_{R}:L^2_g(\Omega)\mapsto \R^+:=\R\cup\{+\infty\}$,
\beq \label{AR}
a_R(v)=\left\{\begin{array}{cl}
a_R[v,v] \quad \hbox{when}\,\, v \in H^1_0(\Omega),\\
\infty \quad \hbox{otherwise}.
\end{array}\right.
\eeq

For the ideal cloak, \ie,  $R=1$,   define
\beq \label{A R is 1}
a_1(v)=\left\{\begin{array}{cl}
a_1[v,v] \quad \hbox{when}\,\, v \in H^1_{0,g}(\Omega),\\
\infty \quad \hbox{otherwise}.
\end{array}\right.
\eeq

We will  make extensive use of De Giorgi's  $\Gamma-$convergence, see, \eg,
\cite{A,dM}.

\begin{definition}
Let $\{J_R: 1\le R\le 2\}$ be a family of functionals on a Hilbert space
$\mathcal H$.
We say that the $J_R$ $\Gamma$-converge to $J_1$, or $
J_1=\hbox{\rm  $\Gamma-$lim}_{R\searrow 1} J_R$ on $\mathcal H$, if 

\noindent
(i) for every $v\in\mathcal H$, and all sequences $v_R$ converging to $v$ in
$\mathcal H$ as $R \searrow 1$, $ J_1(v)\leq \liminf_{R \searrow 1}
J_R(v_R)$; and
 
\noindent
(ii)  for every $w\in \mathcal H$ there exists a sequence $w_R$ converging
to
$w$ in $\mathcal H$  such that $J_1(w)\geq \limsup_{R \searrow 1} J_R(w_R)$.

\end{definition}

We then have the following result:


\begin{lemma}\label{lem: A_R gamma converge}
 The functionals $a_{R}$ $\Gamma$-converge to $a_1$ as $R\searrow 1$, 
\beq\label{eq: gamma limit}
\glim{R \searrow 1} a_R=a_1,\quad\hbox{on }L^2_g(\Omega).
\eeq
\end{lemma}

{\bf Proof.}
By Lemma \ref{lem:closed form}, $H^1_{0, g}(\Omega)$ is a
Hilbert
space, when endowed with the norm $(\|u\|_{L^2_g}^2+a_1(u))^{1/2}$.
Since the functionals $a_{\sigma_R}$ are pointwise 
decreasing 
as $R\searrow 1$,
it follows from \cite[Prop 5.7]{dM}
that the  functionals $a_R$ $\Gamma-$converge on $L^2_g$ to the
\beq\label{A1 formula 1}
{\tilde a}_{1}=\hbox{sc}^-G,
\eeq
that is, the lower semicontinuous envelope of $G$ (see  \cite[Def.
3.1]{dM}), where  $G:L^2_g\to \R^+:=\R\cup\{+\infty\}$ is defined
by
\beq\label{A2 formula 1}\quad\quad
G(v)=\left\{\begin{array}{cl}
\int_{\Omega}(\sigma_1 \nabla^b u\cdotp
\overline{\nabla^b u}
  \,dx +qg^{1/2}|u|^2)\, dx, &\hbox{for }u \in H^1_0
(\Omega),\\
\infty,&\hbox{otherwise}. \end{array}\right.
\eeq
By  \cite[Prop 11.10]{dM}, the function $\tilde a_{1}:L^2_g(\Omega)\to
\R\cup
\{\infty\}$
is a quadratic form.
Moreover, by \cite[Prop 12.16]{dM}, its domain
\ba
{\mathcal D}(\tilde a_{1})=\{u\in L^2_g(\Omega): \tilde a_{1}(u)<\infty\},
\ea
endowed with the norm $(\|u\|^2_g+\tilde a_{1}(u))^{1/2}$, is a Hilbert
space.

Now $H^1_0(\Omega)$ is contained  in both
${\mathcal D}(\tilde a_{1})$ and ${\mathcal D}(a_1)= H^1_{0, g}(\Omega)$,
and
the norms of these Hilbert spaces coincide on $H^1_0(\Omega)$.
Moreover, by the proof of   \cite[Lem.\ 3.3]{GKLU1},
$H^1_0(\Omega)$ is dense in ${\mathcal D}(a_1)$.
Thus ${\mathcal D}(a_1)\subset {\mathcal D}(\tilde a_{1})$.

On the other hand, as $G(v)\geq a_1(v)$ for all $v \in L^2_g(\Omega)$ and
$a_1$ is
lower
semicontinuous (see \cite[Prop.\ 2.16]{dM}), it follows that the lower 
semicontinuous envelope
$\tilde a_{1}$ of
$G$
also satisfies $\tilde a_{1}(v)\geq a_1(v)$. Hence
${\mathcal D}(\tilde a_{1})\subset {\mathcal D}(a_1)$.
Thus, ${\mathcal D}(\tilde a_{1})={\mathcal D}(a_1)$.
\hfill \hfill\proofbox

\smallskip

Let us next consider the resolvent ${\mathcal R}_R(\lambda)$ for $\lambda
<0$.
The following result then holds:
\begin{lemma} \label{week}
For any $\lambda <0,$ the resolvents
${\mathcal R}_R(\lambda)$, $R >1,$
strongly converge on $L^2_g(\Omega)$ to ${\mathcal R}_1(\lambda)$, \ie,
for any $f \in L^2_g(\Omega)$,
\beq \label{13.3}
\lim_{R\searrow 1} {\mathcal R}_R(\lambda)f = {\mathcal R}_1(\lambda)f
\eeq
{strongly in $L^2_g(\Omega)$ and weakly in $H^1_{0, g}(\Omega)$.}
\end{lemma}
{\bf Proof.}
{The quadratic forms
$a_R(u)-\la \|u\|^2_g$, $ R \geq 1$ in the Hilbert space $L^2_g
(\Omega)$
are associated to the unbounded selfadjoint operators ${\ttilde {\mathcal
A}}_R-\la I$.
Thus we can use  \cite[Thm. 13.6]{dM}
to show that the resolvents  $({\ttilde {\mathcal A}}_R-\la I)^{-1}$
satisfy (\ref{13.3}) {in $L^2_g(\Omega)$}. Indeed, to show 
the strong convergence  (\ref{13.3})
in $L^2_g(\Omega)$
it is sufficient to prove the following three properties,
(\ref{P1}), (\ref{P2}) and (\ref{P3}): {
\beq\label{P1}
& & a_R:L^2_g(\Omega)\to \R\cup \{\infty\} \hbox{ are lower
semicontinuous;}\\
\label{P2} 
& &\glim{R\searrow 1} a_R =a_1\quad\hbox{on }L^2_g(\Omega);\\
& &\label{P3}
a_1(u) -\la\|u\|^2_g \leq \liminf_{R \searrow 1} \left(
a_R(u_R)
-\la\|u_R\|^2_g \right)
\hbox{ as $u_R\weakto u$ in $L^2_g(\Omega)$,}
\eeq
where $\weakto$ denotes weak
convergence
in $L^2_g(\Omega)$.}

Clearly, the quadratic forms $a_R(u)-\la \|u\|^2_g$
are lower semicontinuous on $L^2_g(\Omega)$,  see proof of Lemma \ref{lem:
A_R gamma
converge}.
By Lemma \ref{lem: A_R gamma converge},
the quadratic forms $a_R$ $\Gamma-$converge to $a_1$.
Thus (\ref{P1}) and (\ref{P2}) are valid. To see (\ref{P3}),
it suffices to consider the case when
 $u_R \weakto u$ in $L^2_g$ and
\ba\liminf_{R \searrow 1} \left( a_R(u_R)-\la\|u_R\|^2_g \right)
<\infty.
\ea
Next we consider a sequence $R^k \searrow 1$ such that
   \ba
\liminf_{R \searrow 1} \left( a_R(u_R)-\la\|u_R\|^2_g \right)=
\lim_{k \to \infty} \left( a^k(u^k) -\la\|u^k\|^2_g \right),
\ea
where $a^k=a_{R^k}$ and $u^k=u_{R^k} \in H^1_0(\Omega)$.
Since $u^k$ converges weakly in $L^2_g$, the norms $\|u^k\|_{g}$ are
uniformly bounded. Since also
\beq \label{leq}
a_1(u^k) \leq a^k(u^k),
\eeq
we see that the sequence $u^k$ is uniformly bounded
in  $H^1_g(\Omega)$.
Let us now choose a subsequence of $u^k$ (still denoted
by $u^k$) which converges weakly to $u$ in $H^1_{0, g}(\Omega)$. As the
embedding
$H^1_g\hookrightarrow L^2_g$ is compact, this sequence
converges
strongly in $L^2_g(\Omega)$.
Using  the weak convergence in $H^1_{0,g}(\Omega)$,
we see that
\beq\label{A lim}\quad\quad
a_1(u) -\la\|u\|^2_g=
\lim_{k \to \infty}  \int_\Omega \left(\sigma_1 \nabla^b u^k \cdotp
\overline{\nabla^b u}
+g^{1/2}q u^k \overline{u}-\la g^{1/2} u^k \overline u \right) \,dx.
\hspace{-1cm}
\eeq
By the Cauchy-Schwarz inequality,
\ba
& &
\left|\int_\Omega \left(\sigma_1 \nabla^b u^k \cdotp \overline{\nabla^b u}
+g^{1/2}qu^k\overline u-\la g^{1/2}
u^k \overline u
\right)\, dx \right|
\\ \nonumber
&\leq&
\left( \int_\Omega \left(\sigma_1
   \nabla^b u^k \cdotp \overline{\nabla^b u^k}+g^{1/2}q|u^k|^2-\la g^
{1/2} |u^k|^2  \right)\,
dx\right)^{\frac 12}\cdotp\\
& &\quad \cdotp
\left( \int_\Omega \left( \sigma_1 \nabla^b u \cdotp \overline{\nabla^b u}
+g^{1/2}q|u|^2-\la g^{1/2}
|u|^2
\right)\, dx\right)^{\frac 12}.
\ea
Using (\ref{leq}) and (\ref{A lim}), this implies that
\ba
& &
a_1(u)-\la\|u\|^2_g \leq
\liminf_{k \to \infty} \left(a_1(u^k)-\la\|u^k\|^2_g\right)^{1/2} \cdot
\left(a_1(u)-\la\|u\|^2_g\right)^{1/2}
\\ \nonumber
& &
\leq \liminf_{k \to \infty} \left(a_k(u^k)-\la\|u^k\|^2_g\right)^{1/2}
\cdot
\left(a_1(u)-\la\|u\|^2_g\right)^{1/2} .
\ea
The desired inequality  (\ref{P3}) follows immediately, proving
(\ref{13.3}) in  $L^2_g(\Omega)$.

Finally, to prove the (\ref{13.3}) holds weakly in $H^1_{0,g}(\Omega)$,
observe that, since $\sigma_R \geq
\sigma_1$,
\beq
\label{19a}\quad\quad
\int_\Omega (\sigma_R \nabla^b v \cdot \overline{\nabla ^b v}+g^{1/2}
q|v|^2)\, dx \geq
\int_\Omega (\sigma_1 \nabla^b v \cdot \overline{\nabla ^b v}+g^{1/2}
q|v|^2)\, dx\hspace{-1cm}
\eeq
for $v \in
H^1_0(\Omega).$
On the other hand, denoting $u_R= {\mathcal R}_R(\la) f$
{and using $\la\leq 0$},
we have
\ba
\int_\Omega (\sigma_R \nabla^b u_R \cdot \overline{\nabla^b u_R}+g^
{1/2}q|u_R|^2)\, dx \leq \|f\|_g \,
\|u_R\|_g.
\ea
The above two inequalities, together with the strong convergence
(\ref{13.3}) in $L^2_g(\Omega)$,
 show that the
$\|u_R\|_{H^1_g}$
are uniformly bounded. Thus, if weak convergence (\ref{13.3})  in
$H^1_{0,g}(\Omega)$
is not valid, there
is  a $v\not =u$ and a subsequence  $R_n \searrow 1$ such that,
$
u_{R_n} \weakto v$ in $H^1_g$. Then $u_{R_n} \to v$ in $L^2_g(\Omega)$,
which
is in contradiction with the strong convergence (\ref{13.3}) in
$L^2_g(\Omega)$. Thus
 (\ref{13.3}) holds weakly  in $H^1_{0,g}(\Omega)$.}
\hfill \hfill\proofbox

In some applications, \eg, dealing with scattering of plane waves
$e^{i<k,x>}$, $k \in \R^3$, by the cloaking device,
the main interest concerns not  values $\lambda<0$, but rather
$\la=|k|^2 >0$. To analyze this case, let us first consider the behavior
of the spectra, $\hbox{spec}({\ttilde {\mathcal A}}_R)$, as $R\searrow 1$.

\begin{lemma}
\label{spectrum_R}
Let $K$ be a compact set with $K \cap \hbox{spec}({\ttilde {\mathcal A}}_1)
=
\emptyset$.
Then, for $R$ sufficiently close to $1$,
$K \cap \hbox{spec\,}({\ttilde {\mathcal A}}_R) = \emptyset$.
\end{lemma}
{\bf Proof.}  {It suffices to 
consider $K=[a,b]\subset\R$.} Assume, to the contrary, the existence of a
sequence $R_n
\searrow 1,\,
\mu_n \in [a,b]$, and functions $u_n \in H^1_0 \subset H^1_{0, g},\,
\|u_n\|_g=1$, such that
\beq \label{13.1}
\ttilde {\mathcal A}_{R_n} u_n = \mu_n u_n.
\eeq
Then,
\ba
\int_\Omega \left( \sigma_{R_n} \nabla^b u_n \cdot\overline{\nabla^b u_n}
+g^{1/2}q|u_n|^2 \right)\, dx
=\mu_n \,\int_\Omega g^{1/2}|u_n|^2\,dx =\mu_n \leq b.
\ea
Therefore, {as $\sigma_R\geq \sigma_1$ and $q\geq 0$, this implies}
  $\|u_n\|^2_{H^1_g} \leq b+1$. Thus there exists a subsequence of
$u_n$
and $\mu_n$,
which we relabel as the original sequence,  $u \in H^1_{0, g}$ and $ \mu \in
[a,b]$,
such that
\ba
& &
u_n \weakto u \quad \hbox{weakly} \, \hbox{in} \,\, H^1_g;\quad
u_n \rightarrow u \quad \hbox{strongly} \,\hbox{in} \,\, L^2_g; \quad \hbox{
and }
\mu_n \rightarrow \mu
\ea
as $n\to \infty$.
Thus, in particular, $\|u\|_{g}=1$.

Taking, \eg, $\la=-1$ in Lemma \ref{week}, we know that
${\mathcal R}_R(-1) \rightarrow {\mathcal R}_1(-1)$ as $R\searrow 1$
in the strong operator
topology
on $L^2_g$. Consider ${\mathcal R}_1(-1)u$; then
\ba
{\mathcal R}_1(-1)u = \lim_{n \to \infty} {\mathcal R}_{R_n}(-1)u
=\lim_{n \to \infty} {\mathcal R}_{R_n}(-1)u_n,\quad\hbox{in }L^2_g(\Omega)
\ea
where, in the last step we have used fact that $\ttilde {\mathcal A}$ are
non-negative
operators yielding the estimate
$\|{\mathcal R}_{R_n}(-1)\|_{L^2_g\to L^2_g}\leq 1$.

However, taking into account (\ref{13.1}),
\ba
{\mathcal R}_{R_n}(-1) u_n = \frac{1}{1+\mu_n}u_n \rightarrow
 \frac{1}{1+\mu}u\quad\hbox{in }L^2_g(\Omega)\hbox{ as }\ {n \to \infty}.
\ea
Thus, ${\mathcal R}_1(-1) u=(1+\mu)^{-1}u$
  with $ \|u\|_{L_g^2}=1$, implying that
$\mu \in \hbox{spec\,}({\ttilde {\mathcal A}}_1)$
{with $u$ being an associated eigenfunction.} 
This contradiction proves
the
statement.
\hfill \hfill\proofbox

\begin{lemma}
\label{cor:13.1}
Let $K\subset \C$ be compact subset such that
$K \cap  \hbox{spec}({\ttilde {\mathcal A}}_1)= \emptyset$.
Then, for any $f \in L^2_g(\Omega)$ and $\la \in K$,
\beq
\label{13.4}
\lim_{R\searrow 1} {\mathcal R}_R(\la)f = {\mathcal R}(\la)f
\eeq
strongly in $ H^1_{0, g}(\Omega)$,
and the convergence is uniform for $\la\in K$.
\end{lemma}

{\bf Proof.} Let $\delta>0$ satisfies
\ba
K_\delta \cap  \hbox{spec}({\ttilde {\mathcal A}}_1)= \emptyset,
\quad K_\delta=\{z \in\C: \hbox{dist}(z, K) \leq \delta\}.
\ea
It then follows from Lemma \ref{spectrum_R} that, for $R$
sufficiently close to $1$, $K_{\delta/2}  \cap  \hbox{spec}({\ttilde
{\mathcal A}}_R)= \emptyset$.
As all ${\ttilde {\mathcal A}}_R,\, R\geq 1$ are self-adjoint 
{in $L^2_g(\Omega)$,
we see that, for $R$
sufficiently close to $1$,
\ba
\|{\mathcal R}_R(\la)\|_{L^2_g\to L^2_g} \leq \frac{2}{\delta}, \quad
\hbox{when}\,\, \la \in K.
\ea
}
This implies that $\C \setminus \hbox{spec}({\mathcal A}_1) = \Delta_{b}$, 
where   $\Delta_{\it b}$  denotes the {\it region of boundedness}
for the
family of operators ${\mathcal A}_R,\, R \geq 1$, \ie,
{the set of those $\la\in \C$ for which the
norms $\|({\mathcal A}_R-\la)^{-1}\|_{L^2_g\to L^2_g}$ are bounded by some 
constant $C_\lambda>0$ for all $R>1$,}
see \cite[Sec.
VIII.1.1]{Kato}.
As $ \hbox{spec}({\mathcal A}_1)$ is countable, $\Delta_{b}$ is connected.
On the other hand, by Lemma \ref{week}, $\R_- \subset \Delta_s$, where $
\Delta_s$
is the {\it region of strong convergence} for the above family, \ie, the set
of $\la \in
\C$ such that for $f \in L^2_g(\Omega)$
\beq
 \label{05.10.5}
\lim_{R\to 1}{\mathcal R}_R(\la)f = {\mathcal R}_1(\la)f\quad\hbox{in }
L^2_g(\Omega), 
\eeq
Therefore, by \cite[Thm.
  VIII.1.2]{Kato},
\beq
  \label{05.10.6}
\Delta_s=\Delta_b=\C \setminus \hbox{spec}({\mathcal A}_1).
  \eeq

{By \cite[Lemma 3]{GKLU1},
$H^1_{0}(\Omega)$ is dense
in $H^1_{0, g}(\Omega)={\mathcal D}(a_1)$. Thus $H^1_{0}(\Omega)$ 
is a core of the quadratic form $a_1[\cdotp,\cdotp]$.} 
Now $H^1_0(\Omega)= {\mathcal D}(a_R),$ for $R>1$ and
$a_R$ are monotonically increasing with $R \geq 1$
on $ H^1_{0}(\Omega)$. Thus it follows from  \cite[Th. 
VIII.3.6]{Kato}, that
  \beq
\label{05.10.4}
\lim_{R \searrow 1} a_1({\mathcal R}_R(\la)f- {\mathcal R}_1(\la)f)=0,
\eeq
  uniformly
for $\la \in K$ where $K$ is an arbitrary compact subset of $\Delta_s$. 
By Lemma \ref{lem:closed form},
the desired convergence (\ref{13.4})
now follows from (\ref{05.10.5})--(\ref{05.10.4}).
\hfill \hfill\proofbox

{Let $\mu\not \in \hbox{spec}(\A_1)$.
It follows from Lemma
\ref{spectrum_R} that $\mu\not \in \hbox{spec}(\A_R)$ for $R>1$ sufficiently
close to 1.  For $\mu\not \in \hbox{spec}(\A_R)$, we 
denote by $N_{R}(\mu)$ the subspace of $L^2_g(\Omega)$
spanned by the eigenfunctions of $\A_{R}$ with eigenvalues $\lambda_j<\mu$.
We also denote by $P_R^\mu,$ the orthoprojectors in $L^2_g(\Omega)$ onto
$N_{R}(\mu)$.
By \cite[Th.
 III.6.17]{Kato}, these (Riesz)  projectors $P_R$
have the representation
\beq\label{Riesz}
  P_R^\mu u = \frac{1}{2 \pi i} \int_{\Gamma} (\A_R-z)^{-1} u\, dz,
\eeq   where the contour $\Gamma\subset \C$ surrounds {all the eigenvalues
$\lambda_j$
of $\A_R$ satisfying $\la_j<\mu$ and only those.}

\begin{lemma} \label{R-dimension}
{Let $\mu\not \in \spec(\A_1)$.}
For $R$ sufficiently close to $1$
\beq \label{RR-dimension}
\hbox{dim}(N_{R}(\mu))=\hbox{dim}(N_{1}(\mu)).
\eeq
Moreover, 
\beq
\label{R_projectors}
\lim_{R \searrow 1}\|P_R^\mu - P_1^\mu\|_{L^2_g \to L^2_g}=0.
\eeq
\end{lemma}
{\bf Proof.}
{Recall that $\D(a_R)=H^1_0(\Omega)$ are independent of $R>1$,
$H^1_0(\Omega)\subset \D(a_1)=H^1_{0,g}(\Omega)$, and 
$a_R[u,u]$ are decreasing, as $R\searrow 1$, for 
all $u\in \D(a_R)$. 
Thus the identity (\ref{RR-dimension}) follows directly by \cite[Th.
 VIII.3.15]{Kato}.

Using representation (\ref{Riesz}) and Lemma \ref{cor:13.1}, 
we see that 
\ba\lim_{R \searrow 1} P_R^\mu = P_1^\mu\ \quad \hbox{strongly
  in} \,\, L^2_g(\Omega).
\ea
As $P_R^\mu$ and $P_1^\mu$ are orthoprojectors, {this and
(\ref{RR-dimension}) yields
(\ref{R_projectors}) by}   \cite{Kato} (see Lemmas VIII.1.23 and
VIII.1.24).}
\hfill \hfill\proofbox

We remark that in the course of this paper we need a number of 
results concerning convergence of orthoprojectors that appear similar to
(\ref{R_projectors}), \eg, (\ref{06.10.7}), (\ref{08.10.4}) and
(\ref{final-estimate2}), but these are for different operators or with
respect
to different operator norms and require separate proofs.

\subsection{Approximating the singular bulk modulus $g^{1/2}$
by nonsingular densities} \label{density}

{Above, in the operator ${\ttilde {\mathcal A}}_R=-g^{-1/2}\nabla^b\cdotp
\sigma_R\nabla^b+q$ there appears
the determinant $g$ of the metric {(also denoted $g$!)},
which vanishes at the cloaking surface $\Sigma$.
We now consider how to approximate {the scalar function} $g$ by functions
$g_m$ that are bounded from below with positive
constants. To this end, we introduce the functions
\beq \label{2.03.11}
g_m(x)=\max\left(g(x),1/m \right),\quad m\in \Z_+.
\eeq
{\mllltext Then $L^2(\Omega, g_m^{1/2} dx)\subset L^2_g(\Omega)$ and}
\beq\label{g g_m inclusion}
\|f\|_g \leq \|f\|_{L^2(\Omega, g_m^{1/2} dx)}, \quad \hbox{for} \,\, f \in L^2(\Omega, g_m^{1/2} dx).
\eeq
The multiplication
map $f\mapsto g^{1/2}f$ is unitary from $L^2_g=L^2(\Omega,
g^\frac12 dx)$ onto
$L^2(\Omega, g^{-\frac12} dx)$.
Note that $L^2(\Omega, g^{-\frac12} dx) \subset L^2(\Omega)\subset
L^2_g(\Omega)$.
Next we will consider operators
${g}^{1/2} {\ttilde {\mathcal A}}_Ru$.
{For $f\in L^2_g(\Omega)$, we have
\beq\label{eq: non-modified}
(\ttilde {\mathcal A}_R-\lambda)u=f,
\eeq
where both sides are in $L^2_g(\Omega)$, if and only if
$u$ is a solution to the acoustic equation with  mass density
tensor $\sigma_R$, bulk modulus $g^{1/2}$, and potential $q g^{1/2}$),
\beq\label{eq: modified}
(g^{1/2}{\mathcal A}_R-\la g^{1/2})u=F,
\eeq
where $F=g^{1/2}f\in L^2(\Omega, g^{-\frac12} dx)$.
By the above considerations, we have for $f\in L^2_g(\Omega)$}
\beq \label{28a}
(\ttilde {\mathcal A}_R-\lambda)^{-1}f=  (g^{1/2}{\mathcal A}_R-\lambda
g^{1/2})^{-1}(g^{1/2}f).
\eeq

Later in this section we keep $R>1$  fixed.
Define an unbounded selfadjoint operator ${\mathcal B}_{R}$ in
$L^2(\Omega)$,
having the same differential expression as
the operator $g^{1/2}{\mathcal A}_R$,
but
with different domain,
\beq
\label{Laplace-type 2}
& &{\mathcal B}_{R}u:= -\nabla^b \cdot \left(\sigma_R \nabla^b  u
\right)+g^{1/2}qu,\\ \nonumber
& &{\mathcal D}({\mathcal B}_{R})=\{u \in H^1_0(\Omega):\,
\nabla^b \cdot \left(\sigma_R \nabla^b  u \right)
  \in L^2(\Omega) \}.
\eeq
{Since $\mathcal D(\A_R)\subset H_0^1(\Omega)$,  see (\ref{AR-operator}),
${\mathcal B}_{R}$ is an
extension
of
$ g^{1/2}{\mathcal A}_R$ and,
in particular,
$\nabla^b \cdot(\sigma_R\nabla^b u)
\in L^2(\Omega, g^{-\frac12} dx)$ for $u\in {\mathcal
D}({\mathcal A}_{R})$;
however, $\nabla^b \cdot( \sigma_R\nabla^b u) \in L^2(\Omega)$ for
$u\in
{\mathcal
D}({\mathcal B}_{R})$.
{Note that by (\ref{28a}),
\beq \label{eq: Matti A-formula}
(\A_R-\lambda)^{-1}f=  (\B_R-\lambda
g^{1/2})^{-1}(g^{1/2}f)\quad\hbox{for }f\in L^2_g(\Omega),
\eeq
where $\la\not\in \hbox{spec}\,(\A_R)$. We will use this formula extensively
later for $ f\in L^2(\Omega)\subset
L^2_g(\Omega)$.}
\begin{lemma} The operator ${\mathcal B}_R-\lambda g^{1/2}$ has a bounded
inverse
if and only if $\la \not \in \hbox{spec}\,(\ttilde {\mathcal A}_{R})$.
\end{lemma}

{\bf Proof.} {For
$\la <0$} the operator ${\mathcal B}_R-\lambda g^{1/2}$ is  strictly
positive and, since ${\mathcal D}(\B_R) \subset H^1_0(\Omega)$,
has a compact resolvent. Therefore, the operator $({\mathcal B}_R-\lambda
g^{1/2})^{-1}$
exists for $\la < 0$ and  is bounded in
$L^2(\Omega)$.
Since  the multiplication, $u \mapsto g^{1/2}u$, is bounded in
$L^2(\Omega)$,
  by  the analytic
Fredholm theory  \cite{ReedSimon1} the operator-valued function
\ba
\la \mapsto ({\mathcal B}_R-\lambda g^{1/2})^{-1}=
\left[I -(\la +1)({\mathcal B}_R+g^{1/2})^{-1} g^{1/2}  \right]^{-1}
({\mathcal
B}_R+g^{1/2})^{-1}
\ea
is  a meromorphic operator-valued function of $\la\in \C$.
Therefore,   if the inverse
$({\mathcal B}_R-\lambda g^{1/2})^{-1}$ does not exist for a given $\la \geq
0$,
then
{there is  $v\in H^1_0(\Omega)$} such that
\ba
({\mathcal B}_R-\lambda g^{1/2})v=0.
\ea
In this case
${\mathcal B}_R v=\lambda g^{1/2}v\in L^2(\Omega, g^{-\frac12} dx)$ and we
see that
$v\in {\mathcal D}({\ttilde {\mathcal A}}_R)$,
\ie, $\la \in \hbox{spec}({{\mathcal A}}_R)$.
On the other hand, if $\la \in \hbox{spec}({{\mathcal A}}_R)$, \ie
\ba
-g^{-1/2}\nabla^b \cdot \left( \sigma_R\nabla^bu \right)+qu =\la u,
\ea
then $\nabla^b\cdot \left( \sigma_R \nabla^b u \right)= q g^{1/2}u -\la
g^{1/2}u \in
L^2(\Omega)$, \ie, $u \in {\mathcal
D}({\mathcal B}_R)$, so that $\B_R-\la g^{1/2}$ does not have a
bounded
inverse. \hfill \hfill\proofbox

Next we consider the uniform convergence of resolvents.
{To this end we introduce  operators ${\mathcal B}_{R, m},\, m \in
\Z_+$, 
in $L^2(\Omega)$, of the form
\beq \label{03.10.1}
& &
{\mathcal B}_{R, m} u:= -\nabla^b\cdot \left( \sigma_R \nabla^b u \right)
+ q g_m^{1/2}u,
\\ \nonumber
& &{\mathcal D}({\mathcal B}_{R, m})=\{u \in H^1_0(\Omega):\,
\nabla^b \cdot \left(\sigma_R \nabla^b  u \right)
  \in L^2(\Omega) \}={\mathcal D}({\mathcal B}_{R}).
\eeq
The operator $\B_{R, m}$ is associated with the operator $\A_{R, m}$ in the
same way 
that ${\mathcal A}_R$ is
  associated with ${\mathcal B}_R$,
where the operator  $\A_{R, m}$
is the self-adjoint operator in $L^2(\Omega, g^{1/2}_mdx)$ defined by
\beq \label{ARm}
& &{\mathcal A}_{R, m}u:= -g_m^{-1/2}\nabla^b \cdot \left(\sigma_R \nabla^b
u
\right)+qu,\\ \nonumber
& &{\mathcal D}({\mathcal A}_{R,m})=\{u \in H^1_0(\Omega):\,
\nabla^b \cdot \left(\sigma_R \nabla^b  u \right)
 \in L^2(\Omega) \}.
\eeq
{Note that ${\mathcal D}({\mathcal A}_{R,m})={\mathcal D}({\mathcal
B}_{R,m})$.}

\begin{lemma}\label{lem A} Let $R>1$ and  $K\subset  \C$ be  compact and
such that
$K\cap \hbox{spec}\,(\ttilde {\mathcal
A}_{R}) = \emptyset$.
Then {there is an  $m_R\in \Z_+$ such that
$K\cap \hbox{spec}\,(\ttilde {\mathcal
A}_{R,m}) = \emptyset$ for $m>m_R$, and}
 \beq \label{convergence for A_{R,m}}
\lim_{m\to \infty}\|\left(\A_{R, m}-\la \right)^{-1}-
\left(\A_{R}-\la \right)^{-1}\|_{L^2(\Omega) \to H^1_0(\Omega)}=0,
\eeq
uniformly for $\la \in K$.
\end{lemma}

{\bf Proof.} By the assumptions on $K$, 
the inverse $({\mathcal B}_R-\lambda g^{1/2})^{-1}$
exists and is a continuous function of $\la\in K$ with respect to the
$L^2(\Omega)$-operator norm topology. Let
\ba
d=\max_{\la \in K} \|({\mathcal B}_R-\lambda
g^{1/2})^{-1}\|_{L^2(\Omega)\to L^2(\Omega)} < \infty.
\ea
Denote $V_{R, m}(\la)= (\la-q) (g_m^{1/2}-g^{1/2}) \in L^\infty(\Omega)$, 
{so
that
\beq \label{06.10.1}
\lim_{m\to \infty}\|V_{R, m}(\la)\|_{L^\infty(\Omega)}=0
\eeq
uniformly for $\la \in K$. Thus there is $m(K)>0$ such that
$\|V_{R, m}(\la)\|_{L^\infty} \leq (2d)^{-1}$ for $m >m(K)$, $\la\in K$. 
Therefore, $(\B_{R, m}-\la g_m^{1/2})^{-1}$ exists for $\la\in K$ and is
given by
\beq \label{06.10.2}\quad\quad
\left(\B_{R, m}-\la g_m^{1/2}\right)^{-1}=\left(\B_{R}-\la
g^{1/2}\right)^{-1}
\left[I+ V_{R, m}(\la) (\B_{R}-\la g^{1/2})^{-1}\right]^{-1}\hspace{-1cm}
\eeq 
where the right hand side can be written as a Neumann series.
{This also shows that there is an $m_R$ such that $K\cap
\hbox{spec}\,(\ttilde {\mathcal
A}_{R,m}) = \emptyset$ for $m >m_R$.}

For any  $\la'\in K$,
  $(\B_{R}-\la' g^{1/2})^{-1}$ is a
bounded operator from 
$L^2(\Omega)$ to $H^1_0(\Omega)$,  and for if $|\la-\mu|<(2d)^{-1}$ we have
\ba
\left(\B_{R}-\la g^{1/2}\right)^{-1}=
\left(\B_{R}-\la' g^{1/2}\right)^{-1} \left[I+ (\la'-\la) g^{1/2}
(\B_{R}-\la'
g^{1/2})^{-1}\right]^{-1}.
\ea
Using this we see that the norm of  $(\B_{R}-\la g^{1/2})^{-1}:
L^2(\Omega)\to H^1_0(\Omega)$ is uniformly bounded in $\la\in K$.
Using  formulae (\ref{06.10.1}) and (\ref{06.10.2}), we see that
\beq \label{06.10.3}\quad\quad
\lim_{m\to \infty}\|\left(\B_{R, m}-\la g_m^{1/2}\right)^{-1}-
\left(\B_{R}-\la g^{1/2}\right)^{-1}\|_{L^2(\Omega) \to H^1_0(\Omega)}=0
\eeq 
uniformly  for $\la \in K$, and that the norms of
operators
$(\B_{R, m}-\la g_m^{1/2})^{-1}:L^2(\Omega) \to H^1_0(\Omega)$ are uniformly
bounded
for
$\la \in K$. This proves 
\ba
\lim_{m\to \infty}\|\left(\B_{R, m}-\la g_m^{1/2} \right)^{-1}-
\left(\B_{R}-\la g^{1/2} \right)^{-1}\|_{L^2(\Omega) \to H^1_0(\Omega)}=0.
\ea

Additionally,  the multiplication operators $g_m^{1/2},\, g^{1/2}$ are 
bounded on $L^2(\Omega)$, uniformly in
$m$,  and
\ba
\|g_m^{1/2}-g^{1/2}\|_{L^2\to L^2} \leq m^{-1/2} \to 0, \quad
\hbox{as}\,\, m \to
\infty.
\ea
Together with equation (\ref{06.10.3}) and the
boundedness of 
$(\B_{R, m}-\la g_m^{1/2})^{-1}$ as operators
from 
$L^2(\Omega)$ to $H^1_0(\Omega)$,
this implies
that 
\ba
\lim_{m\to \infty}\Big\|\left(\B_{R, m}-\la g_m^{1/2} \right)^{-1}g_m^{1/2}-
\left(\B_{R}-\la g^{1/2}\right)^{-1} g^{1/2}\Big\|_{L^2(\Omega) \to
H^1_0(\Omega)}=0.
\ea
This in turn implies equation (\ref{convergence for A_{R,m}}), due to 
{formula (\ref{eq: Matti A-formula}) and} the
relations between $\B_{R, m},\, \B_R$ and $\A_{R, m},\, \A_R$,
which follow from their definitions (\ref{AR-operator}), (\ref{Laplace-type
2}-\ref{03.10.1}), and (\ref{ARm}).
\hfill \hfill\proofbox

\smallskip
}

{Let $R>1$, $\mu \not\in \hbox{spec}(\A_R)$.
 {It follows from Lemma
\ref{lem A}  that $\mu\not \in \hbox{spec}(\A_{R,m})$ for $m$ sufficiently
large.}
Denote by $N_{R, m}(\mu)$ the subspace of $L^2(\Omega, g^{1/2}_mdx)$
spanned by the eigenfunctions of $\A_{R.m}$ with eigenvalues in 
$(-\infty,\mu)$, \cf the definition of $N_R(\mu)$.
  Also
denote by
$P^\mu_{R, m}$ the orthogonal eigenprojectors  onto
$N_{R,m}(\mu)$ in $L^2(\Omega, g^{1/2}_m dx)$. 
}

Clearly, since, for $m \in \Z_+$, $L^2(\Omega, g^{1/2}_m
dx)=L^2(\Omega)$ as
sets, we can consider $P^\mu_{R, m}$ as projectors, although not
orthogonal,
on $L^2(\Omega)$.
Recall that $P^\mu_R$ is an orthoprojector in $L^2_g(\Omega)$ onto
$N_R(\mu)\subset
H^1_0(\Omega)$.
Restricting it to $L^2(\Omega)$, we obtain a projector, which we still call
$P^\mu_R$,
on $L^2(\Omega)$. Again, $P^\mu_R$ is not an orthoprojector on
$L^2(\Omega)$.
However, we can compare these projectors, as well as spaces 
$N_{R, m}(\mu),\, N_R(\mu)$.

\begin{corollary} \label{m-dimension}
{Let $R>1$ and  $\mu \not\in \hbox{spec}(\A_R)$.
  The Riesz projectors $P_{R,m}^\mu$ corresponding to the operators
  $\A_{R,m}$ satisfy
\beq \label{06.10.7}
\lim_{m\to \infty}\|P^\mu_{R, m}-P_R^\mu\|_{L^2(\Omega) \to
H^1_0(\Omega)}=0.
\eeq
Moreover, for $m$ sufficiently large,} 
\beq \label{06.10.8}
\hbox{dim}(N_{R, m}(\mu))=\hbox{dim}(N_{R}(\mu)).
\eeq
\end{corollary}

{\bf Proof.} 
{Let $\Gamma \subset \C$ be a contour surrounding
only $\la_0$ from $\hbox{spec}(\A_R)$. When $m$ is large enough,
the Riesz projectors $P_{R,m}^\mu$ have a representation analogous to 
(\ref{Riesz}), obtained by replacing $\A_{R}$ by $\A_{R,m}$ and using
the contour $\Gamma$.
Thus\ba
& &\lim_{m\to\infty} \|P_{R, m}^\mu -P_{R}^\mu\|_{L^2(\Omega) \to
H^1_0(\Omega)} \leq
\\ \nonumber
& &
\lim_{m\to\infty} \frac{1}{2\pi} \int_{\Gamma} 
\|(\A_{R, m}-z )^{-1}-(\A_{R}-z )^{-1}\|_{L^2(\Omega) \to H^1_0(\Omega)}
\, dl(z), 
\ea
where $dl$ is the arclength measure on $\Gamma$.
Taking into account (\ref{convergence for A_{R,m}}), this formula
implies (\ref{06.10.7}).

Using equation (\ref{06.10.7}), we see that there
exists an 
$m_0$ such that for
$m \geq m_0$ we have
$
\|P_{R, m}^\mu -P_{R}^\mu\|_{L^2(\Omega) \to L^2(\Omega)} <1.
$
Using, \eg,\ \cite[Cor.\ IV.2.6]{Kato}, we see that this
 proves (\ref{06.10.8}).
\hfill \hfill\proofbox
}

\section{Approximating   anisotropic 
by  isotropic mass densities}
\label{sec-approx second}

We now show, using techniques from
homogenization theory,  \cf \cite{Allaire,Cherka,dM},  that we can
approximate arbitrarily closely,
on the level of the operators,
 the nonsingular anisotropic approximate  mass densities
$\sigma_R$,
{for any fixed  $R>1$, by}
a family of nonsingular \emph{ isotropic} mass densities
$\sigma_{R, \e}, \, \e >0$, which will thus also function as approximate
cloaks.
This can be considered as the reverse of the traditional homogenization
theory.

\subsection{Inverse homogenization with magnetic potential}

Observe that, as all the approximate cloaks $\sigma_R,\,
R > 1$, are rotationally invariant, it is natural to use
 spherical coordinates.
Namely,
 we will use either the Euclidian coordinates
$x=(x^1, x^2, x^3), $ or the spherical coordinates
$s=s(x)= (r(x), \theta(x), \phi(x))$.
Note that we use the same notation,
$x$, for a point inside $\Omega$ and its Euclidian coordinates,
$x=(x^1, x^2, x^3)$.  Which meaning is intended will be always clear from
the
context.  We denote by
$X: \, (x^1, x^2, x^3) \mapsto (r, \theta, \phi)$
the corresponding coordinate transformation.
To exploit the rotational invariance,
we will employ
 in the  homogenization process, \cf \cite{Allaire,Cherka},
cells which are {\it cubes}
in these spherical coordinates.

{To approximate the anisotropic mass densities in spherical coordinates in
the
ball $\Omega=B(0,3)$,
let us consider isotropic mass densities of the  form
\beq\label{basic model 1}
\sigma_\e(x)=\sigma(x, r(x)/\e),\quad \sigma(x, r')=h(x, r')I\in \R^{3\times
3}.
\eeq
Here $h(x,r')$ is  a
scalar valued function, to be chosen later,  that is periodic in $r'$ with
period 1 and is bounded from above and below,
\ie,
\beq\label{basic model 2}
h(x,r'+1)=h(x,r'),\quad 0<c_1\leq h(x,r')\leq c_2.
\eeq
We will consider densities for which $h(x,r')$ is independent of $r'$ for
$x$ with 
$r(x)<1$ and $5/2<r(x)<3$, that is,
\beq\label{basic model 3}
h(x,r')=h(x)\quad \hbox{if} \,\, |x|<1 \hbox{ or}\,\, 5/2<|x|<3.
\eeq
We make this  assumption since later we will
use the isotropic mass densities to approximate the non-singular anisotropic
mass densities $\sigma_R,\,
R>1$ that are
isotropic for
$r(x)<1$ and $5/2<r(x)<3$.}

Let
$(r,\theta,\phi)$ and $(r',\theta',\phi')$ be spherical coordinates
corresponding to
two different scales.
Then, in these coordinates,
\ba
(X_*(\sigma_\e))(s)=\sigma^*(s,  s/\e),
\ea
where
\ba
\sigma^*(s,r')
&=&
h(s,r')
\det(D X (x))^{-1}DX (x) DX^t(x)|_{X(x)=s}
\\
&=&h(s,r')\left( \begin{array}{ccc}
r^2\sin \theta & 0 & 0 \\
0 &\sin \theta & 0  \\
0 & 0 &\frac{1}{\sin \theta}  \\
\end{array}\right) .
\ea
Here and later we denote by $\sigma$,  sometimes with various indices,
various  mass tensors (or matrices) in the Euclidian coordinates,
while  $\sigma^*$ always stands for their
representation in the spherical coordinates.

{In the following, the material on homogenization is a quite straightforward
generalization
of  known results \cite{Allaire,Cherka}.  However, as we need to introduce
changes due both to  the presence of a magnetic potential
and the use of spherical coordinates, for completeness we give details of
the arguments.}

In the small-scale coordinates $t=(r',\phi',\theta')$, we denote by $e^1=
(1,0,0)$, $e^2=(0,1,0)$, and
$e^3=(0,0,1)$
the vectors corresponding to the differential forms $dr'$, $d\theta'$ and
$d\phi'$,
respectively.
Let    $W^j(s,t),\, j=1, 2, 3,$ be
the solutions of
\beq\label{cell equation}
\nabla _t \cdot \sigma^*(s,t) (\nabla_t  W^j(s,t)+e^j)=0,\quad
t=(r',\theta',\phi')\in \R^3,
\eeq
that are  $1$-periodic functions in $r',\theta'$ and $\phi'$ variables
(noting that the periodicity in $r',\theta'$ and $\phi'$
has no relation to periodicity in the ``{large-scale}''
spherical coordinates
$\theta,\, \phi$),
  and satisfy, 
$
\int_ {[0,1]^3}W^j(s,t)dt=0,
$
for all $s$, where $dt =dr' d\theta' d\phi'$.
Since $\sigma^*$ is independent of $\theta',\, \phi'$,
the above conditions imply that $W^j=0$ for $j=2,3$.
As for $W^1$, it satisfies
\bfo
\frac{\p}{\p r'} \left( h(s, r')\frac{\p W^1}{\p r'}\right)=
-\frac{\p  h(s, r')}{\p r'},
\efo
with $W^1$ being $1$-periodic with respect to $(\theta', \phi')$. These
imply that $W^1$ is independent of $(\theta', \phi')$ with
\bfo
\frac{\p W^1}{\p r'} =-1 + \frac{C_0}{h(s, r')}.
\efo
To find the constant $C_0$ we use the periodicity of $W^1$,
now  with respect to $r'$, to get that $C_0$ is given by the harmonic means
$h^{harm}$ of $h$,
\beq \label{harmonic1}
C_0 = h^{harm}(s):=\left(\int_0^1  \frac {dr'}{h(s, r')}\right)^{-1}.
\eeq

Define the corrector matrices \cite{Allaire} as
\beq \label{corrector_matrix}
P_{j}^k(s,t)=\frac {\p}{\p t^j}W^k(s,t)+\delta^k_j.
\eeq
Then the homogenized mass density in the spherical coordinates,
$\sigma^*_{hom}$,  is
given by
\beq \label{corrector}
\left( \sigma^*_{hom} \right)^{jk}(s)=\sum_{p=1}^3
\int_ {[0,1]^3} \left( \sigma^* \right)^{jp}(s,t)P^k_p(s,t)\,dt.
\eeq
We note that,   applying integration by parts and using  definition
(\ref{corrector_matrix}) and equations
(\ref{cell equation}), formula (\ref{corrector})
can be written also in a more symmetric form
\ba
\left( \sigma^*_{hom} \right)^{jk}(s)=\sum_{p,q=1}^3
\int_ {[0,1]^3}
\left( \sigma^* \right)^{pq}(s,t)P^j_p(s,t)P^k_q(s,t)\,dt.
\ea

Using the above formulae for the $W^i$, it follows from (\ref{corrector})
that
\ba
\sigma^*_{hom}(s)
=\left(\begin{array}{ccc}
h^{harm}(s) r^2 \sin(\theta)& 0 & 0 \\
0& h^a(s) \sin(\theta)& 0  \\
0 & 0 &h^a(s) \sin^{-1}(\theta)\\ \end{array}\right).
\ea
Here $h^a(s)$ denotes the arithmetic means of $h$ in the second variable,
\ba
h^a(s) =\int_{[0,1]} h(s,r')\,dr'.
\ea

Returning  to the Euclidian coordinates, one sees that
  the conductivity,
$\sigma_{hom}(x)= X_* \sigma^*_{hom}(x),$
has the form
\beq\label{simga hom}
\sigma_{hom}(x)&=&\omega_{1}(x)\Pi(x)+\omega_{2}(x)(I-\Pi(x)),
\eeq
with
\beq\label{85a} \omega_{1}(x)=
h^{harm}(x),
\quad \omega_{2}(x)= h^a(x)
\eeq
and $\Pi(x):\R^3\to \R^3$ being the projection to the radial direction,
\ba
\Pi(x)\,v=\left(v\,\cdotp \frac{x}{|x|}\right)\frac{x}{|x|},
\ea
represented by the matrix $(|x|^{-2}x^jx^k)_{j,k=1}^3$.

Next, we  analyze the Dirichlet
problems for  elliptic
equations
\beq\label{non-homogenized eq 1}
-\nabla^b \cdot \sigma_\e \nabla^b u_\e+Qu_\e  =f, \quad
u_\e|_{\p \Omega}= h.
\eeq
{Here,  $b(x)=(b_1(x),b_2(x),b_3(x))$
is the magnetic potential and
$Q(x)$ is a scalar function, with
$Q,b_j\in L^\infty(\Omega;\R)$,
and $\sigma_\e$ are isotropic
mass densities bounded from above and below by  positive
constants independent of $\e$. Moreover, $f \in H^{-1}(\Omega)$, and
$h\in H^{1/2}(\p \Omega)$}.
Our goal is to show that the solutions $u_\e$
  convergence to the solution of the
equation
\beq\label{homogenized eq 1}
-\nabla^b \cdot  \sigma_{hom}\nabla^b u+Qu  =f, \quad
u|_{\p \Omega}= h.
\eeq
By adapting 
the technique of  Allaire
\cite{Allaire},
we can prove the following result.

\begin{proposition}\label{prop. homogenization}
Let $\sigma_\e$, $\e>0$ be mass densities in $\Omega$  satisfying
(\ref{basic model 1}), (\ref{basic model 2}), and (\ref{basic model 3}),
$\sigma_{hom}$ be the mass density defined by (\ref{simga hom}), $Q\in
L^\infty(\Omega)$,
$Q(x)\geq 0$,
and  $b=(b_1(x),b_2(x),b_3(x))$
be a vector field, $b\in L^\infty(\Omega;\R^3)$.
Then the solutions
$u_\e$ of equations (\ref{non-homogenized eq 1})
and solution $u$ of equation (\ref{homogenized eq 1}) satisfy
\beq \label{convergence}
\lim_{\e\to 0} u_\e =u\quad \hbox{weakly in} \,\,
H^1(\Omega).
\eeq
\end{proposition}

{\bf Proof.} Let $E_h \in H^1(\Omega),\,\hbox{supp}(E_h) \subset \{ 5/2 \leq
|x| \leq
3\}$ be an extension of $h$, \ie, $E_h |_{\p \Omega}=h$. Writing
$u_\e =E_h +v_\e$ and $u=E_h + v,$ we see that
the functions $v_\e$ and $ v$ satisfy
\beq \label{10.1}
& &-\nabla^b \cdot \sigma_\e \nabla^b v_\e+Qv_\e  ={\tilde f}, \quad
\hbox{on
}\Omega,\quad
v_\e|_{\p \Omega}= 0,\\
& &-\nabla^b \cdot  \sigma_{\hom} \nabla^b v+Qv ={\tilde f}\quad \hbox{on
}\Omega,\quad
v|_{\p \Omega}= 0. \label{10.1b}
\eeq
Here
\beq \label{10.2}
{\tilde f}=f -  \nabla^b \cdot \sigma_\e \nabla^b E_h+QE_h\in
H^{-1}(\Omega),
\eeq
is independent of $\e$. For proving (\ref{convergence}) it is enough
to show that $v_\e$ converges  to $v$ weakly in $H^1_0(\Omega)$.

{Let us recall that by Lemma \ref{a_R_closed},  {$a_R$ is (strictly)
positive
definite.} 
Since $Q(x)\geq 0$, we see that there exists a $c_0>0$ such that 
\ba
\int_{\Omega} \left( |\nabla^b u(x)|^2+Q(x) |u(x)|^2\right) \,dx \geq c_0
\|u\|^2_{H^1_0(\Omega)}\quad\hbox{for }u\in H^1_0(\Omega).
\ea
As the mass densities $\sigma_{\e}$ are uniformly bounded
from above and below by positive constants, 
{it follows from the proof of Lemma \ref{a_R_closed} that}
there is also  $c_1>0$ such that 
\ba
\int_{\Omega} \left( \nabla^b u\cdotp \sigma_\e  \overline {\nabla^b u}
  +Q(x) |u|^2\right) \,dx \geq c_1
\|u\|^2_{H^1_0(\Omega)}\quad\hbox{for }u\in H^1_0(\Omega).
\ea
Thus, using the Lax-Milgram lemma we see that 
the solutions $v_\e(x)$ of equations
(\ref{10.1})
satisfy
\beq
\label{no_0_eigenvalue}
\|v_\e\|_{H^1_0(\Omega)} 
\leq c \|{\tilde f}\|_{H^{-1}(\Omega)},
\eeq
where $c>0$ is  independent of $\e >0$.}
Therefore, the solutions $v_\e(x)$
 are }
uniformly bounded in $H^1_0(\Omega)$. Thus, for
an arbitrary
{sequence $\e_n \to 0$, the corresponding  $v_{\e_n}$ have }
 a subsequence
that converges weakly to some function $w(x)$ in $H^1_0(\Omega)$. Let us
show that $w$ coincides with
the solution $v$ of equation (\ref{10.1b}).

To this end we consider convergence in a finer,
two-scale, sense on local coordinate neighborhoods.
Let $U\subset \S^2$ be an open set on which we can define,
{in a regular manner,  spherical
coordinates.
For example, by choosing
two antipodal points as the South and North poles and connecting those
by a meridian $\gamma$,
we can take $U$ so that
${\overline U}\subset\subset \S^2\setminus \gamma$ and define polar
coordinates on $U$.
Let
$\Omega'=\{r\omega:\ \omega \in U,\ r\in (r_1,r_2)\} \subset \Omega$
with some  $r_2>r_1>0$.  Clearly,
$X:\Omega'\to \R^3$ defines the spherical coordinates, 
$x\mapsto (r,\theta,\phi)$,
with domain $W:= X(\Omega')= (r_1, r_2) \times U$.}

Rewrite now  the equations (\ref{10.1}) and (\ref{10.1b}) on $W$
in these spherical coordinates and multiply the  equations so obtained
by $\det(D X (X^{-1}(s)))$. The resulting equations are
\beq \label{10.1sp}
& &-(\nabla_s+ib^*(s))\cdot \sigma^*(s,\frac s\e)(\nabla_s+ib^*(s))v_
\e(s)+Q^*(s)v_\e(s)={\hat f}(s),\\
& &
-(\nabla_s+ib^*(s))\cdot \sigma^*_{hom}(s)(\nabla_s+ib^*(s))v(s)+Q^*(s)v(s)
={\hat f}(s), \label{10.1bsp}
\eeq
where $s\in W$.  Here
\ba
 Q^*(s)&=&\det(D X (X^{-1}(s))) Q(X^{-1}(s)) \in L^\infty(W),\\
b^*_j(s)&=&
\p_j X^k (X^{-1}(s))  b_k(X^{-1}(s)) \in L^\infty(W; \R^3),
\ea
correspond to
the electric potential  $Q$ and magnetic potential $b$ in the spherical
coordinates

and
$\hat f(s)=\det(D X (X^{-1}(s))) \tilde f(X^{-1}(s)) \in H^{-1}(W)$.
To simplify notations, we continue to denote the functions $v$ and $v_\e$ in
the
spherical coordinates by  $v(s)=((X^{-1})^*v)(s)$,  $v_\e(s)=((X^
{-1})^*v_\e)(s)$.
Finally,  $\nabla_s$ is the vector field $(\p_{r},\p_{\theta},\p_{\phi})$.

In the following, even though in the forthcoming applications the
mass densities $\sigma^*_\e(x,r(x)/\e)$ will depend only on the
{small-scale variable $r'= r(x)/\e$,
  we consider  the general case when the mass densities  depend on
all small-scale coordinates
$t=(r',\theta',\phi')$. }
Let $T=[0,1]^3$ and $C^m_\#(T)$ denote those $C^m(T)$ functions that can be
continued
as $\Z^3$-periodic functions in $\R^3$ which are in $C^m(\R^3)$. By
definition, a family
$v_\e(s)$, $v_\e\in L^2(W)$ is
said to {\it two-scale converge}, as $\e\to 0$,  to a function $v_0(s,t) \in
L^2(W\times T)$
if, for all test functions $\psi(s,t)$ in $C^\infty_0(W;C^\infty_\#(T))$, we
have
\beq\label{def:two scale}
\lim_{\e\to 0}\int_{W} v_\e(s)\psi(s,\frac s\e)\,ds=\int_{W}\int_{T}
v_0(s,t)\psi(s,t)\,dsdt.
\eeq
By \cite{Allaire}, the two-scale convergence of $v_\e$ implies the weak
convergence
of $v_\e(s)$ in $L^2(W)$ to the function $w(s)=\int_T v_0(s,t)\,dt$, so that
the
two-scale
convergence gives  finer information on the convergence than the weak
convergence.
For example, functions of the form $u(s,s/\e)$ two-scale converge to
$u(s,t)$.

By \cite{Allaire}, every bounded family $v_\e(s)\in L^2(W)$ contains
a two-scale converging sequence. Moreover, if $v_\e(s)$ is a bounded family
in $H^1(W)$
that converges in $L^2(W)$ to $w(s)$ as $\e\to 0$, then
$v_\e(s)$ also two-scale converges to $w(s)$ and there is a function
$w_1(s,t)\in L^2(W\times T)$,
so that $\nabla_sv_\e(s)$ two-scale converge to $\nabla_s w(s)+\nabla_t
w_1(s,t)$.
For example, if  functions $v_\e(s)$ have the form $v_\e(s)=u_0(s)+\e
u_1(s,s/\e)$,
then $\nabla_s v_\e(s)$ two-scale converge to $u(s,t)=\nabla u_0(s)+\nabla_t
u_1(s,t)$.

As noted above, the solutions $v_\e(x)$ of equations (\ref{10.1})
are uniformly bounded in $H^1_0(\Omega)$. Consider the
restrictions of these functions on $\Omega'$  and rewrite them  in the
spherical coordinates
on $W$. Then
  any sequence of $v_\e(s)$ has a subsequence
  $v_{\e_j}(s)$, $j\in\mathbb Z_+,\, \e_j \to 0$ as $j \to \infty$,
that weakly converges, in $H^1(W)$,  to some function $w(s)$.
By \cite{Allaire}
$v_{\e_j}(s)$ also two-scale converge to $w(s)$, as $\e\to 0$,
and there is a function $w_1(s,t)\in L^2(W\times T)$
so that $\nabla v_{\e_j}(s)$ two-scale converge to $\nabla_s w(s)+\nabla_t
w_1(s,t)$.

Let us now multiply both sides of equation (\ref{10.1sp}) by a test function
$\phi(s)+\e\phi_1(s,s/\e)$,
where $\phi(s)\in C^\infty_0(W)$ and  $\phi_1(s,t)\in C^\infty_0(W;C^
\infty_\#(T))$,
and integrate over $W$. Using integration by parts, we obtain
\beq\label{test 1}
& &\int_{W}\sigma^*(s,\frac s\e)[(\nabla_s +ib^*(s)) v_\e(s)]
\,\cdotp \overline {(\eta_1(s,\frac s\e)+\e\eta_2(s,\frac s\e))}\,ds+\\
\nonumber
& &\hspace{-1cm}+\int_W Q^*(s)v_\e(s)\overline {[\phi(s)+\e \phi_1(s,\frac
s \e)]}\,ds
=\int_{\Omega'} \hat f(s)
\overline {[\phi(s)+\e \phi_1(s,\frac s \e)]}\,ds, \label{test 1sec}
\eeq
where
\ba
& &\eta_1(s,t)=
\nabla_s \phi(s)+ib^*(s)\phi(s) +\nabla_t \phi_1(s,t),\\
& &\eta_2(s,t)=\nabla_s \phi_1(s,t)
+ib^*(s)\phi_1(s,t).
\ea
Substitute  $\e=\e_j$ in (\ref{test 1}) and let $j\to \infty$. First, we
observe that since
$\eta_2(s,s/\e)$, $\e\in (0,1)$ are uniformly bounded in $L^2(W)$,
the terms in (\ref{test 1}),
involving $\eta_2$, tend to zero. Second,
{as
$v_{\e}(s)$ converge in strong topology
of $L^2(W)$ to $w$, the
first integral in (\ref{test 1sec})
tends to $\int_W Q^*(s) w(s)\overline {\phi(s)}\, ds$. }
Third, the inner product of $\hat f(s)$ and $\e \phi_1$ in the
last integral in (\ref{test 1sec}) goes to zero.
Fourth, the
functions $\psi_\e(s)=\sigma^*(s,s/\e)\eta_1(s,s/\e)$ two-scale converge to
the function
$\psi(s,t)=\sigma^*(s,t)\eta_1(s,t)$. Since
$\psi(x,y)\in L^2(\Omega; C^0_\#(T))$, we have by
\cite[Lem.\ 1.3]{Allaire} that
\beq\label{A 8}
\lim_{\e\to 0}\|\psi_\e\|_{L^2(W)}=\|\psi\|_{L^2(W\times T)}.
\eeq
Furthermore, as $\nabla v_{\e_j}(s)$ two-scale converges
to $\nabla_s w(s)+\nabla_t w_1(s,t)$,
{it follows from (\ref{A 8}) and \cite[Thm.\ 1.8]{Allaire}, that
\ba
& &\int_{W}\sigma^*(s,\frac s\e_j)[(\nabla_s +ib^*(s)) v_{\e_j}(s)]
\,\cdotp \overline {\eta_1(s,\frac s\e_j)}\,ds \rightarrow \\
\nonumber
& &\int_{W}\int_T\sigma^*(s,t)[\nabla_s w(s)+\nabla_t w_1(s,t)]\cdot
{\overline {\eta_1(s, t)}}\, ds dt,
\ea
as $j \to \infty$.
Summarizing, we see that }
\beq\label{test 2}
& &\int_{W}\int_T\sigma^*(s,t)[\nabla_s w(s)+\nabla_t w_1(s,t)+ib^*(s)
w(s)]\\
\nonumber
& &\quad\quad
\,\cdotp \overline {[\nabla_s \phi(s)+ib^*(s)\phi(x) +(\nabla_t
\phi_1)(s,t)]}\,dsdt+\\
& &+\int_{W}\int_T Q^*(s)w(s) \overline {\phi(s)}\,dsdt=\int_{W} \hat f(s)
\overline {\phi(s)}\,ds . \nonumber
\eeq

Taking $\phi(s)=0$ in (\ref{test 2}) and varying $\phi_1(s,t)$
over all test functions,
integration by parts 
{with respect to $t$} shows that
\beq \label{10.1 B}\quad\quad\quad
\nabla_t \cdot \sigma^*(s,t)
[\nabla_s w(s)\hspace{-1mm}+\hspace{-1mm}\nabla_t w_1(s,t)\hspace{-1mm}+
\hspace{-1mm} ib^*(s)w(s)]\hspace{-1mm}=\hspace{-1mm}0\ \hbox{a.e.\, in}\ W\times
T.\hspace{-1cm}
\eeq
Since $\nabla_s w(s)+ib^*(s)w(s)$ depends only on $s$ and thus can be
considered
as a constant vector in the $t$-variable, we see that
\beq \label{u1}\quad\quad
w_1(s,t)= \sum_{j=1}^3 \left(\frac{\p w}{\p s^j}(s)+ib^*_j(s) w(s)\right)\,
W^j(s, t)
\quad\hbox{a.e. in}\ W\times T,\hspace{-1cm}
\eeq
where $W^j$ are solutions of the cell equations (\ref{cell equation}).

On the other hand, taking $\phi_1(s,t)=0$  in (\ref{test 2}) and
varying
$\phi(s)$ over all test functions, we see, using
integration by parts 
{with respect to $s$,} that
\beq \nonumber
& &\hspace{-1cm}-(\nabla_s+ib^*(s)) \cdot \left(
\int_T \sigma^*(s,t)
[\nabla_s w(s)+\nabla_t w_1(s,t)+ib^*(s)w(s)]dt\right)+
\\ & &\quad\quad\quad +Q^*(s)w(s)=\hat f(s) \,\, \hbox{a.e. in}\,\, W.
\label{10.1 CA}
\eeq Comparing the above integral with formula (\ref{corrector}) for
$\sigma_{hom}$,
we see that (\ref{u1}) and
(\ref{10.1 CA}) imply that
\ba
-(\nabla_s+ib^*(s)) \cdot \sigma^*_{hom}(s)(\nabla_s+ib^*(s))w(s)+Q^*
(s)w(s)=\hat f(s)\quad\hbox{in }{W},
\ea
or equivalently, in the Cartesian coordinates
\beq
  \label{10.1 C2A cart}\quad\quad
-(\nabla+ib) \cdot \sigma_{hom}(\nabla+ib)w+Qw=\tilde
f\quad\hbox{in }{\Omega'}.\hspace{-1cm}
\eeq

Since $\Omega\setminus \{{\it O}\}$ can be covered with coordinate
neighborhoods $\Omega'$ used above,  the equation
(\ref{10.1 C2A cart}) is valid in the domain $\Omega\setminus \{{\it O}\}$.
Moreover, since all the mass densities $\sigma_{hom}$ and $\sigma_\e$
are the same near the origin we see that all functions $v_{\e_j}(x)$ satisfy
the equation  
(\ref{10.1 C2A cart}) near the origin. As $v_{\e_j}(x)$ converge weakly in
$H^1_0(\Omega)$
to $w(x)$, the equation (\ref{10.1 C2A cart}) is valid in $\Omega$.
This means that $w(x)$, as well as $v(x)$,
is the solution of the equation (\ref{10.1b}).
By the Lax-Milgram Theorem, equation  (\ref{10.1b}) has a  unique solution,
and thus $w=v$.

Summarizing, we have now shown that an arbitrary sequence of
the original family $v_\e(x)$ has a subsequence that  weakly converges, in
$H^1_0(\Omega)$, to the solution $v(x)$ of
equations (\ref{10.1b}). This shows that
 the whole family  $v_\e(x)$ of the solutions of the equations
(\ref{10.1}) converge weakly  to $v(x)$ in $H^1_0(\Omega)$.
\hfill \hfill\proofbox

\subsection{Approximation of the mass densities $\sigma_R$.}
\label{isotropic}

{Next we apply the above results to approximate the mass densities
$\sigma_R$
for a fixed
$R>1$.
In the forthcoming analysis, we will choose the function $h$ to be
\beq
\label{13a} h(s, r')=\frac {a(s)} {1+b(s)p(r')},
\eeq
where $p(r')$ is a fixed positive, {smooth} $1$-periodic function of $r'$
and $a(s),b(s)$ are positive. For any $x \in \Omega$ and a given
function $p(r')$,
the pair $(\omega_{1}(x), \omega_{2}(x))$ in (\ref{simga hom}), (\ref{85a})
depends only on the values of functions
$(a(s(x)), b(s(x)))$.  By choosing $(a(s), b(s))$ appropriately, it
is
possible to achieve
  any value
\ba
(\omega_{1}(x), \omega_{2}(x))\in \R_+^2, \quad \hbox{with}\,\,
\omega_{1}(x)\leq \omega_{2}(x).
\ea

We can
choose $a(x)=a_R(x)$ and $b(x)=b_R(x)$ so that $\sigma_{hom}(x)$
corresponds to the approximate cloaking mass density, that is,
\beq\label{eq: right homogenization}
\sigma_{hom}(x)=\sigma_R(x),\quad x\in \Omega.
\eeq

In sequel,  we will denote by $\sigma_R(x,r')=h_R(x,r')I\,$ the isotropic
mass densities
for which the homogenized mass densities $\sigma_{hom}$ satisfy
(\ref{eq: right homogenization}).}
Observe that, for  $R>1$, we 
{can choose $ h_R(s, r')$ so that }
\beq \label{estimate}
0<c_1(R) \leq h_R(s, r') \leq c_2, \quad c_1(R) \to 0 \,\, \hbox{as}\,\, R
\searrow 1.
\eeq

Let $\la <0$
and
  ${\mathcal B}_{R, m, \e}$
be the operators
\ba
& &{\mathcal B}_{R, m, \e}u=-\nabla^b\cdotp \sigma_{R,\e}\nabla^b
u+g_m^{1/2}qu,\\
& & {\mathcal D}({\mathcal B}_{R, m, \e})=\{
u\in H^1_0(\Omega):\,  -\nabla^b\cdotp \sigma_{R,\e}\nabla^b u \in L^2
(\Omega)
\}.
\ea

\begin{lemma} \label{conv}
For any $\la<0$ and $f\in L^2(\Omega)$,
\ba
\lim_{\e\to 0}
({\mathcal B}_{R, m, \e}-\la g_m^{1/2})^{-1}f=
({\mathcal B}_{R, m}-\la g_m^{1/2})^{-1}f,
\ea
{where ${\mathcal B}_{R,m}$ is as in (\ref{03.10.1}),} and  the limit holds
 both in the sense of the strong topology
in $L^2(\Omega)$ and weak topology in $H^1(\Omega)$.
\end{lemma}

{\bf Proof.}
{This  follows immediately from Prop. \ref{prop.
homogenization} by taking
$
Q= (q-\la) g_m^{1/2}
$.
\hfill \hfill\proofbox
}

As  in the previous section, one can analyze the
convergence
of the resolvents in more detail.
{To this end, introduce  operators 
${\mathcal A}_{R, m,  \e}=g_m^{-1/2}{\mathcal B}_{R, m,  \e}$
that is,
\beq \label{A_Re}
& &\quad{\mathcal A}_{R, m, e}=-g_m^{-1/2}\nabla^b \cdot \sigma_{R, \e} \nabla^b
u+q u,\\
& &\nonumber \D({\mathcal A}_{R, m, \e})={\mathcal D}({\mathcal B}_{R, m,
\e})=\{
u\in H^1_0(\Omega):\,  -\nabla^b\cdotp \sigma_{R,\e}\nabla^b u \in L^2
(\Omega)
\}.
\eeq
Clearly, the operators ${\mathcal A}_{R, m, e}$ are self-adjoint in
$L^2(\Omega, \,g_m^{1/2}dx)$
and
\beq
\label{1.03.11}
(\A_{R,m,\e}-\la)^{-1} f=
(\B_{R,m,\e}-\la g_m^{1/2})^{-1}( g_m^{1/2} f)
\eeq
for $\la\not\in\hbox{spec}\,(\A_{R,m,\e})$ and $f\in L^2(\Omega)$,
{\it cf.} (\ref{03.10.1}), (\ref{ARm}).

\begin{lemma} \label{weak}
Let $R>1$, $m\in \Z_+$ and $K \subset \C$ be  compact and  such that 
$\dist(K,\hbox{spec}({\ttilde{\mathcal A}}_{R,m})) \geq \delta_0>0$. Then

\noindent
(i) There is  an $\e_0=\e_0(K,R, m)$ such that such that for $0<\e<\e_0$ 
\ba
\|({\mathcal A}_{R, m, \e}-\la )^{-1}\|_{L^2(\Omega,  g_m^{1/2} dx)\to
L^2(\Omega,  g_m^{1/2} dx)}
\leq \frac {2}{\delta_0},\quad
\la \in K,
\ea

\noindent
(ii) 
For any $f\in L^2(\Omega)$,
\beq\label{08.10.7}
 \lim_{\e \to 0}({\mathcal A}_{R, m, \e}-\la )^{-1}f
=
({\mathcal A}_{R,m}-\la )^{-1}f
\eeq
in the strong topology of $L^2(\Omega)$ and  weak topology of
$H^1_0(\Omega)$, with convergence
being uniform for $\la \in K$.
\end{lemma}

{\bf Proof.}
(i) Since the 
${\mathcal A}_{R, m, \e}$ are self--adjoint operators
on  $L^2(\Omega, g_m^{1/2} dx)$, it is enough to show that
there is $\e_0>0$ such that the
operators ${\mathcal A}_{R, m, \e}-\mu$
are invertible for $|\la-\mu|\leq \delta_0/2$ if $\e <\e_0$.

Assume that there are $\e_n \searrow 0,\, \mu_n \to \mu,\, |\la-\mu| \leq
\delta_0/2$, 
and
  $u_n\, $, with
$\|u_n\|_{L^2(\Omega, g_{m}^{1/2} dx)}=1$,
such that
\ba
{\mathcal A}_{R, m, \e_n} u_n-\mu_n u_n =0.
\ea
{Rewrite this as
\beq \label{49}\quad\quad
-\nabla\cdotp \sigma_{R,  \e_n} \nabla u_n=f_n:=-g_{m}^{1/2} q u_n+\mu_n
g_{m}^{1/2}
u_n, \quad
 u_n|_{\p \Omega}=0.
\eeq
Taking into account the uniform boundedness of the right hand side of
(\ref{49})
in $L^2(\Omega)$,
it follows from (\ref{10.1}), (\ref{no_0_eigenvalue}), with $Q=0$, that
$
\|u_n\|_{H^1_0(\Omega)}  \leq C
$,
for some $C>0$ independent  $\e_n$. }

Thus, up to a subsequence, there exists $u \in H^1_0(\Omega)$
{such that $
u_n \to u$ as $n \to \infty$
in the weak topology of $H^1(\Omega)$} and strong topology of $L^2(\Omega)$.
This shows, in particular, that $\|u\|_{L^2(\Omega, g_{m}^{1/2} dx)}=1$.

Let us show that $u$ is an eigenfunction of $\A_{R, m}$, see
(\ref{ARm}),
corresponding to { the eigenvalue} 
$\mu$; as $|\la- \mu|\leq \delta_0/2$ with $\la \in K$
and $\dist(K,\hbox{spec}({\ttilde{\mathcal A}}_{R,m}))>\delta>0$, this would
yield a contradiction.} 
{We compare $u_n$ with $v_n$, 
the solution to
\beq \label{51}\quad
-\nabla^b \cdotp \sigma_{R,\e_n} \nabla^b v_{n} ={\tilde f}_n:=
-g_{m}^{1/2}q u+\mu g_{m}^{1/2} u,
\quad
  v_n|_{\p \Omega}=0.
\eeq
Letting $ \mu_n \to \mu$,
(\ref{51}) and 
{the weak convergence 
$u_n \to u$ in $H^1(\Omega)$ imply
that 
\ba
\lim_{n\to \infty}\|f_n -{\tilde f}_n\|_{L^2(\Omega)}=0 .
\ea
Appealing again to  (\ref{estimate}), with $R>1$  fixed, 
}
we see that
$u_n -v_n \to 0$ in the
strong topology of   $H^1_0(\Omega)$.

Using Proposition\ \ref{prop. homogenization} with $Q=0$, we see that $v_n
\weakto v$
in $H^1(\Omega)$,
where
\ba
-\nabla^b \cdotp \sigma_{R} \nabla^b v=-g_m^{1/2}qu +\mu g_m^{1/2} u,\quad
v|_{\p \Omega}=0.
\ea
Summarizing, we have that $u=v$, showing that $u$ is an eigenfunction of
${\ttilde{\mathcal A}}_{R}$
corresponding
to the eigenvalue $\la$.
This
proves claim (i).

(ii)
{Compare the solutions of 
\ba
-\nabla^b \cdotp \sigma_{R,\e} \nabla^b u_{m, \e}
=f_{m, \e}:=f -g_{m}^{1/2}q u_{m, \e}+\la g_{m}^{1/2} u_{m, \e},
\ \
u_{m, \e}|_{\p \Omega}=0.\hspace{-1cm}
\ea
with those of
\ba 
-\nabla^b \cdotp \sigma_{R} \nabla^b u
=f_m:=f-g_m^{1/2}q u+\la g_m^{1/2} u,
\quad
  u|_{\p \Omega}=0.
\ea
It follows from claim (i)
together with (\ref{no_0_eigenvalue}), that  the $u_{m, \e}$ are uniformly
bounded in
$H^1_0(\Omega)$. Choosing a weakly converging subsequence, we show, by
similar reasoning
to the  above,
that the limit is $u_m$. This proves 
\ba
\lim_{\e \to 0}({\mathcal B}_{R, m, \e}-\la g_m^{1/2})^{-1}f
=
({\mathcal B}_{R,m}-\la g_m^{1/2})^{-1}f
\ea
{This,  
(\ref{1.03.11}), and the boundedness of the multiplication operator
$f \mapsto g_m^{1/2}f$ in $L^2(\Omega)$ yields (\ref{08.10.7}).}
\hfill
\hfill\proofbox
}

{For $\mu\not \in\spec(\A_{R,m,\e})$ 
we denote by  $N_{R,m,\e}(\mu)$
the subspace of $L^2(\Omega)$ spanned by the
eigenfunctions
of $\A_{R,m,\e}$ corresponding to the eigenvalues in $(-\infty,\mu)$,
and by $P^\mu_{R,m,\e}$  the orthoprojectors in $L^2(\Omega, g_m^{1/2}
dx)$ onto $N_{R,m,\e}(\mu)$.}

\begin{lemma} \label{again_projectors}
{
Let $R>1$, $m\in \Z_+$, and  $\mu\not \in    \hbox{spec\,}({\mathcal
A}_{R,m})$.
Then

\noindent (i)
There is an $\e_{R,m}>0$ such that
$\mu\not \in \hbox{spec\,}({\mathcal A}_{R,m,\e})$ for $0<\e<\e_{R,m}$.

\noindent(ii)
For $f\in  L^2(\Omega)$ we have
\beq
\label{08.10.4}
\lim_{\e\to 0} P^{\mu}_{R,m,\e}f=P_{R,m}^\mu f, 
\eeq 
in the strong topology of $L^2(\Omega)$ and  weak topology of
$H^1_0(\Omega)$. Moreover,
\beq\label{08.10.3}
\lim_{\e\to 0}\dim(N_{R,m,\e}(\mu))=\dim(N_{R,m}(\mu)).
\eeq 
}
\end{lemma}
{\bf Remark.} We note that the strong convergence of the resolvents does
\emph{not}
generally imply the stability of the resolvent set, \cf \cite[Sect.\
VIII.1.2]{Kato}. 

{\bf Proof.} (i) 
{Take $K=\{\mu\}$ and $\delta_0=\dist(K,\hbox{spec}({\ttilde{\mathcal
A}}_{R,m}))$.
Lemma \ref{weak} (i) implies that, for sufficiently small $\e$, 
$\dist(K,\hbox{spec}({\ttilde{\mathcal A}}_{R,m,\e})) >\delta_0/2$.
}

(ii) Relation (\ref{08.10.4}) follows immediately from
(\ref{08.10.7}) and the
Riesz formula  for the projectors, \cf (\ref{Riesz}).
For the proof of (\ref{08.10.3}),  let us consider $\lambda_0\in
\spec(\A_{R,m})$
and $a,b\in \R$, $a<\la_0<b$  such that $[a,b]\cap
\spec(\A_{R,m})=\{\la_0\}$.
Then by (i) there are $\theta(\e),\, \theta(\e) \to 0$ as $\e \to 0$ such
that 
\ba
& &\hbox{spec\,}({\ttilde {\mathcal A}}_{R, m, \e})\cap [a+ \theta(\e)
,\la_0-\theta(\e))=\emptyset,\\
& &\hbox{spec\,}({\ttilde {\mathcal A}}_{R, m, \e})\cap (\la_0+ \theta(\e)
,b-\theta(\e)]=\emptyset.
\ea

Let $P^\e=P^{b-\theta(\e)}_{R,m,\e}-P^{a+\theta(\e)}_{R,m,\e}$
and $P^0=P^{b}_{R,m}-P^{a}_{R,m}$.
Then, by (\ref{08.10.4}),
$
\lim_{\e\to 0} P^\e f=P^0 f
$
in the strong topology of $L^2(\Omega)$ and  weak topology of
$H^1_0(\Omega)$.

To establish (\ref{08.10.3}), it is sufficient to show that
\beq\label{aux result}
\lim_{\e \to 0} \hbox{dim}(\hbox{Ran\,}(P^\e))=
\hbox{dim}(\hbox{Ran\,}(P^0)).
\eeq
Indeed, if we show (\ref{aux result}) for each eigenvalue of $\A_{R,m}$
smaller than $\mu$, equality (\ref{08.10.3}) follows.

To show (\ref{aux result}), we note that the ranges of the 
orthoprojectors $P^\e,\,P^0$ are also subspaces of $H^1_0(\Omega)\subset
L^2(\Omega)$.
We show first that 
\beq\label{dimension claim 1}
\liminf_{\e \to 0} \hbox{dim}(\hbox{Ran\,}(P^\e))
\geq \hbox{dim}(\hbox{Ran\,}(P^0)).
\eeq
On the contrary, if this does not hold, there is a sequence $\e(n)\to 0$
such that
\beq \label{08.10.10}
\hbox{dim}(\hbox{Ran\,}(P^{\e(n)})) <\hbox{dim}(\hbox{Ran\,}(P^0))=k_0.
\eeq
Denote by $\eta_k,\, k=1, \dots, k_0$, an $L^2(\Omega)$--orthonormal basis 
in $\hbox{Ran\,}(P^0)$. 
(Note that as $\hbox{Ran\,}(P^0) \subset H^1_0(\Omega)$ is finite
dimensional  all norms are equivalent). 
Introduce $\eta_{n,k} = P^\e \eta_k\in \hbox{Ran\,}(P^\e),\, k=1, \dots,
k_0$.
Consider the
Gram-Schmidt matrices,
$G^n=[G^n_{kl}]_{k,l=1}^{k_0}$,
\ba
G^n_{kl}= \int_\Omega \eta_{n, k}(x) {\overline \eta_{n, l}(x)}\,  dx.
\ea
By  (\ref{08.10.4})
{and orthonormality of   $\eta_k$, the matrix $G^n$ is
invertible for sufficiently large $n$. Thus, $\eta_{n,k},\, k=1, \dots, k_0$
are linearly independent. }
 This implies that
$\hbox{dim}(\hbox{Ran\,}(P^{\e(n)}))\geq k_0$, contradicting
(\ref{08.10.10}). This proves (\ref{dimension claim 1}).

Assume next that
\beq \label{02.10.0}
\limsup_{\e \to 0}\, \hbox{dim}(\hbox{Ran\,}(P^\e))
>\hbox{dim}(\hbox{Ran\,}(P^0)).
\eeq
Then there is a sequence $\e(n)$, 
such that 
\ba
\hbox{dim}(\hbox{Ran\,}(P^{\e(n)})) >\hbox{dim}((\hbox{Ran\,}(P^0)).
\ea
Thus, there are  $u_n$ such that 
\beq \label{02.10.1}
& &u_n\in \hbox{Ran\,}(P^{\e(n)}), \quad  \|u_n\|_{L^2(\Omega,
g^{1/2}_{m}dx)}=1,
\\
 \label{02.10.1 b}
& &
(u_n, v)_{L^2(\Omega,g_m^{1/2}dx)}=0,\,\, \hbox{for all}\,\, v \in
\hbox{Ran\,}(P^0).
\eeq
Then 
\ba
u_n= \sum_{k=1}^{k_n} u_{n, k} \psi_{n, k},\quad \sum_{k=1}^{k_n} |u_{n,
k}|^2=1,
\quad k_{n}=\hbox{dim}(\hbox{Ran\,}(P^{\e(n)})),
\ea
where $\psi_{n, k}$ are $L^2(\Omega, g^{1/2}_{m}dx)$--orthonormal
eigenfunctions of $\A_{R, m, \e(n)}$,
\ba
\A_{R, m,\,\e(n)} \psi_{n, k}= \la_{n, k} \psi_{n, k},\quad |\la_{n, k}
-\la_0| \leq \theta(\e(n))
\ea
Therefore,
\beq \label{02.10.3}\quad\quad
\A_{R, m, \e(n)} u_n -\la_0 u_n = f_n:=\sum_{k=1}^{k_n} u_{n, k} (\la_{n, k}
-\la_0)
\psi_{n, k}
\rightarrow 0 \,\, \hbox{as}\,\, n\to \infty\hspace{-1cm}
\eeq
in $L^2(\Omega, g^{1/2}_{m}dx)$,
so that
\ba
&&\int_\Omega  \left( \sigma_{R, \e(n)} \nabla^b u_n \cdot \overline {
\nabla^b
u_n} + g^{1/2}_{m} (q-\la_0)|u_n|^2\right)
\, dx\\
& &\ \  =  \left(f_n,
u_n \right)_{L^2(\Omega, g^{1/2}_{m}dx)} \to 0 \hbox{ as } n \to
\infty.\nonumber
\ea
 Together with (\ref{02.10.1})
and the fact that  $q \in L^\infty(\Omega)$,  this implies that the forms
$\int_\Omega \sigma_{R, \e(n)} \nabla^b u_n \cdot \overline {  \nabla^bu_n}
\,dx$ 
are uniformly bounded in $n$. 
Recalling (\ref{no_0_eigenvalue}), we see that
  with some $c>0$, 
\ba
\|u_n\|_{H^1_0(\Omega)} <c, \quad \hbox{for all}\,\, n \in \Z_+.
\ea 
{Restricting  to a  subsequence of the  $\e(n)$,
  assume there exists $u_0 \in H^1_0(\Omega)$}
such that
\beq \label{02.10.5}
\lim_{n\to \infty} u_n=u_0 \quad \hbox{weakly in $H^1_0(\Omega)$ and
strongly in
$L^2(\Omega)$,}
\eeq
so that also $\|u_0\|_{L^2(\Omega, g^{1/2}_{m}dx)} =1$.
Let us show that
$
{\mathcal A}_{R, m}u_0= \la_0 u_0,
$, 
contradicting {the fact that $u_0 \perp \hbox{Ran\,}(P^0)$
in $L^2(\Omega,g_m^{1/2}dx)$, which follows from (\ref{02.10.1}).}
Let $v_n$ be the solutions to
\ba
- \nabla^b \cdot \sigma_{R, \e(n)} \nabla^b v_n=g^{1/2}_{m} (\la_0-q)
u_0,\quad 
v_n|_{\p \Omega}=0.
\ea
From (\ref{no_0_eigenvalue}), together with (\ref{02.10.3}),
(\ref{02.10.5}),
{we see that} $
u_n-v_n \to 0$ as $n\to \infty$ in $L^2(\Omega, g^{1/2}_{m}dx).$
On the other hand, by Prop. \ref{prop. homogenization} with $Q=0$,
$
v_n \to v_0 \hbox{ in }
L^2(\Omega)  \hbox{ as }\,\, n\to \infty 
$,
where $v_0$ is the solution to
\ba
- \nabla^b \cdot \sigma_{R} \nabla^b v_0=g^{1/2}_{m} (\la_0-q) u_0,\quad 
v_n|_{\p \Omega}=0.
\ea
Hence, $v_0=u_0 \neq 0$ is an eigenfunction of ${\mathcal A}_{R, m}$ 
{and belongs in $\hbox{Ran\,}(P^0)$. This proves} the claim.
}}
\hfill \hfill\proofbox

\section{Approximate acoustic cloaking}
\label{isotropic-to-singular}

  In this section, we  show that, for $\la \notin
\hbox{spec}({\ttilde{\mathcal
A}}_1)$, it is possible to approximate the solutions to the singular,
anisotropic acoustic
equation
\ba
{{\mathcal A}}_{1} u-\la  u =f,\quad f \in L^2(\Omega),
\ea
by  solutions to certain non-singular, isotropic acoustic equations. Namely,

\begin{theorem}
\label{approximation A}
 Let $K\subset \C$ be a compact set such that
$K \cap  \hbox{spec\,}({\mathcal A}_1)= \emptyset$, and  $\la \in K$.
Then, for $f\in L^2(\Omega)$,
\ba
\lim_{R\to 1}\lim_{m\to \infty}\lim_{\e\to 0}
({\mathcal A}_{R,m,\e}-\la)^{-1}f=
( {\mathcal A}_{1}-\la )^{-1}f,
\ea
in the strong topology of $L^2_g(\Omega)$ and weak topology of
$H^1_g(\Omega)$,
uniformly with respect to $\la\in K$.
\end{theorem}

{\bf Proof.}
By  Lemma \ref{weak},  (\ref{08.10.7}), it follows that
\beq\label{M1 B}
\lim_{\e\to 0}
({\mathcal A}_{R,m,\e}-\la)^{-1}f=
({\mathcal A}_{R,m}-\la )^{-1}f
\eeq
in the strong topology of $L^2(\Omega)$
and weak topology of $H^1(\Omega)$, uniformly in $\la\in K$.}
Using  Lemma \ref{lem A} we obtain
\beq\label{M2}
\lim_{m\to \infty}
({\mathcal A}_{R,m}-\la )^{-1}f=
({\mathcal A}_{R}-\la )^{-1}f
\eeq
in the strong topology  of $H^1(\Omega)$, uniformly in $\la\in K$.
Using (\ref{inclusion}) and (\ref{inclusionH}), the convergences
in (\ref{M1 B}) and  (\ref{M2}) are valid  also in $L^2_g(\Omega)$ and
$H^1_g(\Omega)$.

As  $f\in L^2(\Omega)\subset L^2_g(\Omega)$ we have 
  by Lemma \ref{cor:13.1}
\ba
\lim_{R\searrow 1}
({\mathcal A}_{R}-\la )^{-1}f=
({\mathcal A}_{1}-\la)^{-1}f,
\ea
in the strong topology of $L^2_g(\Omega)$
and weak topology of $H^1_g(\Omega)$, uniformly with respect to $\la\in K$.
This equation, together with (\ref{M1 B}),  (\ref{M2}), proves the claim.
\hfill \hfill\proofbox

Let us turn our attention to the convergence of the DN maps.
For $R>1$ and $\e>0$, the DN map for the acoustic equation,
\beq \label{acoustic3}
- \nabla^b \cdot \sigma_{R, \e} \nabla^b u+g^{1/2}q u-\la g^{1/2}_m u =0,
\quad u|_{\p \Omega}=h,
\eeq
is the operator
\beq \label{acousticDN}
\Lambda_{R, \e, m}^\la: h \mapsto (\p_\nu + i \nu\,\cdotp b)u_{R, \e,
m}^h|_{\p \Omega}.
\eeq
Here $u_{R, \e, m}^\la$ is the unique solution to (\ref{acoustic3}) assuming
$\la \notin
\hbox{spec}({\mathcal A}_{R, \e,m})$ and $\p_\nu$ is the normal (radial)
derivative at $\p \Omega$.

{\nnntext Recall that by Lemma \ref{lem: DN equivalence} the DN map $
\Lambda_{out}^\la$
defined in (\ref{Dirichlet-to-Neumann map R=1}) corresponds  to  the
boundary
measurements for equation (\ref{a4}). 
{In the following we assume, for technical simplicity, that $b \in
C^1(\Omega,\, \R^3)$.}

\begin{corollary} \label{cor;acousticDN}
{Assume that the magnetic field $b$ is $C^1$-smooth on $\overline \Omega$}
and $h \in H^{3/2}(\p \Omega),\, [a, b] \cap \hbox{spec}({\ttilde{\mathcal
A}}_{1}) = \emptyset$.
Then, uniformly with respect to $\la \in [a, b]$,
\beq \label{ontheboundary}
\lim_{R\to 1}\lim_{m\to \infty}\lim_{\e\to 0}
\left(\Lambda_{R,m,\e}^\la h -
\Lambda_{1}^\la h \right) =0,
\eeq
where the convergence is  strong in the $H^{1/2}(\p \Omega)$ topology.
\end{corollary}

{\bf Proof.}
Let $h^e \in H^2(\Omega),\, \hbox{supp}(h^e) \subset \{x: 7/3 \leq |x| \leq
3\}$, be an extension
of $h$, that is, 
\beq
\label{10.09.11}
h^e|_{\p \Omega}=h, \quad
\|h^e\|_{H^2(\Omega)} \leq C,\,\, \hbox{ if} \,\, \|h^e\|_{H^{3/2}(\p
\Omega)} \leq 1.
\eeq
Then the solution $u_{R,m,\e}^\la(h)$ to (\ref{acousticDN}) may be
represented as
\ba
u_{R,m,\e}^\la(h) =h^e +v_{R,m,\e}^\la(h), \quad u_{1}^\la(h)
=h^e+v_1^\la(h);
\ea
where
\beq \label{20.09.11}& &\quad v_1^\la= - ({\mathcal A}_{1}-\la)^{-1} F^\la(h),\\
\nonumber & &v_{R,m,\e}^\la(h)=- ({\mathcal B}_{R,m,\e}-\la g_m^{1/2})^{-1} F^\la(h)=
- ({\mathcal A}_{R,m,\e}-\la)^{-1} F^\la(h),\\ \nonumber
& &F^\la(h)= - \nabla^b\cdotp\nabla ^b h^e-\la h^e \in L^2(\Omega),\quad 
\hbox{supp}(F^\la(h)) \subset \overline\Omega\setminus B(\frac 73).\nonumber
\eeq
Here we use the fact that, in the layer $\overline\Omega\setminus B(\frac 73)$,
$ \sigma^{ij}_{R, \e} = \sigma^{ij}_1=\delta^{ij}, \quad g_m =g =1.
$
By Theorem \ref{approximation A},
\ba
\lim_{R\to 1}\lim_{m\to \infty}\lim_{\e\to 0}\|v_{R,m,\e}^\la(h)-
v_1^\la(h)\|_{L^2(\{ 7/3 \leq |x| \leq 3\})} =0.
\ea
It then follows from \cite[Thm. 9.13] {GT} that
\ba
\lim_{R\to 1}\lim_{m\to \infty}\lim_{\e\to 0}\| v_{R,m,\e}^\la(h)-
v_1^\la(h)\|_{H^2(\{ 8/3 \leq |x| \leq 3\})} =0,
\ea
so that, by trace theorem for Sobolev spaces,  we have
\beq \label{b-regularity}
\lim_{R\to 1}\lim_{m\to \infty}\lim_{\e\to 0} (\p_\nu+ i \nu\,\cdotp b)
\left(u_{R,m,\e}^\la(h)-
u_1^\la(h) \right)|_{\p \Omega}=0,
\eeq
strongly in $H^{1/2}(\p \Omega)$.
\hfill \hfill\proofbox

Our next goal is to show the existence of a sequence of non-singular
isotropic
mass densities  and non-singular bulk moduli,
uniformly bounded (in $n$)
from above but not from below, with the corresponding acoustic operators 
approximating the singular
acoustic operator
${\mathcal A}_1-\la $.

\begin{theorem}
\label{sequence-approximation}

There exist  sequences $R(n)\to 1$, $m(n)\to \infty$, and $\e(n)\to 0$ 
such that, for any $f \in L^2(\Omega)$, $\la \in \C,\, \mu \in \R,\, \la,
\mu \notin 
\hbox{spec}({\ttilde {\mathcal A}}_1)$,
 \beq
  \label{final-estimate1}
& &\lim_{n\to\infty}
({\mathcal A}_{R(n),m(n) \e(n)}-\la)^{-1} f=
({\mathcal A}_{1}-\la)^{-1} f \quad\hbox{in }\,\,L^2_g(\Omega),\\
& &  \label{final-estimate2}
\lim_{n\to \infty}P^\mu_{R(n), m(n),\e(n)} f
=P^\mu_1 f \quad\hbox{in }\,\, L^2_g(\Omega),\\
& &  \label{final-estimate3}
\lim_{n\to \infty} \hbox{dim}(N_{R(n), m(n),\e(n)}(\mu))
=\hbox{dim}(N_{1}(\mu)).
\eeq
{For compact $K \subset \C$ with $ K \cap \hbox{spec}({\mathcal
A})=\emptyset$,  for 
$n$ sufficiently large one has $ K \cap \hbox{spec}({\mathcal A}_{R(n),m(n)
\e(n)})=\emptyset$ and   the limits are   uniform in $\la \in K $ and
$\mu\in K\cap \R$.}
\end{theorem}

{\bf Proof.}
Let $\{f_p:\, p=1, 2, \dots\}$ be a dense set in $L^2(\Omega)$.
Let $K_n\subset \C,\, n=1, 2, \dots,$ be a sequence of compact sets so that
\beq
\label{93a}
& &K_n \cap \hbox{spec}({\ttilde {\mathcal A}}_1)=\emptyset,
\quad K_n \subset K^{int}_{n+1}, \\ \nonumber
& &\bigcup_{n=1}^{\infty} K_n = \C \setminus
\hbox{spec}({\ttilde {\mathcal A}}_1), \ \
\hbox{dist}( \hbox{spec}({\ttilde {\mathcal A}}_1),K_n) \geq n^{-1/2}.
\eeq
As $L^2(\Omega)\subset L^2_g(\Omega)$, it follows from
 Lemmas 
\ref{cor:13.1} and \ref{R-dimension}, 
that for any $n \in \Z_+$,  there exists a $R(n) >1$
such that
\beq \label{5.03.11}
& &\quad \|({\mathcal A}_{R(n)}-\la)^{-1} f_p - ({\mathcal A}_1-\la)^{-1} f_p \|_g
< \frac{1}{n},\\
& & \nonumber
\hbox{dim}(N_{R(n)}(\mu))=\hbox{dim}(N_{1}(\mu)), \quad
\dist(\spec({\mathcal A}_{R(n)}),K_n)\geq n^{-1/2}/2,\\ \nonumber
& &\|P^\mu_{R(n)} - P^\mu_1\|_{L^2_g \to L^2_g}\leq \frac 1n,\quad
\,\, \|P^\mu_{R(n)} f_p - P^\mu_1 f_p\|_{L^2_g} \leq \frac 1n,
\eeq
for all $p=1, \dots, p_n,\, \la \in K_n,\, \mu \in K_n \cap \R$. Here $p_n$
is
defined so that
\ba
\{f \in L^2(\Omega): \|f\|_{L^2(\Omega)} \leq n\} \subset
N_{1/n}(\{f_p\}_{p=1}^{p_n}), 
\ea
where $N_\e(S)$ denotes the $\e-$neighborhoods of $S$ in
$L^2(\Omega)$.

Having chosen $R(n)$, using (\ref{inclusion}) and Lemma  \ref{lem A}
and Corollary \ref{m-dimension}, we choose $m(n)$ such
that
\beq   \label{BBB}
& &\quad \|({\mathcal A}_{R(n),m(n)}-\la)^{-1} f_p
-({\mathcal A}_{R(n)}-\la)^{-1} f_p \|_g
< \frac{1}{n},
\\ \nonumber
& &\hbox{dim}(N_{R(n),m(n)}(\mu))=\hbox{dim}(N_{1}(\mu)),
\\ \nonumber
& & \dist(\spec({\mathcal A}_{R(n),m(n)}),K_n)\geq n^{-1/2}/3,\\
 \nonumber
& &  \|P^\mu_{R(n),m(n)} - P^\mu_{R(m)}\|_{L^2 \to L^2_g}\leq \frac 1n,
  \quad \|P^\mu_{R(n), m(n)} f_p - P^\mu_{R(n)} f_p\|_{L^2_g} \leq \frac 1n,
\eeq
for all $p=1, \dots, p_n,\, \la \in K_{n},\, \mu \in K_n \cap \R$. 
Having chosen $R(n)$ and $m(n)$, one can now use (\ref {inclusion}) and
Lemmas
\ref{weak}
and \ref{again_projectors} to
choose $\e(n)>0$ 
so that, 
\beq\label{eq: alte1}
& &\hspace{-5mm}\|({\mathcal A}_{R(n),m(n), \e(n)}\hspace{-1mm}
-\hspace{-1mm}\la )^{-1} f_p
\hspace{-1mm} -\hspace{-1mm} ({\mathcal A}_{R(n),m(n)}
\hspace{-1mm}-\hspace{-1mm}
\la)^{-1} f_p \|_{g}\hspace{-1mm}
<\hspace{-1mm} \frac{1}{n},\hspace{-5mm}\\
\nonumber
& &\hspace{-5mm}\hbox{dim}(N_{R(n),m(n),\e(n)}(\mu))=\hbox{dim}(N_{R(n),
m(n)}(\mu)),\\
\nonumber
& &\hspace{-5mm}\dist(\spec({\mathcal A}_{R(n),m(n),\e(n)}),K_n)\geq 
n^{-1/2}/4,\\ \nonumber
& &\hspace{-5mm} \|P^\mu_{R(n),m(n),\e(n)}f_p
-P^\mu_{R(n),m(n)}f_p\|_{L^2_g}\leq \frac 1n,
\eeq
for all $p=1, \dots, p_n,\, \la \in K_{n}$ and $ \mu \in K_n \cap \R$;
{and,
\beq
\label{3.03.11}\quad\quad\quad\quad
\|({\mathcal A}_{R, m, \e}-\la )^{-1}\|_{L^2(\Omega, g_m^{1/2} dx)\to
L^2(\Omega,  g_m^{1/2} dx)}\leq
6n^{1/2},\quad
\la \in K_n,\hspace{-1cm}
\eeq
\cf (\ref{BBB}) and Lemma \ref{weak} (i).}

Clearly, equations (\ref{5.03.11}), (\ref{BBB}) and (\ref{eq: alte1}) imply
that
\beq \label{1.04.11}
& &\|({\mathcal A}_{R(n),m(n), \e(n)}\hspace{-1mm}
-\hspace{-1mm}\la )^{-1} f_p
\hspace{-1mm} -\hspace{-1mm} ({\mathcal A}_{1}
\hspace{-1mm}-\hspace{-1mm}
\la)^{-1} f_p \|_{g}\hspace{-1mm}
<\hspace{-1mm} \frac{3}{n},
\\ 
\nonumber
& &\hbox{dim}(N_{R(n),m(n),\e(n)}(\mu))=\hbox{dim}(N_{1}(\mu)),\\
\nonumber
& &\|P^\mu_{R(n),m(n),\e(n)}f_p -P^\mu_{1}f_p\|_{L^2_g}\leq \frac 3n,
\eeq
for all $p=1, \dots, p_n,\, \la \in K_{n}, \, \mu \in K_n \cap \R$.

In particular, (\ref{1.04.11}) implies (\ref{final-estimate3}) due to
$\cup_n K_n = \C\setminus
\hbox{spec}(\A_1)$.
It remains to show that,  for $f \in L^2(\Omega)$ 
and a compact set $K$ such that 
$K\cap \hbox{spec}({\ttilde{\mathcal A}}_1)=\emptyset$,
one has that 
$({\mathcal A}_{R(n),m(n), \e(n)}-\la )^{-1} f$ converge to
$({\mathcal A}_{1}-\la)^{-1} f$ in  $L^2_g(\Omega)$ and that convergence
is uniform for all $\lambda\in K$. Indeed, this will provide equation
(\ref{final-estimate1}).
Using the Riesz formula for $P^\mu_{R(n),m(n),\e(n)},\, P^\mu_1$, this will
also prove
equation  (\ref{final-estimate2}).

{Let $n_0 \in \Z_+$ be such that $K \subset K_n$ for all $n\geq n_0$.
Furthermore,} for $n \geq n_0$, there exists $f_{p(n)}$ with $ p(n) \leq
p_n$ such
that 
\beq
\label{D.16.2}
\|f -f_{p(n)}\|_{L^2(\Omega)} <\frac{1}{n}.
\eeq
Clearly, for $\la \in K$, 
\beq \label{D.16.3}
& &\quad\quad \| ({\mathcal A}_{R(n),m(n),\e(n)}   - \la)^{-1} f -
({\mathcal A}_{1} -\la )^{-1} f\|_g \\ \nonumber
& &\,\, \leq
\|({\mathcal A}_{R(n),m(n),\e(n)} -\la)^{-1} f _{p(n)}-
({\mathcal A}_{1} -\la)^{-1} f_{p(n)}\|_g \\ \nonumber
& &\quad + \|({\mathcal A}_{R(n),m(n),\e(n)} -\la)^{-1} (f _{p(n)}-f)\|_g+
\|({\mathcal A}_{1} -\la )^{-1} (f_{p(n)}-f)\|_g.
\eeq
By (\ref{1.04.11}), the first term in the right-hand side of (\ref{D.16.3})
is
bounded by $3/n$.
To estimate the second term, we use the estimates (\ref{g g_m inclusion}),
(\ref{3.03.11}),
and (\ref{D.16.2}), which imply
that this term s bounded by 
$12(2/n)^{1/2}$ for $n\geq n_0$.
Finally, by (\ref{inclusion}) and (\ref{93a}), 
\ba\hspace{-1mm}
\|({\mathcal A}_1- \la)^{-1}\|_{L^2 \to L^2_g}
\hspace{-1mm}\leq
 \hspace{-1mm} \sqrt 8\,\|({\mathcal A}_1-\la)^{-1}\|_{L^2_g
\to L^2_g}\hspace{-1mm} \leq\hspace{-1mm}
 \frac { \sqrt 8}{\hbox{dist}\,(\la,{\rm spec}\,({\mathcal A}_1))}
\hspace{-1mm}\leq\hspace{-1mm}
\sqrt {8 n},
\ea
for $\la \in K$. Thus by  (\ref{D.16.2}),   the third  term on the
right-hand side of (\ref{D.16.3}) is
bounded
by $(8/n)^{1/2}$ for $n\geq n_0$.

Summarizing, we see that the left-hand side of (\ref{D.16.3}) tends to
$0$, as
$n \to \infty$,
uniformly for $ \la \in K,$ thus proving (\ref{final-estimate1}).
\hfill \hfill\proofbox

In the sequel, we will use abbreviated  notations
\beq\label{isotropic objects}
& &\sigma_{(n)} := \sigma_{R(n), \e(n)},\quad g_{(n)}:=g_{m(n)},
\quad
{\mathcal A}_{(n)}:={\mathcal A}_{R(n),m(n), \e(n)}
\eeq
for the sequences
$R(n), m(n), \e(n)$  obtained in Theorem \ref{sequence-approximation}.
{For simplicity, denote by $\Lambda_{(n)}^\la$, rather than
$\Lambda_{R(n),m(n), \e(n)}^\la$,
the corresponding DN map.}

Regarding the convergence of the DN-maps, by the same
arguments
as in proof of Corollary \ref{cor;acousticDN}, we obtain

\begin{corollary}
\label{cor:sequence}
{Assume that the magnetic field $b$ is $C^1$-smooth in $\overline \Omega$}.
{Let $K\subset \C$ be  compact and such that $K \cap
\hbox{spec}({\mathcal A}_1) =\emptyset$.
Then, 
for any $h \in
H^{3/2}(\p \Omega)$,
\beq \label{8.07.11}
\lim_{n\to \infty}\Lambda_{(n)}^\la h =\Lambda_{out}^\la h,
\eeq
where  the limit is  in $H^{1/2}(\p \Omega)$, uniformly with
respect to $\la\in K$.}
\end{corollary}

{\mllltext In particular, when $b$ and 
$q|_{\Omega\setminus B(1)}$ vanish, $\Lambda_{out}^\la$ corresponds to
the measurements on the boundary of a homogeneous ball. Thus (\ref{8.07.11})
means that the isotropic material parameters $\sigma_{(n)}$ and $g_{(n)}$ approximate
an acoustic invisibility cloak as $n\to \infty$.}

\section{Approximate quantum cloaking}
\label{schrodinger}

The results of the previous sections can now be used to obtain approximate
quantum cloaking at
a fixed energy,
for any potential $Q\in L^\infty(\Omega)$ supported inside
the cloaked region $B(1)$.

{\mltext In sequel, we assume that
the conductivities $\sigma_R$ satisfy (\ref{eq: sigma_R conditions})
and that also $\sigma_R\in C^2(\overline \Omega;\R^2)$.}
This yields that also  $\sigma_{R,\e},\sigma_{(n)}\in C^2(\overline
\Omega;\R^2)$.

{Let $E\in \R$ be a given energy level and
\beq\label{Aux potenital}
q^E(x):=\frac 14 Q(x)+\frac 34 E \chi_1(x), 
\eeq
where $\chi_1$ is the indicator function of the ball $B( 1)$.
In this section we assume that $E$ and $Q$ are such that $E$ 
is not in the spectrum of the operator
$\A_1=\A_1^E$ defined in (\ref{a13}) using the potential $q^E$.
{Observe that  
$\hbox{spec}(\A_1^E) = \hbox{spec}(A_{out}) \cup 
 \hbox{spec}(A^{E}_{in}),$
where $A_{out}$ and  $A^{E}_{in}$ are defined by (\ref{A_out}) in $(\ref{A_in})$
using potential $q^E$.
{\mllltext Note that $A_{out}$ is independent of $E$ and 
$A^E_{in}=\frac 14 S_{in}+\frac 34 E$, where
$S_{in}$ is the
Schr\"odinger operator
\beq \label{AchangedS}\nonumber
S_{in}=-\Delta+Q,\,
\D(S_{in})=\{v_2\in H^2(B( 1)):\ \p_\nu v|_{\p B( 1)}=0\}.
\eeq
Observe that}
\beq \label{1.5.12} \quad\quad
E\not \in  \hbox{spec}(\A^E_1)\quad\Longleftrightarrow\quad \hbox{$E\not \in \hbox{spec}(S_{in})$ 
and $E\not\in \hbox{spec}(A_{out})$.} 
\eeq

Then for $n$ large enough,
$E \notin \hbox{spec}(\A_{(n)})$. Next, consider the solutions $u_{n}$ of
\beq\label{non-flat Schrodinger equations}
\quad\quad\quad\left( -g_{(n)}^{-1/2}\nabla^b \cdotp
\sigma_{(n)} \nabla^b+q^E  - E \right) u_n
=0\quad\hbox{in }\,\,\Omega,
\quad u_n|_{\p \Omega}=h.
\eeq
This equation can be converted to a Schr\"odinger equation
with magnetic potential
using
the  gauge transformation,
\beq \label{gauge}
\psi_n(x)=\sigma_{(n)}^{1/2}(x) u_n(x).
\eeq
Then
\beq \label{12.07.11}
& &-\sigma^{-1/2}_{(n)} \nabla^b \cdotp \sigma_{(n)}
\nabla^b(\sigma_{(n)}^{-1/2}\psi_n)=
-\nabla^b\cdotp\nabla^b \psi_n+W_{(n)} \psi_n,
\eeq
where
\beq \label{2.21.11}
\, W_{(n)} =\sigma_{(n)} ^{-1/2}\Delta \,(\sigma_{(n)} ^{1/2}), \quad
\hbox{supp}(W_{(n)}) \subset {\overline {B( 2)}} \setminus B( 1).
\eeq
Thus, using the transformation (\ref{gauge}) we see that 
the acoustic
equation (\ref{non-flat Schrodinger equations}) for $u_n$ is 
equivalent to the 
Schr\"odinger equation for $\psi_n$,
\beq\label{flat equation}
\quad\quad \left(-\nabla^b\cdotp \nabla^b+W_{(n)} +\frac{
g_{(n)}^{1/2}}{ \sigma_{(n)}}\left(q^E-E \right) \right)\psi_n=0, 
\quad \psi_n|_{\p
\Omega}=h,
\eeq
where for the boundary condition we use
$\psi_n|_{\p\Omega}=u_n|_{\p\Omega}=h$, 
since $\sigma_{(n)}=1$ near $\p \Omega$.

Next,  define  the {\it cloaking potential}
\beq
\label{effective-potential}
V^E_{n}(x)&:=&W_{(n)}(x) +\frac{
g_{(n)}^{1/2}(x)}{ \sigma_{(n)}(x)}\left(\frac{3E}{4} \chi_1(x)-E\right)+E\\
&=&W_{(n)}(x) +E(1-\chi_1(x))\left(1-\frac{
g_{(n)}^{1/2}(x)}{ \sigma_{(n)}(x)}\right), \nonumber
\eeq
where we have used the fact that 
$g_{(n)}^{1/2}(x)=8, \, \sigma_{(n)}(x) =2$ for $|x|<1$.
Thus $V^E_{n}=0$ in $B( 1)$. Clearly, $V^E_{n}$ vanishes also near $\p
\Omega$.
Then (\ref{flat equation}) can be written as
\beq\label{flat equation 2}
\left(-\nabla^b\cdotp \nabla^b+V^E_{n}(x)+Q-E\right)\psi_n=0, 
\quad \psi_n|_{\p
\Omega}=h.
\eeq
Now we are ready to prove our main result concerning approximate cloaking in
quantum mechanics.

\begin{theorem}
\label{quantum-cloaking}
Assume that $Q\in L^\infty(\Omega)$ is a function supported in $B( 1)$,
$b \in C^1(\Omega, \R^3)$,
and $E\in \R$ are {\mllltext such that $E \notin \hbox{spec}(A_{out}) 
\cup \hbox{spec}(S_{in})$.} 
Then for any
$h \in H^{3/2}(\p \Omega)$,
\beq \label{appr-quantum}
\lim_{n\to \infty} { \Lambda}^{E}_{V^E_{ n}+Q} h = \Lambda_{out}^E
h\quad\hbox{in } H^{1/2}(\p \Omega).
\eeq
Here $ { \Lambda}^E_{V^E_{ n}+Q}$ are the DN maps,
\ba
{  \Lambda}_{V^E_{ n}+Q}^E: h \mapsto \p_\nu \psi_n|_{\p \Omega}; \,\,
\hbox{where}
 \,
(-\hspace{-1mm}\nabla^b\cdotp\nabla^b\hspace{-1mm}+\hspace{-1mm}
V^E_{ n}\hspace{-1mm}+\hspace{-1mm}Q\hspace{-1mm}-\hspace{-1mm}E)\psi_n=0,\,  \psi_n|_{\p
\Omega}=h,
\ea while the DN-map $\Lambda_{out}^E $ corresponds to the
operator
$A_{out}$ {\mllltext with  $\kappa_1=0$, see (\ref{A_out}), (\ref{Dirichlet-to-Neumann map R=1}).}
\end{theorem}

{\bf Proof.}
{ By the hypotheses of the Theorem, it follows that
$E \notin \hbox{spec}((\A_1^E)$,
where $\A_1$ defined by (\ref{a13}) with magnetic potential $b$ and
the potential $q^E$ 
given by (\ref{Aux potenital}). Thus, the Dirichlet problem (\ref{non-flat Schrodinger equations})
is uniquely solvable for large $n$.
}
 As the gauge transformation (\ref{gauge}) is the
identity map near $\p \Omega$, we see that 
$\p_\nu \psi_n|_{\p\Omega}= \p_\nu u_n|_{\p\Omega}$ and
$\psi_n|_{\p\Omega}= u_n|_{\p\Omega}$.
The DN maps for the Schr\"odinger equation (\ref{flat equation 2})
and for  equation (\ref{non-flat Schrodinger equations}) thus coincide,
and the assertion follows from Corollary \ref{cor:sequence}.\hfill \hfill\proofbox
}

Note that Theorem \ref{quantum-cloaking} is of a very different nature than
the well-known
results
from the classical theory of  spectral convergence, since the cloaking
potentials
$V^E_{ n}$ do not tend to $0$ as $n \to \infty$.
{\mmltext On the contrary as seen from the construction of $\sigma_{R, \e}$
in Sec. \ref{isotropic} and definition (\ref{2.21.11}), (\ref{effective-potential}),
$\sup_x |V_n^E(x)|\to\infty$ as $n\to\infty$. Moreover, $V_n^E$ is of a highly oscillatory nature
in $B( 2) \setminus B( 1)$ with quasiperiod tending to $0$ as $n \to \infty$. }

Theorem \ref{quantum-cloaking} has two important physical
consequences; see \cite{GKLU6,GKLU7} for further discussion and
applications.
Consider separately two cases:

\begin{itemize}
\item
{\mllltext Suppose that $b=0$. Since $Q$ is supported in $\overline B(1)$,
the operator}  $A_{out}= - \Delta$
is the free Schr\"odinger operator. Then,
from a physical point of view, the potentials $V^E_{ n}+Q$
can be considered as {\it almost transparent potentials} at energy $E$. 
 Also, the
{ $V^E_{ n}$, which depend on $E$ but} are independent of $Q$,  serve as  
{\it approximate invisibility cloaks}
for two-body scattering in quantum mechanics. As all measurement devices
have
limited precision, we can interpret this as saying that, given
a specific  device, one can design,
{for a given energy level $E$,}  
a potential to
  cloak an object \ie, an arbitrary potential,  from
any single-particle measurements made  at this energy $E$.

\smallskip

\item
Now suppose that $Q=0$, while $b \in C^1(\Omega, \R^3), \, b \neq 0
\, $.
We now have $
A_{out}= - \nabla^{\beta_1} \cdot \nabla^{\beta_1}$, and, due to the
transformation
rules for the magnetic potentials,
$\beta_1$ is
in general
no longer bounded near ${\it O}$.  
{Thus,  the
potentials $V^E_{ n}$ act as devices which give an external observer the
illusion
that, as $n \to\infty$, the magnetic field is unbounded  near
${\it O}$.
In particular, for $b(x)=B_0 \times x$, 
$B_0\in \R^3$, corresponding to a homogeneous magnetic field, the illusion
$\beta_1(x)$ has a singularity of the order $|x|^{-1}$ at  ${\it
O}$;
see \cite{GKLU6} for details.
}
\end{itemize}

\section{DN map near exceptional values of $E$}\label{sec exceptional}

{Theorem \ref{quantum-cloaking} shows that the behavior of 
$ \Lambda^{E}_{V^E_{ n}+Q}$, {\mltext when $E$ is far from } 
$\hbox{spec}(\A_{in}^E)$ and $n$ is large,
well
approximates the behavior of 
$ \Lambda^{E}_{out}$.
This situation changes dramatically when $E$ is close
to an eigenvalue {\mllltext of the cloaked region.} 

For simplicity, let us consider the case of an $E^0\in \R$ which,
while being in  the resolvent set of $A_{out}$, is also a simple eigenvalue
of $A_{in}^{E^0}${\mllltext , that is, $E^0\in \hbox{spec}(S_{in})$,
see (\ref{1.5.12}).}
This implies that $E^0$ is a simple eigenvalue of $\A_1^{E^0}$.
 The corresponding 
eigenfunction $u^0$ then satisfies
$\hbox{supp}\,(u^0)\subset \overline B( 1)$, \ie, $u^0$
is  a trapped state supported in the cloaked region, \cf  
Lemma \ref{lem: DN equivalence} and 
 (\ref{eq:
interior eigenstate}).}

{\mltext In the following, let  $d\hspace{-1mm} \in \hspace{-1mm}(0,1)$ be such that $\hbox{dist}(E^0,
{spec}(\A_1^{E^0})\hspace{-1mm} \setminus \hspace{-1mm E^0)\hspace{-1mm} >\hspace{-1mm} d$. }

\begin{theorem} \label{trapped}
{\mltext Let $Q\in L^\infty(\Omega)$ be a function supported in $\overline
B( 1)$,
$b \in C^1(\Omega, \R^3)$,
and $E^0\in \R$ be an eigenvalue of $\A_1^{E^0}$ corresponding to
potentials $b$ and $q^{E_0}$ defined in (\ref{Aux potenital}).
Assume that $E^0$ has multiplicity one, and let $u^0$ be the 
corresponding eigenfunction  supported in $\overline B( 1)$. Then

(i) There
{\mmltext  is a sequence $E_{(n)},\, E_{(n)} \to E^0$ as $n\to \infty$,
such that $E_{(n)}$ are  }
simple Dirichlet eigenvalues    
of the Schr\"odinger operators $-\nabla^b\cdotp
\nabla^b+V^{E_{(n)}}_{(n)}+Q$.
Moreover,  the $L^2(\Omega)$-normalized eigenfunctions
$\phi_{(n)}$ of 
{\mmltext these Schr\"odinger operators}
 for the eigenvalues $E_{(n)}$ satisfy,
for any  $\rho>2$,
\beq
\label{4.06.11}
\lim_{n\to \infty}\phi_{(n)}|_{\overline \Omega\setminus B( \rho)}=0, 
 \quad \hbox{in}\,\, \,C^1(\overline \Omega\setminus B(\rho)).
\eeq

(ii) Let $h \in H^{3/2}(\p \Omega)$
and $\psi_{(n)}(h)$ be the solution to  (\ref{flat equation 2})
for some $E$ with  $0<|E-E_{(n)}|<d/2$.
Then 
\beq \label{1.09.11}\quad \quad
\left(\p_\nu +i \nu \cdot b \right)\psi_{(n)}(h)|_{\p \Omega} &=& 
\frac {\a^0_{(n)}(h)}{E- E_{(n)}}
\left(\p_\nu +i \nu \cdot b \right) \psi_{(n)}|_{\p \Omega}
+p^E_{(n)}(h),\hspace{-1cm}\\
 \label{coefficient}
\a^0_{(n)}(h)&=&
\int_{\p \Omega} h\frac {\p \overline {\psi_{(n)}}}{\p \nu}\,dS,
\eeq
and functions $p^E_{(n)}(h)$ are uniformly bounded,  as
$n\to \infty$, in $H^{1/2}(\p \Omega))$ for
$\|h\|_{H^{3/2}}\leq 1$  and $E\in (E^0-d/4,E^0+d/4)$.}
\end{theorem}
}

{\bf Proof.}
For a given potential $Q$, the potential $q^E$ 
defined in (\ref{Aux potenital}) depends on
$E$. Thus we start by analyzing how the eigenvalues and eigenfunctions of
$\A_1^E$ and $\A_{(n)}^E$ change relative to 
{\mmltext the variation of  $E$.
Denote by $\lambda(k, E), \,k=1, 2,\dots, $ the eigenvalues of $ \A_1^E$, numbered
in  increasing order and taking multiplicity into account. The dependence of these eigenvalues on $E$
then follows from  (\ref{1.5.12}).
Similarly, let
$\lambda_{(n)}(k, E)$ be
the eigenvalues of $\A_{(n)}^E$, where $\A_{(n)}^E$ is the operator of form
(\ref{isotropic objects}) with $q=q^E$. Observe that}
\beq\label{Aux potenital3}
\A_1^E=\A_1^{E^0}+(q^E-q^{E^0}),\quad \A_{(n)}^E=\A_{(n)}^{E^0}+(q^E-q^{E^0}).
\eeq
Using Kato-Rellich formula \cite[Thm.\ VII.3.6]{Kato}  and the fact that
 $\|q^{\tilde E}-q^E\|_{L^\infty}=\frac 34|\tilde E-E|$,
 see (\ref{Aux potenital}),  we obtain that 
\beq\label{eq: iteration}
|\lambda_{(n)}(k,\tilde E)-\lambda_{(n)}(k, E )|\leq \frac 34|\tilde E-E|.
\eeq

{In the future we will consider only the value $k=k^0$ such that 
$\lambda(k^0, E^0)=E^0$, writing, \eg, $\lambda_{(n)}( E)$ for $\lambda_{(n)}(k^0, E)$.}
Next,  consider the spectral projectors for  $\mu=E^0-d/2$ or $\mu=E^0+d/2$.
Let $P_1^\mu(E)$  be the Riesz projectors
for the operators $\A_1^E$ 
and $P_{(n)}^\mu(E)$  be the  projectors
for  $\A_{(n)}^E$. They are
defined analogously to (\ref{Riesz}) using
a contour $\Gamma\subset \C$ that surrounds all  of the eigenvalues smaller
than $\mu$. We  can assume that $\Gamma$ is such a contour
that, for  $n$ large enough, 
the distance from $\Gamma $ to the eigenvalues of the operators $\A_1^{E^0}$ and
$\A_{(n)}^{E^0}$ 
is more than  $d/4$. Then the norm of $(\A_{(n)}^{E^0}-z)^{-1}$ in
$L^2(\Omega,g_{(n)}^{1/2}dx)$
is bounded by $4/d$.
Thus, assuming that $|E-E^0|<d/8$, we obtain, using the 
formula
\ba
& &( \A_{(n)}^E-z)^{-1}-(\A_{(n)}^{E^0}-z)^{-1}\\
&=&
(\A_{(n)}^{E^0}-z)^{-1}\bigg(\big(I-(
q^E-q^{E^0})(\A_{(n)}^{E^0}-z)^{-1}\big)^{-1}-I\bigg)
\ea
in (\ref{Riesz}), the estimate
\beq\label{eq: difference of Riesz}
& &\quad\|P^\mu_{(n)}(E)-P^\mu_{(n)}(E^0)\|_{L^2(g_{(n)}^{1/2}dx)\to
L^2(g_{(n)}^{1/2}dx)}\leq C_\mu |E-E^0|,
\eeq
where $C_\mu$ depends only on the choice of $\Gamma$.

As $E^0$ is the only eigenvalue  of $\A_1^{E^0}$ in the interval $(E^0-d,E^0+d)$
and has the multiplicity one, 
it follows from
 Theorem \ \ref{sequence-approximation}  that when 
$n$ is large enough, 
then $\A_{(n)}^{E^0}$
has only  one eigenvalue $E^0_n$ in the interval $(E^0-3d/4,\, E^0+3d/4)$
 and $|E^0_n-E^0|<d/4.$ 
Moreover, the eigenvalue $E^0_n$  is simple. 

{ Let us show that there are $E_{(n)}, \, E_{(n)} \to E^0$,
such that 
\beq\label{2.5.12}
\lambda_{(n)}(E_{(n)})=E_{(n)}.
\eeq
Observe that, by  Theorem \ \ref{sequence-approximation}, $\lambda_{(n)}(E^0) \to E^0$
as $n \to \infty$. Together with (\ref{eq: iteration}) this implies that for any $\e >0$ there is $n(\e)$ such
that for $n>n(\e)$, 
\ba
\lambda_{(n)}(\cdot):\,[E^0-\e,\, E^0+\e] \to [E^0-\e,\, E^0+\e].
\ea
As $\lambda_{(n)}(\cdot)$ is a contraction, see (\ref{eq: iteration}), we conclude 
by the Banach fixed point theorem that there is a unique $E_{(n)}$ satisfying 
(\ref{2.5.12}).

Returning to equation (\ref{eq: difference of Riesz}) 
we see that
\beq\label{proj. estimetes}
\lim_{n\to \infty} \|P^\mu_{(n)}(E_{(n)})  -P^\mu_{(n)} (E^0)\|
_{L^2(g_{(n)}^{1/2}dx)\to (g_{(n)}^{1/2}dx)}=0,
\eeq
Combining this  with (\ref{final-estimate2}) (for $q=q^{E^0}$)}
 and embedding (\ref{g g_m inclusion}),
we see
\beq\label {eq: difference of Riesz 3}
\lim_{n\to \infty}P^\mu_{(n)}(E_{(n)}) f
=P^\mu_1(E^0) f \quad\hbox{in }\,\, L^2_g(\Omega)
\eeq

Let $\mu=E+d/2$ and $\nu=E-d/2$.
For large $n$, the operators
$ P_{(n)}^\mu(E_{(n)}) - P_{(n)}^\nu(E_{(n)})$ are the orthoprojectors,
in $L^2(\Omega,g_{(n)}^{1/2}dx)$ onto
 the eigenspace of $ \A_{(n)}^{E_{(n)}}$ corresponding to the 
eigenvalue $E_{(n)}$. 
{\mllltext Using  (\ref{proj. estimetes}) we see that for $n$ large enough, 
the eigenvalue $E_{(n)}$ has multiplicity one.
Denote by $\tilde u_{(n)}$ the eigenfunction corresponding
to $E_{(n)}$,} normalized 
in $L^2_g(\Omega)$. Now $u^0$ is supported in $B( 1)$ and thus 
\ba
\|u^0\|_{g} =\|u^0\|_{ L^2(\Omega,g_{(n)}^{1/2}dx)} =1,\,\,n>0, \quad
\|u^0\|_{L^2(\Omega)}= \frac{1}{{\sqrt 8}}.
\ea
Using (\ref {eq: difference of Riesz 3}) we see that
\ba
\left( P_{(n)}^\mu(E_{(n)}) - P_{(n)}^\nu(E_{(n)})\right) u^0=a_n  \tilde
u_{(n)}\underset{n\to \infty}\longrightarrow
\left(P_1^\mu(E^0) - P_1^\nu(E^0)\right) u^0=u^0,
\ea
in $L^2_g(\Omega)$,
where $a_n =|a_n| e^{i \a_n}$ and  $|a_n| \to 1$  as $ n \to \infty.$

Denoting $u_{(n)} = e^{i \a_n}\tilde   u_{(n)}$, we see that
\beq \label{15.09.11}
\lim_{n\to \infty} {u}_{(n)}= u^0 \quad \hbox{in}\,\, L^2_g(\Omega).
\eeq
Since $u^0(x)=0,\, |x| >1$, this implies
\beq \label{5.06.11}
\lim_{n\to\infty}\|u_{(n)}|_{\Omega\setminus
B( 2)}\|_{L^2(\Omega\setminus B( 2))}=0.
\eeq
Observe that
$\sigma_{(n)} = \gamma_0,\, g_{(n)}=1$ and $V^E_{(n)}=0$ in $\Omega\setminus
B( 2)$. Thus
it follows from (\ref{5.06.11}) that
the functions $u_{(n)}$ satisfy
\ba
-\nabla^b \cdotp \nabla^b u_{(n)}=E_{(n)}u_{(n)} \quad\hbox{in
}\Omega\setminus
B( 2), \quad u_{(n)}|_{\p \Omega}=0,
\ea
where the right side $E_{(n)}u_{(n)} \to 0$ in $L^2(\Omega\setminus
B( 2))$ as $ n \to \infty$.
Since $b \in C^1(\Omega\setminus B( 2))$, standard elliptic regularity
results \cite{GT} 
imply that 
\beq \label{3.06.11}
\lim_{n\to \infty}\|u_{(n)}\|_{C^1(\Omega\setminus B( \rho))}=0, 
\eeq
 for any $2< \rho<3$.
Using the transformation
(\ref{gauge}) to define 
$\phi_n(x)=\sigma_{(n)}^{1/2}(x) u_n(x)
$,  we see that 
\ba
(-\nabla^b\cdotp\nabla^b+V^{E_{(n)}}_{ n}+Q) \phi_n =E_{(n)}  \phi_n, \quad
\phi_n|_{\p
\Omega}=0.
\ea
This proves 
{that $\phi_n$ is an eigenfunction  of the Schr\"odinger operator
$-\nabla^b\cdotp\nabla^b+V^{E_{(n)}}_{ n}+Q$ for the eigenvalue $E_{(n)}$.}
Moreover, since $\sigma_{(n)}=2$ in $B( 1)$, it follows from
(\ref{inclusion}) and
(\ref{15.09.11}) that 
\ba
\liminf_{n\to \infty} \|\phi_n\|_{L^2(\Omega)}\geq {\sqrt 2}\liminf_{n\to \infty} 
\|u_n\|_{L^2(B( 1))}
 \geq {\sqrt 2} \|u^0\|_{L^2(B( 1))}= \frac 12.
\ea
This inequality  and equation (\ref{3.06.11}) together
imply (\ref{4.06.11}). Thus we have  proven (i).
}

Next, consider (ii).
We start with the boundary-value 
problem for the acoustic equation
\ba
\left( - g^{-1/2}_{(n)} \nabla^b\cdotp \sigma_{(n)} \nabla^b+V^{E}_{ n}+q^E+E
\right) v^E_n(h) =0, \quad
v^E_n(h)|_{\p
\Omega}=h.
\ea
Clearly
\beq
\label{3.09.11}\quad\quad\quad
v^E_n(h)= a^E_n(h) u_{(n)} +w^E_n(h), \quad\hbox{with }
\left(w^E_n(h),\,  u_{(n)}  \right)_{L^2(\Omega, g_{(n)}^{1/2} dx)}=0,
\hspace{-1cm}
\eeq
{\mllltext where, using the notations introduced in 
(\ref{10.09.11}), (\ref{20.09.11}), we have} 
\beq \label{6.09.11}
& &a^E_n(h)=\left(h^e,\,  u_{(n)}  \right)_{L^2(g_{(n)}^{1/2} dx)}+
\frac{1}{E-E_{(n)}} \left(F^E(h),\,  u_{(n)}  \right)_{L^2(g_{(n)}^{1/2}
dx)},\\
\nonumber
& &w^E_n(h)=h^e-(h^e,\,  u_{(n)})_{L^2(g_{(n)}^{1/2} dx)}\, u_{(n)} -
{\tilde w}^E_n(h),
\\ \nonumber
& & {\tilde w}^E_n(h)= (\A_{(n)}-E)^{-1} \left( F^E(h) - (F^E(h),\,
u_{(n)})_{L^2(g_{(n)}^{1/2} dx)}\, u_{(n)} \right).
\eeq
Since $F^E(h) - (F^E(h),\,  u_{(n)})\, u_{(n)} $ and $u_{(n)}$
are orthogonal in $L^2(\Omega, g_{(n)}^{1/2} dx)$,
 and $\hbox{dist}(E, \hbox{spec}(\A_{(n)}) \setminus \{E_{(n)}\}) >d/2$,
it follows from (\ref{10.09.11}) that, for $n$ large enough,
\ba
\|{\tilde w}^E_n(h)\|_{L^2(\Omega, g_{(n)}^{1/2} dx)} \leq C,
\ea                 
where  $C$ is independent of $E\in (E^0-d/4,E^0+d/4)$ and $h$ satisfying 
$\|h\|_{H^{3/2}(\p \Omega)} \leq 1.$
Note that in $\Omega \setminus B( 2)$ the function
${\tilde w}^E_n(h)$ satisfies the equation
\ba
-\nabla^b \cdot \nabla^b {\tilde w}^E_n(h)&=&  E {\tilde w}^E_n(h) +
F^E(h) - (F^E(h),\,  u_{(n)})_{L^2(g_m^{1/2}dx)}\, u_{(n)},
\\ {\tilde w}^E_n(h)|_{\p \Omega}&=&0.
\ea
Thus, by boundary elliptic regularity,  \cite[Thm. 9.13] {GT}, 
\ba
\|{\tilde w}^E_n(h)\|_{H^2(\Omega\setminus B(\rho))} \leq C_\rho, \quad 2<\rho<3.
\ea
This inequality,  (\ref{10.09.11}) and (\ref{20.09.11}) together  imply that
$p^E_{(n)}(h)= (\p_\nu +i \nu \cdot b) w^E_n(h)$ satisfies
\beq \label{4.09.11}
\|p^E_{(n)}(h)\|_{H^{1/2}(\p \Omega)} \leq C_0,
\quad \hbox{for}\,\, \|h\|_{H^{3/2}(\p \Omega)} \leq 1,
\eeq
if $n\, $ is large and $|E-E_{(n)}| \leq d/2$.
Finally, integration by parts shows that
\ba
a^E_n(h)=\frac{1}{E-E_{(n)}}\int_{\p \Omega} h\frac {\p \overline
{u_{(n)}}}{\p \nu}\,dS=
\frac{1}{E-E_{(n)}}\int_{\p \Omega} h\frac {\p \overline {\phi_{(n)}}}{\p
\nu}\,dS.
\ea
The desired equation (\ref{1.09.11}) follows from the above equation
together with (\ref{3.09.11}), 
(\ref{4.09.11}) and (\ref{coefficient}), if we take into the account the
relation 
(\ref{gauge})  between
$\psi^E_{(n)}(h)$ and $v^E_n(h)$.
\hfill \hfill\proofbox

\smallskip

\noindent{\bf Remark.} Theorem \ref{trapped} means that, away from
the cloaking structure, the eigenfunctions $\phi_{(n)}$ converge to zero
as
$n\to\infty$, \ie, the  $\phi_{(n)}$ represent {\it almost trapped modes}, 
effectively vanishing near $\p \Omega$.
Physically speaking, we can say that
if $E^0 $ is an eigenvalue of the Schr\"odinger operator with
Neumann boundary condition 
in the cloaked region $B( 1)$,
and we connect the interior and the exterior via
the cloaking potential $V^E_n$
in  the layer $B( R_{(n)}) \setminus B( 1)$, with $ R_{(n)} \to 1$
as $n \to \infty$, }
 a particle under the influence of the combined potential on $B(3)$ is still largely  confined (modulo standard tunneling) to 
 the interior region, with a slight shift of the energy of 
the eigenmode from
$E^0$ to $E_{(n)}$. 

Moreover, for energies $E$ close to the values $E_{(n)}$
the presence of the cloaked region is
very clearly seen in the boundary measurements
of the DN map, so that 
the invisibility  effect 
is compromised.
 On the other hand,  at energies which are away from
the  $E_{(n)}$, the DN map for the potential $V^E_{(n)}+Q$
 well approximates  $\Lambda^E_{out}$, and thus the potential $Q$ is
approximately cloaked.

\section{Numerical results}\label{sec numerics}

\vspace{-5mm}

\begin{figure}[htbp]
\centerline{
\includegraphics[width=0.5\linewidth]{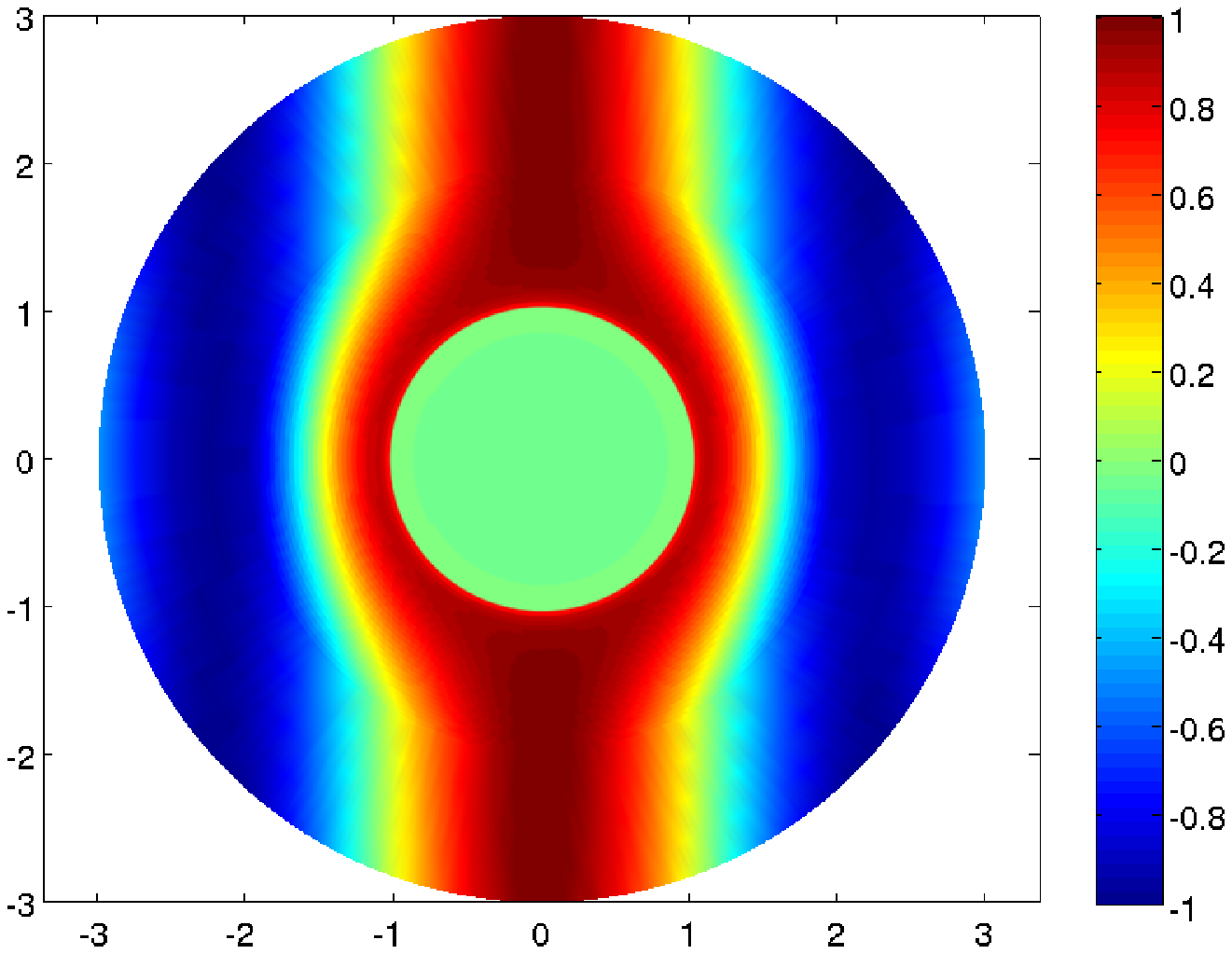}\ \ 
\includegraphics[width=0.5\linewidth]{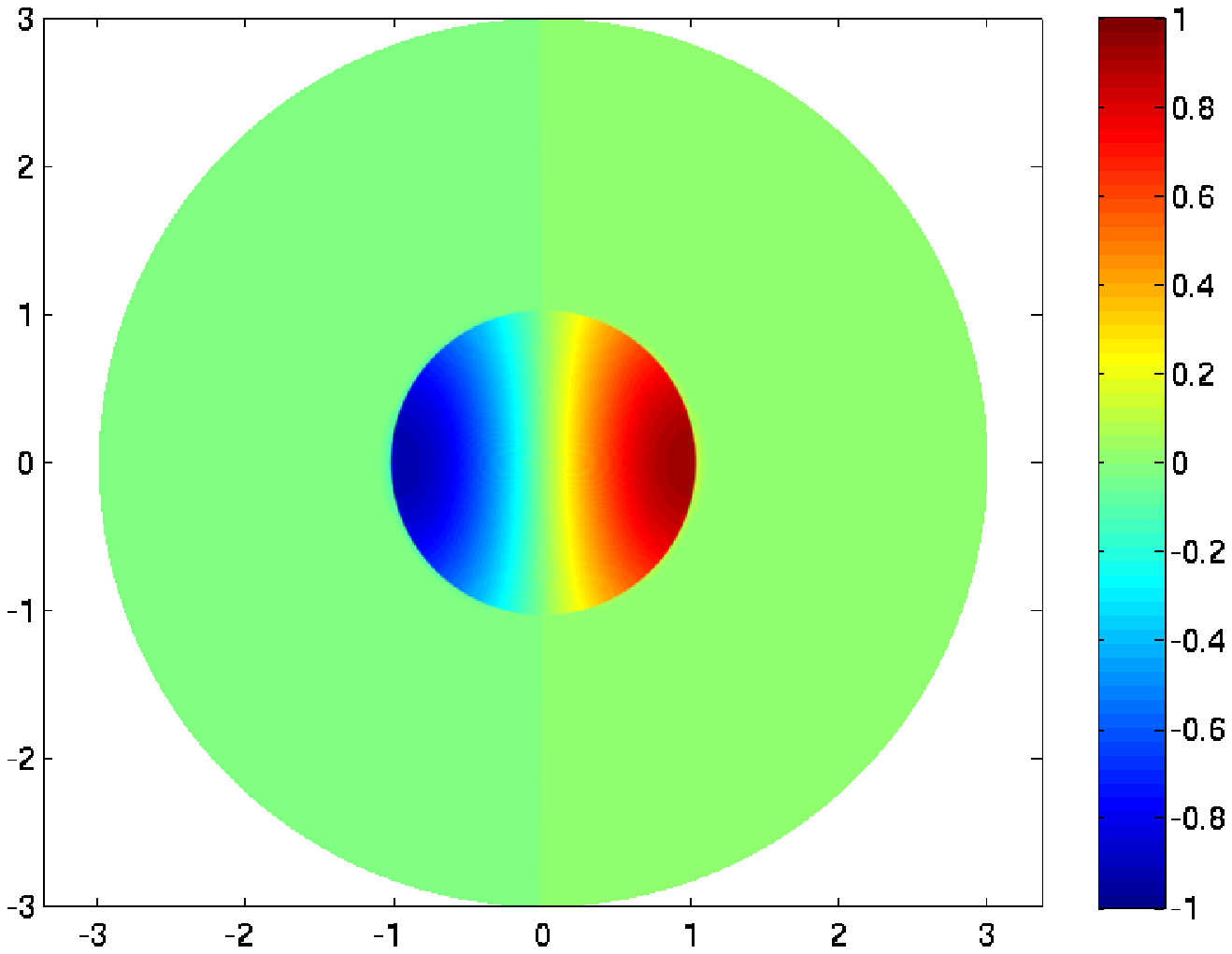}
}

\caption{{\bf Solutions of acoustic equations.}
Left: Solution $u^{tot}$ of the scattering problem (\ref{eq: sc1})
restricted to
ball $B(3)$, when a plane wave scatters
from an approximate cloak in the case (\ref{numerical case 1}), \ie, when
$k^2$
is away from the   exceptional values
$E_{(n)}$.
Right:   Almost trapped {\mmltext eigenfunction $u$ of the acoustic operator
(\ref{eq: bvp1}),
with Dirichlet boundary condition, $h=0$,} in the case 
 (\ref{numerical case 2}), \ie,
when $k^2$ is  equal to the exceptional value
$E_{(n)}$.}
\end{figure}

Next we consider scattering problems for the Helmholtz and Schr\"odinger
equations in the case when the magnetic potential vanishes, \ie, $b=0$.

The scattering problem for Helmholtz equation is 
\beq\label{eq: sc1}
& &(\nabla \cdotp
\sigma_{(n)} \nabla+k^2(1+\alpha(x)) g_{(n)}^{1/2}) u^{tot}
=0\quad\hbox{in }\,\,\R^3,\\
& &\quad u^{tot}(x,k)=u^{in}(x,k)+u^{sc}(x,k)  \nonumber
\eeq
and for the Schr\"odinger equation
\beq\label{eq: sc2}
& &(-\nabla\cdotp\nabla+V^E_{n}+Q-E)\psi^{tot}=0\quad\hbox{in
}\,\,\R^3,\\
& &\psi^{tot}(x,E)=\psi^{in}(x,E)+\psi^{sc}(x,E),  \nonumber
\eeq
where $E>0$, $k=E^{1/2}$, 
the incident fields are $u^{in}(x,k)=\psi^{in}(x,E)=e^{ik\omega\cdotp x}$,
$|\omega|=1$ and
the scattered fields satisfy the radiation condition
\ba
\lim_{r\to \infty}r(\frac {\p}{\p r}-ik)u^{sc}(x,k)=\lim_{r\to
\infty}r(\frac
{\p}{\p r}-iE^{1/2})\psi^{sc}(x,E)=0,\quad
r=|x|.
\ea 
In following we consider $\a(x)$ that corresponds
to a real bounded potential 
$Q$ supported in $B(1)$, that is, $\alpha(x)=-(E^{-1}Q(x)+3)/4$.
We assume that $Q(x)$ is such that $1+\alpha(x)\geq c_0>0$.

We consider also the solutions of the boundary value
problems in $\Omega$ (note that we denote these solutions
by $u$ and $\psi$, without using superscripts),
\beq\label{eq: bvp1}
\quad\left(\nabla \cdotp
\sigma_{(n)} \nabla+k^2(1+\alpha(x)) g_{(n)}^{1/2}\right) u
=0\quad\hbox{in }\,\,\Omega,\quad
u|_{\p \Omega}=h  
\eeq
and for the Schr\"odinger equation
\beq\label{eq: bvp2}\quad
(-\nabla\cdotp\nabla+V^E_{n}+Q-E)\psi=0
\quad\hbox{in }\,\,\Omega,\quad
u|_{\p \Omega}=h.
\eeq
The solutions of these scattering and boundary value problems
 are related through a gauge transformation,
\beq \label{gauge 2}
\psi^{tot}(x)=\sigma_{(n)}^{1/2}(x) u^{tot}(x),\quad
\psi(x)=\sigma_{(n)}^{1/2}(x)
u(x).
\eeq

The computations are made without reference to physical units;
 for simplicity, we use
$E=2$. 
The cloak corresponds to the  parameter $R=1.005$ and
inside the cloak we have located a spherically symmetric potential;
$$
Q(x)=Q_{in}\chi_{[0,R]}(|x|),\quad\hbox{that is},\quad
\alpha(x)=-Q_{in}(4E)^{-1}\chi_{[0,R]}(|x|)-3/4.
$$
To illustrate the 
approximate cloaking, we used 
\beq\label{numerical case 1}
Q_{in}=1,
\eeq 
and to obtain an
almost trapped state,
\beq\label{numerical case 2}
Q_{in}=-2.576.
\eeq

\begin{figure}[htbp]
\centerline{
\includegraphics[width=0.5\linewidth]{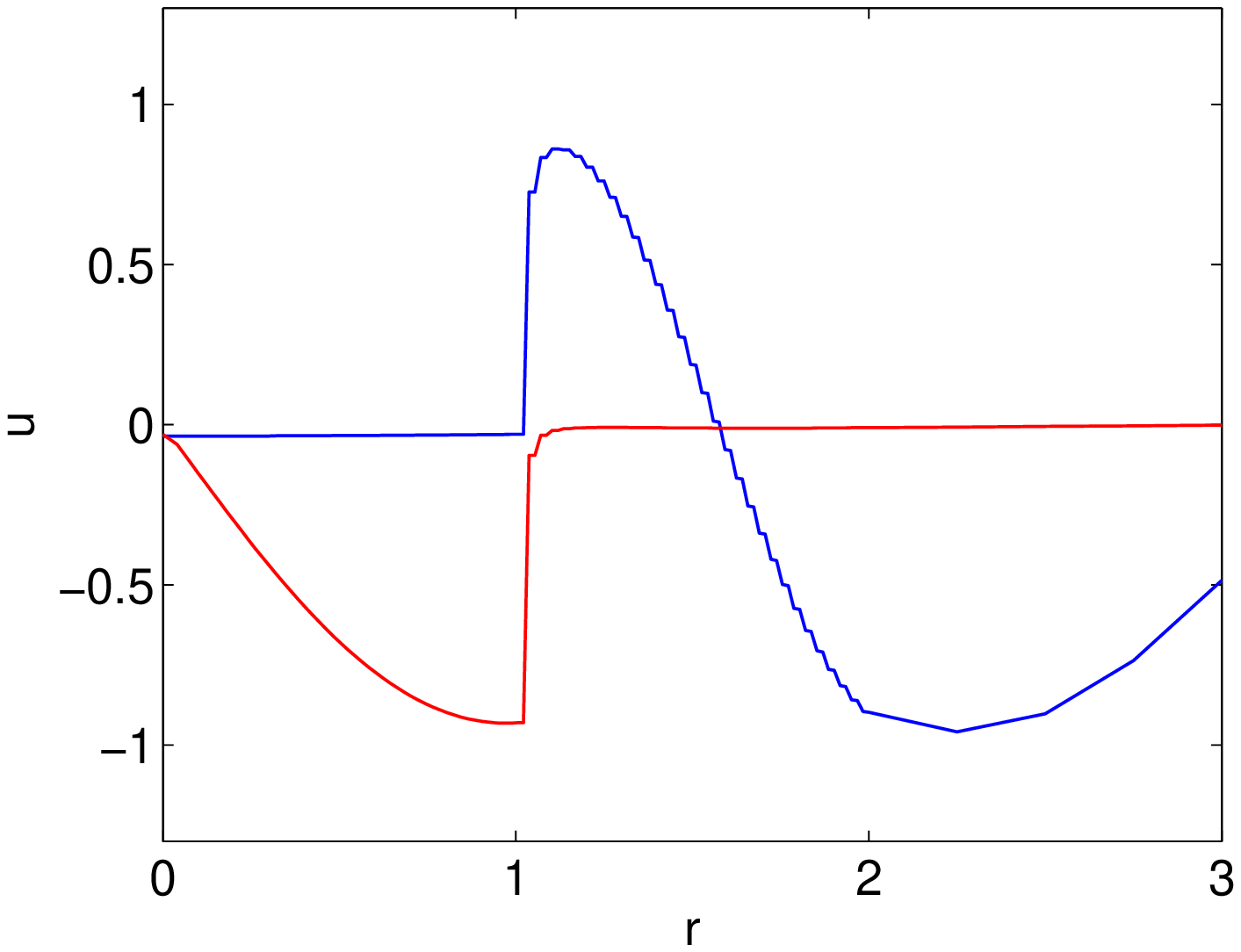}\ \ 
\includegraphics[width=0.5\linewidth]{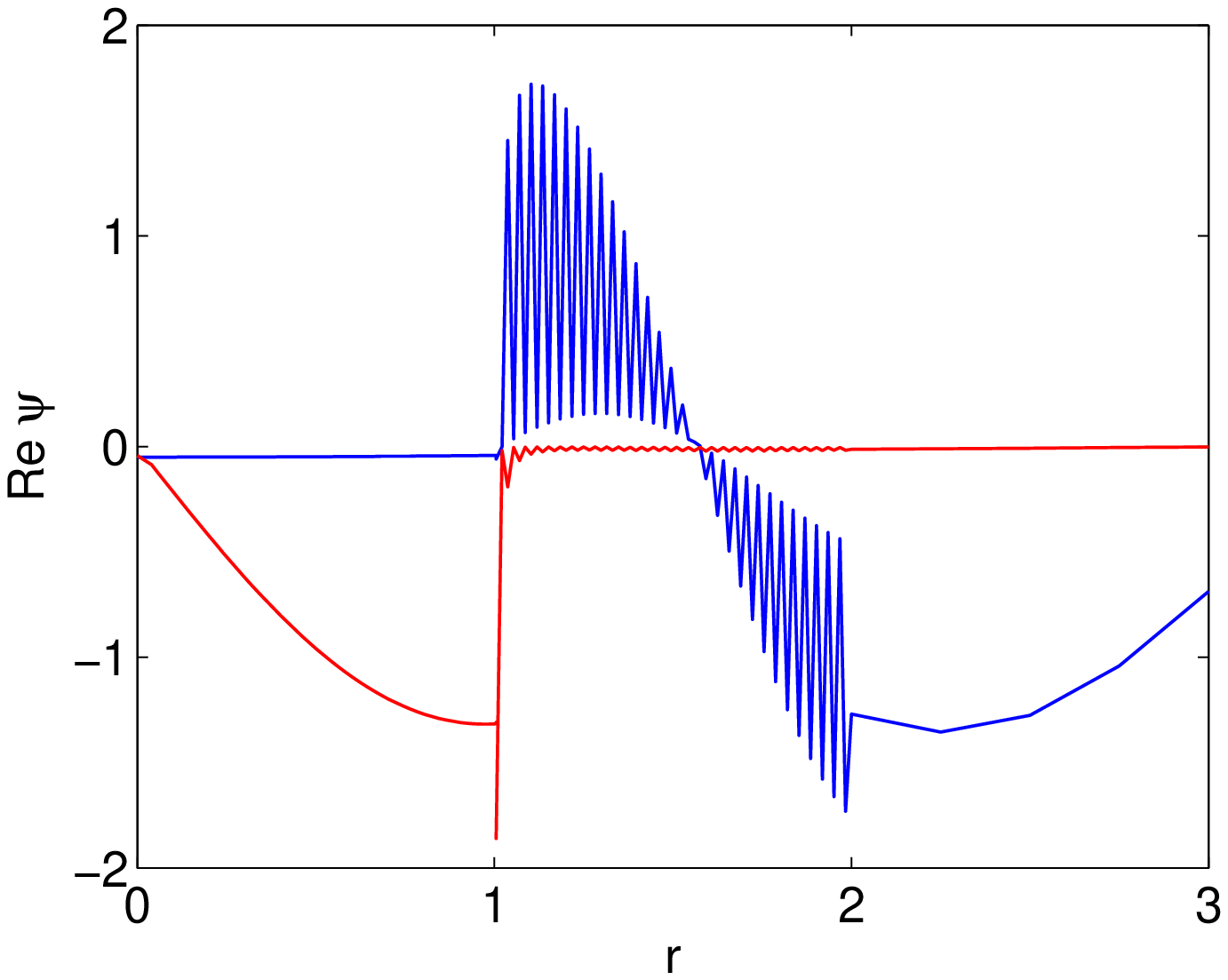}
}

\caption{{\bf Radial profiles of solutions of acoustic and 
Schr\"odinger equations.}
 Left: The solution $u^{tot}$ of the scattering problem
(\ref{eq: sc1}) on the line segment $L=\{(x,0,0):\ x\in [0,3]\}$ in the case 
of Fig. 1(left), where $k^2$ is far from the exceptional values $E_{(n)}$,
is shown with the blue curve.
Also, the {\mmltext eigenfunction $u$ of acoustic operator }
(\ref{eq: bvp2}) on the line segment $L$ in the case 
of Fig. 1(right), where $k^2$ is the exceptional value $E_{(n)}$,
is shown with the red curve.
Right: The solutions $\psi^{tot}$ and $\psi$ of the Schr\"odinger equations
(\ref{eq: sc2}) and (\ref{eq: bvp2})
on the line segment $L$, obtained from the solutions
on the left  via gauge transformations (\ref{gauge 2}).
}
\end{figure}

In our numerical solution we have approximated
$\sigma_{(n)}$ by a piecewise constant function consisting of
30 layers in the region $R<r<2$. The values of the conductivity
in these layers are chosen as in the above sections.
This corresponds to the case when
 the cloaking potential $V_n^E$ is a weighted sum of delta functions,
and their derivatives, 
on spheres.

In the numerical solution of the problem,
we represent the solution $u^{tot}$ and $u$ in terms of spherical harmonics
$Y^n_m$ and Bessel functions up to order
$N=7$ in each
layer where the cloaking conductivity is constant. The 
transmission condition on the boundaries of these layers are
solved numerically by solving linear equations.
After this we compute the solution $\psi^{tot}$ and $\psi$ of the
Schr\"odinger equation using the gauge transformation (\ref{gauge 2}).


\bibliographystyle{amsalpha}

\begin{thebibliography}{A}


\bibitem{Allaire}
G. Allaire, Homogenization and two-scale convergence,
{\it  SIAM J. Math. Anal.} {\bf 23}
(1992), no. 6, 1482--1518.


\bibitem{Allaire2}
G. Allaire, A. Damlamian and U. Hornung,
Two-scale convergence on periodic surfaces and applications,
   In A. Bourgeat, C. Carasso, S. Luckhaus and A.
     Mikelic (eds.), {\it Mathematical Modelling of Flow through Porous
Media},
15-25,    Singapore, World Scientific, 1995.


\bibitem{A}
H. Attouch, {\it Variational convergence for functions and operators}, Appl.
Math. Series, Pitman, Boston, MA, 1984.


\bibitem{Aron}
N.\ Aronszajn,  A unique continuation theorem for solutions of elliptic
partial differential equations or inequalities of second order. {\it J.
Math. Pures Appl.}  {\bf 36}  (1957), 235--249


\bibitem{Ben} A. Bensoussan, J.L. Lions and G. Papanicolaou, {\it Asymptotic
analysis for periodic
structures}, North-Holland, Amsterdam, 1978.



\bibitem{Be} Y.\ Berezanskii, The uniqueness theorem in the inverse problem
of spectral analysis for the Schr\"odinger equation (Russian), {\it Trudy
Moskov. Mat. Obsch.}, {\bf 7} (1958), 1--62.



\bibitem{Buk} A. Bukhgeim, Recovering a potential from Cauchy data, {\it J.
Inverse Ill-Posed Prob.} {\bf 16} (2008), 19--33.



\bibitem{Chanillo} S. Chanillo, A problem in electrical prospection and
an $n$-dimensional Borg-Levinson theorem,
{\it Proc. Amer. Math. Soc.} {\bf 108} (1990),  761--767.


\bibitem{ChenChanRot} H.-Y. Chen and C.T. Chan, Transformation media that
rotate
electromagnetic fields, {\it Appl. Phys. Lett.} {\bf 90}   (2007), 241105.


\bibitem{Ch3} H.-Y.\ Chen and C.T.\ Chan, Acoustic cloaking in three
dimensions using acoustic metamaterials,
{\it Appl. Phys. Lett.} {\bf 91} (2007), 183518.


\bibitem{Ch4} H.-Y.\ Chen and C.T.\ Chan, Electromagnetic wave manipulation
using layered systems, {\it Phys. Rev. B} {\bf 78}  (2008), 054204.



\bibitem{CWZK} H.-S.\ Chen, B.-I.\ Wu, B.\ Zhang and J.A.\ Kong,
Electromagnetic wave interactions with a metamaterial cloak, {\it Phys.
Rev. Lett.}, {\bf 99} (2007), 063903.



\bibitem{Cherka}
A. Cherkaev, {\it Variational methods for structural optimization}, Appl.
Math. Sci.,
{\bf 140}, Springer-Verlag, 2000.



\bibitem{CuSc} S. Cummer and D. Schurig, One path to acoustic cloaking,
{\it New J. Phys.} {\bf 9}  (2007), 45.


\bibitem{Cu}
S.\ Cummer, \etal, Scattering theory derivation of a 3D
acoustic cloaking shell, {\it Phys. Rev. Lett.},
{\bf 100} (2008), 024301.



\bibitem{dM}
G. Dal Maso,
{\it An Introduction to $\Gamma$-convergence},
Prog. in Nonlinear Diff. Eq. and their Appl., {\bf 8}. Birkhauser
Boston, Inc., Boston, 1993.



\bibitem{Dolin} L. Dolin, To the possibility of comparison of
three-dimensional
electromagnetic systems with nonuniform anisotropic filling (in Russian),
{\it Izv.
VUZov. Radiofizika}, {\bf 4} (1961), 964--967.



\bibitem{Far} M. Farhat, S. Guenneau, A.B. Movchan and S. Enoch, Achieving
invisibility over a finite
range of frequencies, {\it Optics Express}, {\bf 16}  (2008), 5656.



\bibitem{GT} D. Gilbarg and N. Trudinger, {\it Elliptic Partial Differential
Equations of Second Order}, 2nd ed., Springer-Verlag, Berlin, 1983.


\bibitem{GKLU1}
A. Greenleaf, Y. Kurylev, M. Lassas and G. Uhlmann, Full-wave invisibility
of
active devices at
all frequencies,  {\it Comm.  Math. Phys.}, {\bf 279} (2007), 749--789.


\bibitem{GKLU2} A. Greenleaf, Y. Kurylev, M. Lassas and G. Uhlmann,
Electromagnetic wormholes and virtual magnetic monopoles from metamaterials,
{\it Phys. Rev. Lett.},  {\bf 99}  (2007), 183901.


\bibitem
{GKLU3}
A. Greenleaf, Y. Kurylev, M. Lassas and G. Uhlmann: Improvement of
  cylindrical cloaking with the SHS lining. {\it Optics
Express} {\bf 15} (2007), 12717-12734.


\bibitem{GKLU4} A. Greenleaf, Y. Kurylev, M. Lassas and G. Uhlmann,
Electromagnetic wormholes via handlebody constructions,
{\it Comm. Math. Phys.}, {\bf 281} (2008), 369-385.



\bibitem{GKLU5} A. Greenleaf, Y. Kurylev, M. Lassas and G. Uhlmann,
Comment on ``Scattering theory derivation of a 3D acoustic cloaking shell'',
arXiv:0801.3279 (2008).


\bibitem{GKLU6} A. Greenleaf, Y. Kurylev, M. Lassas and G. Uhlmann,
Isotropic
transformation optics: approximate acoustic and quantum cloaking,
arXiv:0806.0085 (2008); {\it New J. Phys.} {\bf 10} (2008), 115024.



\bibitem{GKLU7} A. Greenleaf, Y. Kurylev, M. Lassas and G. Uhlmann,
Approximate quantum cloaking and almost trapped states,
arXiv:0806.0368 (2008); {\it Phys. Rev. Lett.} {\bf 101} (2008),  220404.



\bibitem{GLU1}
A.\ Greenleaf, M.\ Lassas, and G.\ Uhlmann, The
Calder\'on problem for conormal
potentials,
I: Global uniqueness and reconstruction, {\it Comm.
Pure Appl. Math.}
{\bf 56} (2003), no. 3,
328--352.


\bibitem{GLU2}
A.\
Greenleaf, M.\ Lassas, and G.\
Uhlmann, Anisotropic conductivities that
cannot detected by
EIT,
{\it Physiolog. Meas.}  (special issue on
Impedance Tomography),
{\bf 24} (2003), 413-420.


\bibitem{GLU3}  A.\
Greenleaf, M.\ Lassas, and G.\
Uhlmann, On
nonuniqueness for
Calder\'on's
inverse problem,  {\it
Math. Res. Lett.} {\bf 10} (2003),
685-693.



\bibitem{GrinNov} P. Grinevich and R. Novikov, Transparent potentials at
fixed energy in dimension two. Fixed-energy dispersion relations for the
fast
decaying potentials, {\it Comm. Math. Phys.} {\bf 174} (1995), 409--446.



\bibitem{HHLe} A. Hendi, J. Henn and U. Leonhardt, Ambiguities in the
scattering
tomography for central potentials, {\it Phys. Rev. Lett.} {\bf 97} (2006),
073902.


\bibitem{HFJ} Y. Huang, Y. Feng and T. Jiang, Electromagnetic
cloaking by layered structure of homogeneous isotropic materials,
{\it Optics  Expresss}, {\bf 15} (2007), 11133-11141.


\bibitem{Kato} T. Kato, {\it  Perturbation theory for linear operators},
Springer-Verlag, 1980.


\bibitem{KochTat} H. Koch and D. Tataru, Carleman estimates and unique
continuation for second-order elliptic equations with
nonsmooth coefficients.
{\it Comm. Pure Appl. Math.}  {\bf 54} (2001), 339--360.


\bibitem{KOVW} R. Kohn, D. Onofrei, M.
Vogelius and M. Weinstein, in preparation.


\bibitem{KSVW} R. Kohn, H. Shen, M.
Vogelius, and M. Weinstein,
Cloaking via change of variables in electrical impedance tomography,
{\it Inver. Prob.}, {\bf 24} (2008), 015016.


\bibitem{Kufner}
   A. Kufner and B. Opic,  {\it Hardy-type inequalities.} Pitman 
Research Notes in
Mathematics Series {\bf 219},  Longman, 1990.



\bibitem{KLS} Y. Kurylev, M. Lassas  and E. Somersalo, Maxwell's equations
with a
polarization independent wave velocity: direct and inverse problems, {\it
Jour.
Math. Pures Appl.} {\bf 86} (2006), 237--270.



\bibitem{LaNa} R. Lavine and A. Nachman,  The Faddeev-Lippmann-Schwinger
equation in multidimensional quantum inverse scattering, in {\it Inverse
problems: an interdisciplinary study (Montpellier, 1986)}, 169--174, Adv.
Electron. Electron Phys., Suppl. 19, Academic Press, London, 1987.



\bibitem{LeSi} H. Leinfelder and C. G. Simader, Schr\"odinger operators with
singular magnetic
vector potentials, {\it Math Zeit.} {\bf 176} (1981), 1-19.



\bibitem{Le}
U.\ Leonhardt,
Optical conformal
mapping, {\it Science} {\bf 312} (2006),
1777-1780.



\bibitem{LePhil}
U.\ Leonhardt and T.\ Philbin, General relativity in electrical engineering,
{\it New J. Phys.} {\bf 8}  (2006), 247.


\bibitem{Lipton}
R. Lipton, Homogenization and field concentrations in heterogeneous media,
{\it SIAM J. Math. Anal.}, {\bf 38} (2006), 1048--1059.



\bibitem{Luo} Y. Luo, \etal,
Design and analytically full-wave validation of the invisibility cloaks,
concentrators, and field rotators created
with a general class of transformations, {\it Phys. Rev. B} {\bf 77} (2008),
125127.


\bibitem{Milt} G. Milton, {\it The Theory of Composites}, Cambridge Univ.
Press, 2001.


\bibitem{MBW} G.\ Milton, M.\ Briane, and J.\ Willis,
On cloaking for
elasticity and physical
equations with a
transformation invariant form, {\it New J.  Phys.}, {\bf 8} (2006), 248.


\bibitem{Nach} A. Nachman, Reconstructions from boundary measurements, {\it
Ann.  Math.} {\bf 128} (1988), 531--576.


\bibitem{Nach2} A. Nachman, Global uniqueness for a two-dimensional inverse
boundary
value problem,
{\it Ann.  Math.} {\bf 143} (1996),  71--96.


\bibitem{NSU} A.\ Nachman, J.\ Sylvester and G.\ Uhlmann,
An $n$-dimensional Borg-Levinson theorem, {\it Comm. Math. Phys.} {\bf 115}
(1988),  595--605.


\bibitem{New} R. Newton, Construction of potentials from the phase shifts at
fixed energy, {\it J. Math. Phys.} {\bf 3} (1962), 75--82.

\bibitem{Ngu} G. Nguetseng, A general convergence result for a functional
related to the theory of
homogenization, {\it SIAM J. Math. Anal.}, {\bf 20}  (1989), 608.


\bibitem{Norris} A. Norris, Acoustic cloaking theory, {\it Proc. Royal Soc.
A}, {\bf 464} (2008), 2411--2434.



\bibitem{OPS} P. Ola, L. P\"aiv\"arinta and E. Somersalo,
An  inverse boundary value problem in electrodynamics,
{\it Duke Math. Jour.} {\bf 70} (1993), 617-653.


\bibitem{PSS}
J.B.\
Pendry, D.\ Schurig, and D.R.\ Smith, Controlling Electromagnetic
Fields,
{\it Science}  {\bf 312} (2006), 1780 -
1782.



\bibitem{Rahm} M. Rahm, \etal, Optical Design of
Reflectionless Complex Media by Finite Embedded Coordinate Transformations,
{\it Phys.
Rev. Lett.} {\bf100} (2008), 063903.



\bibitem{ReedSimon1}
M. Reed and B. Simon, {\it Methods of modern mathematical physics. I.}
Academic Press, New York, 1980. xv+400 pp.



\bibitem{Re} T. Regge, Introduction to complex orbital method, {\it Nuouvo
Cimento} {\bf 14} (1959), 951--976.



\bibitem{RYNQ} Z. Ruan, M. Yan, C. Neff and M. Qiu,
Ideal cylindrical  cloak: perfect but sensitive to tiny perturbations,
{\it Phys. Rev. Lett.} {\bf 99} (2007), 113903.


\bibitem{Sab} P. Sabatier, Asymptotic properties of the potentials in the
inverse-scattering problem at fixed energy, {\it J. Math. Phys.} {\bf 7}
(1966), 1515--1531.



\bibitem{simon} B. Simon, Maximal and minimal Schr\"odinger forms,
{\it J. Operator Theory} {\bf 1} (1979), 37-47.



\bibitem{SuU} Z. Sun and G. Uhlmann, Generic uniqueness for an inverse
boundary value
problem,
{\it Duke Math. Journal} {\bf 62} (1991), 131--155.



\bibitem{SyU} J. Sylvester and G. Uhlmann, A global uniqueness theorem for
an
inverse boundary value problem, Ann. Math. {\bf 125} (1987), 153--169.



\bibitem{ward} A.\ Ward and J.\ Pendry, Refraction and geometry in Maxwell's
equations, {\it Jour. Modern Opt.}, {\bf 43}  (1996), 773--793.


\bibitem{Weder} R. Weder,  A rigorous analysis of
high-order electromagnetic invisibility cloaks,
{\it Jour.  Phys. A: Math. and Theor.}, {\bf  41}  (2008), 065207.

\bibitem{Zhang} S. Zhang, D. Genov, C. Sun and X. Zhang, Cloaking of matter
waves,
{\it Phys. Rev. Lett.} {\bf 100} (2008), 123002.


\end{thebibliography}

{\small 
\noindent
Allan Greenleaf, Department of Mathematics,
University of Rochester, Roch\-es\-ter, NY 14627,USA,
{\tt{allan@math.rochester.edu}}

\noindent
Yaroslav Kurylev,
Department of Mathematical Sciences, 
University College,   London, WC1E 6BT, UK,
{\tt{y.kurylev@ucl.ac.uk}}

\noindent
Matti Lassas, Institute of Mathematics,
Helsinki University of Technology,  FIN-02015, Finland,
{\tt{mjlassas@math.tkk.fi}}

\noindent
Gunther Uhlmann, Department of Mathematics,
 University of Washington, Seattle, WA 98195,
{\tt{gunther@math.washington.edu}}

}

\end{document}